\documentclass[longauth]{aa}
\usepackage{graphicx}

\usepackage{threeparttable}
\usepackage{txfonts}

\usepackage{hyperref}
\usepackage{adjustbox}
\usepackage{rotating}
\usepackage{lscape}
\usepackage{txfonts}

\def \lsun{\ifmmode{{\rm\ L}_\odot}\else{${\rm\ L}_\odot $}\fi}

\def \msun{\ifmmode{{\rm\ M}_\odot}\else{${\rm\ M}_\odot$}\fi}
\def \zsun{\ifmmode{{\rm\ Z}_\odot}\else{${\rm\ Z}_\odot$}\fi}

\def \rsun{\ifmmode{{\rm\ R}_\odot}\else{${\rm\ R}_\odot$}\fi}

\def \mdot{\ifmmode{{\rm\dot{M}}}\else{${\rm\dot{M}}$}\fi}

\newcommand{\ha}{H$\alpha${}}

\newcommand{\hb}{H$\beta${}}

\newcommand{\hii}{H\,{\sc ii}{}}

\begin{document} 

   \title{Optical and near-infrared photometry of 94 type~II supernovae from the Carnegie Supernova Project}
   
   \author{J.~P. Anderson\inst{1}\orcid{0000-0003-0227-3451}, 
   C. Contreras\inst{2}\orcid{0000-0001-6293-9062}, 
   M.~D. Stritzinger\inst{3}\orcid{0000-0002-5571-1833}, 
   M. Hamuy\inst{4}, 
   M.~M. Phillips\inst{2}\orcid{0000-0003-2734-0796}, 
    N.~B. Suntzeff\inst{5}\orcid{0000-0002-8102-181X},
   N. Morrell\inst{2}\orcid{0000-0003-2535-3091}, 
   S. Gonz\'alez-Gait\'an\inst{6}\orcid{0000-0001-9541-0317},
   C.~P. Guti\'errez\inst{7,8}\orcid{0000-0003-2375-2064}, 
   C.~R. Burns\inst{9}\orcid{0000-0003-4625-6629}, 
   E.~Y. Hsiao\inst{10}\orcid{0000-0003-1039-2928}, 
   J. Anais\inst{2}\orcid{0000-0001-9051-1338}, 
   C. Ashall\inst{11}\orcid{0000-0002-5221-7557}, 
   C. Baltay\inst{12}, 
   E. Baron\inst{13,14,15}\orcid{0000-0001-5393-1608},
   M. Bersten\inst{16,17,18},
   L. Busta\inst{2}, 
   S. Castell\'on\inst{2,19}, 
   T. de Jaeger\inst{11}, 
   D. DePoy\inst{6},
   A.~V. Filippenko\inst{20}\orcid{0000-0003-3460-0103}, 
   G. Folatelli\inst{16,17}\orcid{0000-0001-5247-1486}, 
   F. F\"orster\inst{21,22,23,24}, 
   L. Galbany\inst{8,7}\orcid{0000-0002-1296-6887}, 
   C. Gall\inst{25}\orcid{0000-0002-8526-3963}, 
   A. Goobar\inst{26},  
   C. Gonzalez\inst{2}\orcid{0000-0001-5179-980X},  
   E. Hadjiyska\inst{12},  
   P. Hoeflich\inst{10}\orcid{0000-0002-4338-6586}, 
   K. Krisciunas\inst{5}\orcid{0000-0002-6650-694X}, 
   W. Krzemi\'nski\inst{2}$^,$\thanks{Deceased}, W. Li\inst{18}$^{,\star}$, 
   B. Madore\inst{9,27}, 
   J. Marshall\inst{5}, 
   L. Martinez\inst{16,17}\orcid{0000-0003-0766-2798}, 
   P. Nugent\inst{28,20},
   P.~J. Pessi\inst{29}\orcid{0000-0002-8041-8559},
   A.~L. Piro\inst{9}, 
   J-P. Rheault\inst{5}, 
   S. Ryder\inst{30,31}\orcid{0000-0003-4501-8100}, 
   J. Ser\'on\inst{32}, 
   B.~J. Shappee\inst{11}\orcid{0000-0003-4631-1149},
   F. Taddia\inst{3}\orcid{0000-0002-2387-6801}, 
   S. Torres\inst{33}\orcid{0000-0002-2726-6971}, 
   J. Thomas-Osip\inst{34}\orcid{0000-0003-1033-4402}, 
   S. Uddin\inst{5}}

   \institute{European Southern Observatory, Alonso de C\'ordova 3107, Casilla 19, Santiago, Chile \email{janderso@eso.org}
         \and Carnegie Observatories, Las Campanas Observatory, Casilla 601, La Serena, Chile
         \and 
         Department of Physics and Astronomy, Aarhus University, Ny Munkegade 120, DK-8000 Aarhus C, Denmark
         \and 
         "Fundación Chilena de Astronomía", El Vergel 2252 \#1501, Santiago, Chile
                 \and George P. and Cynthia Woods Mitchell Institute for
  Fundamental Physics and Astronomy, Department of Physics and Astronomy,
  Texas A\&M University, College Station, TX 77843, USA
        \and 
        Centre for Astrophysics and Gravitation, Instituto Superior T\'ecnico, University of Lisbon, Av. Rovisco Pais 1, 1049-001, Lisbon, Portugal 
                \and Institut d’Estudis Espacials de Catalunya (IEEC), Gran Capit\`a, 2-4,
     Edifici Nexus, Desp. 201, E-08034 Barcelona, Spain
                \and Institute of Space Sciences (ICE, CSIC), Campus UAB, Carrer de Can Magrans, s/n, E-08193 Barcelona, Spain
          \and The Observatories of the Carnegie Institution for Science, 813 Santa Barbara St., Pasadena, CA 91101, USA
         \and Department of Physics, Florida State University, Tallahassee, FL 32306, USA
         \and Institute for Astronomy, University of Hawaii, 2680 Woodlawn Drive, Honolulu, HI 96822, USA
         \and Department of Physics, Yale University, 217 Prospect Street, New Haven, CT 06511, USA
         \and Planetary Science Institute, 1700 East Fort Lowell Road, Suite 106, Tucson, AZ 85719-2395 USA
         \and Hamburger Sternwarte, Gojenbergsweg 112, 21029 Hamburg, Germany
         \and Homer L.~Dodge Department of Physics and Astronomy, University of Oklahoma, Rm 100 440 W. Brooks, Norman, OK 73019-2061
         \and Instituto de Astrof\'isica de La Plata (IALP), CCT-CONICET-UNLP. Paseo del Bosque S/N, B1900FWA, La Plata, Argentina        
         \and Facultad de Ciencias Astronomicas y Geofisicas, Universidad Nacional de La Plata, Instituto de Astrofisica de La Plata (IALP), CONICET, Paseo del Bosque SN, B1900FWA La Plata, Argentina 
         \and Kavli Institute for the Physics and Mathematics of the Universe (WPI), The University of Tokyo, 5-1-5 Kashiwanoha, Kashiwa, Chiba 277-8583, Japan
         \and Observatoire du Mont-Mégantic,
        Départment de Physique,
        Université de Montréal
         \and Department of Astronomy, University of California, Berkeley, CA 94720-3411, USA
        \and Data and Artificial Intelligence Initiative (D$\&$IA), University of Chile
        \and Millennium Institute of Astrophysics (MAS), Nuncio Monseñor Sótero Sanz, Providencia, Santiago, Chile
         \and  Center for Mathematical Modeling, University of Chile, AFB170001, Chile
         \and Departamento de Astronomía, Universidad de Chile, Casilla 36D, Santiago, Chile
         \and DARK, Niels Bohr Institute, University of Copenhagen, Jagtvej 128, 2200 Copenhagen, Denmark 
      \and The Oskar Klein Centre, Department of Physics, Stockholm University, SE-106 91 Stockholm,
         Sweden
        \and Infrared Processing and Analysis Center, Caltech/Jet Propulsion Laboratory, Pasadena, CA 91125, USA
        \and Lawrence Berkeley National Laboratory, Department of Physics, 1 Cyclotron Road, Berkeley, CA 94720, USA
        \and The Oskar Klein Centre, Department of Astronomy, Stockholm University, AlbaNova SE-10691 Stockholm,
         Sweden 
         \and School of Mathematical and Physical Sciences, Macquarie University, NSW 2109, Australia
        \and Astrophysics and Space Technologies Research Centre, Macquarie University, Sydney, NSW 2109, Australia
        \and Cerro Tololo Inter-American Observatory/NSFs NOIRLab, Casilla 603, La Serena, Chile
        \and NOIRLab, Avda Juan Cisternas 1500, La Serena, Chile
        \and Gemini Observatory/NSF’s NOIRLab, Casilla 603, La Serena, Chile}

\titlerunning{CSP SN~II photometry}
\authorrunning{Anderson et al.}

\date{}

 
  \abstract
   {Type II supernovae (SNe~II) mark the endpoint in the lives of hydrogen-rich massive stars. Their large explosion energies and luminosities allow us to measure distances, metallicities, and star formation rates into the distant Universe. To fully exploit their use in answering different astrophysical problems, high-quality low-redshift data sets are required. Such samples are vital to understand the physics of SNe~II, but also to serve as calibrators for distinct -- and often lower-quality -- samples.} 
   {We present $uBgVri$ optical and $YJH$ near-infrared (NIR) photometry for 94 low-redshift SNe~II observed by the Carnegie Supernova Project (CSP). A total of 9817 optical and 1872 NIR photometric data points are released, leading to a sample of high-quality SN~II light curves during the first $\sim$150 days post explosion on a well-calibrated photometric system.}
   {The sample is presented and its properties are analysed and discussed through comparison to literature events. We also focus on individual SNe~II as examples of classically defined subtypes and outlier objects. Making a cut in the plateau decline rate of our sample ($s_2$), a new subsample of fast-declining SNe~II is presented.}
   {The sample has a median redshift of 0.015, with the nearest event at 0.001 and the most distant at 0.07. At optical wavelengths ($V$), the sample has a median cadence of 4.7 days over the course of a median coverage of 80 days. In the NIR ($J$), the median cadence is 7.2 days over the course of 59 days.
   The fast-declining subsample is more luminous than the full sample and shows shorter plateau phases. Of the non-standard SNe~II highlighted, SN~2009A particularly stands out with a steeply declining then rising light curve, together with what appears to be two superimposed P-Cygni profiles of \ha\ in its spectra. 
   We outline the significant utility of these data, and finally provide an outlook of future SN~II science.}
   {}


   \maketitle
%
\section{Introduction} 
\nolinenumbers
\label{intro}
Supernovae  displaying strong, long-lasting, and broad hydrogen features in their visual-wavelength spectra are known as type II supernovae (SNe~II)\footnote{Throughout this paper we use `SNe~II' to refer to all hydrogen-rich explosions historically referred to as `IIP' or `IIL'.}. 
These events are the most common SNe in the local Universe \citep{li11}, and are believed to result from the core collapse of massive stars ($>$8-10\msun) that arrive to their deathbeds with a significant fraction of their hydrogen envelopes intact. The display of prevalent Balmer spectral lines defines their type II designation \citep{min41}, as opposed to type I events that do not show hydrogen features.
This presence of hydrogen, together with the slowly declining (plateau) nature of their light curves, was the basis for early conclusions of red-supergiant (RSG) progenitors, that is, explosions of massive stars with extended envelopes (e.g. \citealt{gra71,che76,fal77}). Such conclusions received significant support through the almost exclusively star-forming nature of the host galaxies of SNe~II (\citealt{maz76}, see, e.g. \citealt{hak12a} for a modern analysis), implying massive star progenitors.\\
\indent The high relative rates of SNe~II together with their intrinsic brightness mean that they are important objects for our overall understanding of the Universe. 
SNe~II contribute to the chemical enrichment of the interstellar medium, and their energetics
are thought to play a role in both galaxy evolution and in triggering subsequent episodes of star formation (SF).
In addition, given their luminosities, rates, and direct relationship to massive SF, SNe~II allow us to measure key astrophysical parameters across a large volume of the Universe. It is therefore important that we expand our knowledge of these explosive events, both to understand the
progenitors and explosion mechanisms as worthy knowledge in itself, and to bolster confidence in our use of SNe~II to understand the Universe at large.\\
\indent The direct detection of progenitor stars on pre-explosion imaging (e.g. \citealt{van03,sma04}; see \citealt{sma15} for a review), confirmed
earlier modelling conclusions that most SNe~II are the result of RSG star explosions.
While the number remains low, the statistical significance of the high-quality direct detections constrains the zero age main sequence (ZAMS) mass range to be between 8.5 and 18\msun\ \citep{sma15}.
This led to claims of a `RSG problem' \citep{sma09}, given that $>$18\msun\ RSG stars
are known to exist in the local Universe (e.g. \citealt{lev05}), and are expected to explode as SNe~II.
While a number of explanations for this `problem' have been discussed \citep{koc08a,yoo10a,wal12,sma15,dav18a,lim18a}, the statistical significance has also been questioned \citep{dav20a}.
In any case, there is now a general consensus that the majority of SNe~II are the result of RSG explosions. However, direct mapping 
of progenitor properties such as ZAMS mass, metallicity, rotation, and binarity
to the observed diversity of SN~II transient features is still lacking.\\
\indent SNe~II were historically separated into type IIP and IIL \citep{bar79}. The former plateau events show an almost constant luminosity 
lasting several months, while the latter `linear' explosions decline
significantly faster. However, recent large-sample analyses have argued against distinct classes (e.g. \citealt{and14a}, A14 henceforth). 
We therefore refrain from making such subjective separations for the majority of this paper, only doing so
when comparison to the literature is required. 
Another three further subclasses of SNe~II exist in the spectroscopically defined types IIn and IIb (see \citealt{fil97} for a review), and those events showing
similar light curves to the iconic SN~1987A. 
These events are somewhat distinct from the `normal' SN~II class we concentrate on here, and probably arise from somewhat different
progenitors. They are therefore not discussed in any further detail. However, updated photometry for Carnegie Supernova Project-I SNe~IIn and SN~1987A-like events is released here -- readers are referred to the Appendix for more details. (For an overview of core-collapse and hydrogen-rich SN observations and interpretations, the reader is referred to \citealt{arc17a,mod19a}.)\\
\indent Until recently, the majority of SN~II studies concentrated on individual events (see e.g. \citealt{ham01,leo02,pas06}), or on specific correlations for
their use as distance indicators (e.g. \citealt{ham02}). However, to understand the full diversity of events, and hence the full diversity
of progenitor and explosion properties, large samples are required.
A14 presented a characterisation of the $V$-band light curves of 117 SNe~II (a combination of the events from the current data
release and those from \citealt{gal16}). A range of observational parameters was defined and used to investigate correlations between SN~II luminosity, 
decline rates, and light-curve time scales. While some intriguing trends were observed, that work also highlighted the huge
diversity of the SN~II class, with significant variance in both SN brightness and overall light-curve morphology.
A number of other studies have arrived at similar conclusions \citep{psk67,you89,pat94,ric02,arc12,ber13,far14b,far14a,gon15,san15_2,val16,rub16a,rub16b,gal16,dej19a}.
Large diversity is also observed in the spectral properties of SNe~II (see, e.g. \citealt{pas05,ins13,and14b,spi14,gut14,far14b,far14a,gut17b,gut17a,dej19a,dav19a}), where trends are also observed between spectroscopic and photometric parameters.
Attempting to understand this diversity aids in furthering our understanding of
progenitor properties and explosion characteristics. Understanding SN~II diversity is also vital for the current and future
use of SNe~II as astrophysical probes.\\
\indent A number of methods have been developed to enable the use of SNe~II as distance indicators. These 
include the Expanding Photosphere Method  (EPM, \citealt{kir74,sch94_2,ham01_2}), the Standard Candle Method (SCM, \citealt{ham02}), the Spectral-fitting Expanding
Atmosphere Method (SEAM, \citealt{bar04,des08}), the Photospheric Magnitude Method (PMM, \citealt{rod14,rod19a}), and most recently the Photometric Colour Method (PCM, \citealt{dej15}). Recent examples of these methods applied  to samples of SNe~II located at cosmologically interesting distances are presented by \cite{gal16a}, \cite{dej17a}, \cite{gal18a}, \cite{dej17b}, and \cite{dej20a}.\\
\indent It has also been shown, first through modelling \citep{des13,des14} and later through observations \citep{and16},
that SNe~II can be used to measure metallicities. 
Being able to measure both distances and metallicities,
SNe~II show great potential to probe such quantities out to cosmological distances, hence
affording independent constraints on the evolution of the Universe.
This can only be achieved if well observed, nearby samples exist -- against which data can be compared and calibrated. The purpose of the current paper is to release such a data set for widespread use.\\
\indent Many individual photometric and spectral data sets of SNe~II have been published. Recent years have also seen the publication
of SN~II data samples. \cite{far14b}, \cite{far14a}, and later \cite{dej19a} released optical photometric and spectral data for a total of 90 SNe~II obtained by the Lick Observatory
Supernova Search (LOSS; \citealt{fil01}) programme.
\cite{gal16} published optical photometry for 51 events from a range of SN~II follow-up surveys, while \cite{hic17} published optical and near-infrared (NIR) light curves together with optical spectra for another 39 SNe~II from the Center for Astrophysics group. In the current paper, we present optical and NIR photometry of 74 SNe~II\footnote{The optical spectral sequences of CSP-I SNe~II were published by \cite{gut17a}.} obtained by the Carnegie Supernova Project-I (hereafter CSP-I, \citealt{ham06}), together with data from an additional  20 objects observed by the Carnegie Supernova Project-II (CSP-II, \citealt{phi19a}). With a total of 94 SNe~II, this represents a significant increase of data accessible to the community. This particularly holds true with the release of  our high-quality NIR photometry for the majority of the sample. \\
\indent The paper is organised as follows. 
We first discuss in Sect.~2 the different phases that are observed in SN~II visual-wavelength light curves during the first few hundred days post explosion. This sets the scene for our presentation and discussion of high-quality SN~II light curves in later sections.
Then, we give an overview of our sample, the data reduction and photometric
calibration methods, explosion epoch estimations and the sample properties in Sect.~3. Section~4 presents an example SN~II sequence of photometric reference stars within the field, example photometry and a light curve, together with the distributions of two key light curve morphology parameters: $s_2$ (the plateau decline) and $OPTd$ (`optically thick phase duration': the time from explosion to the end of the plateau). In Sect.~5 we discuss a number of individual events and present a new sample of `fast declining SNe~II'.
Finally, Sect.~6 provides conclusions and an outlook of future prospects in the study of SNe~II.\\

\section{SN~II light-curve morphology and physics}
\label{LCphysics}

\begin{figure*}
\includegraphics[width=19cm]{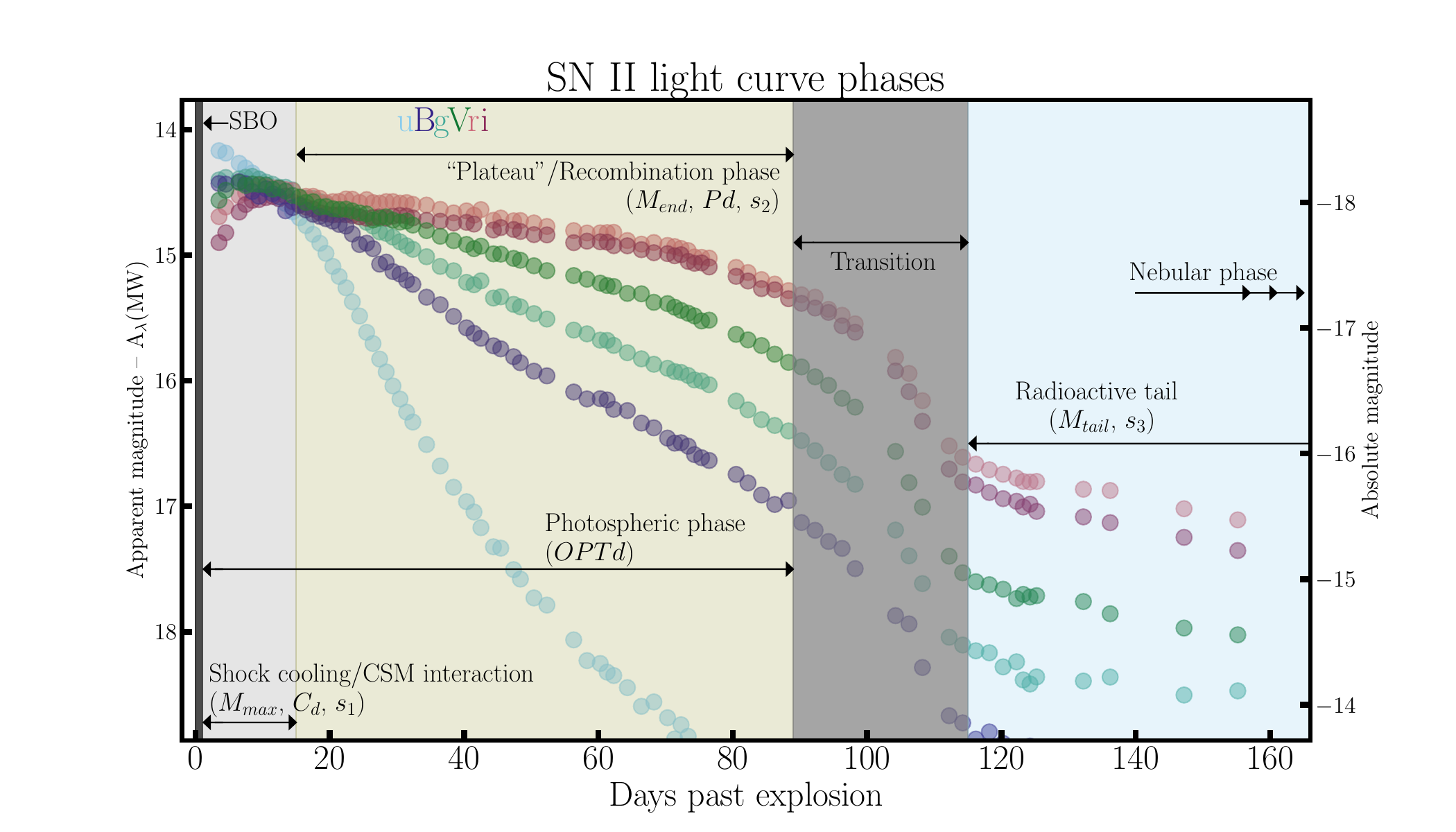}
\caption{Optical wavelength light curves of an example SN~II together with the labelling of the different light curve phases.
The distinct shaded regions are used to define where different physical processes dominate the luminosity of SNe~II (see details in text). The generic naming scheme used in the literature to label these phases is given, together with observational parameters used to characterise those phases in brackets (from A14 and \citealt{gut17b}). It should be noted that phase boundaries are not necessarily well defined -- they change between different SNe~II, and are shown here as distinct abrupt changes for visualisation purposes. (While `SBO' -- shock breakout -- is labelled on this plot and discussed in the text, it is not observed for this example SN~II, nor for any other event in our sample.)}
\label{LCmorph}
\end{figure*}

Standard SNe~II (and the vast majority of SNe showing long-lasting broad hydrogen spectral features) display a well-defined light-curve morphology that can be separated into different observational phases. These phases are understood to be the result of different physical processes affecting the light curve of a SN~II, dominating the observed luminosity at different times.
Figure~\ref{LCmorph} uses example photometry to depict the different light-curve phases of SNe~II, labelling them with the common literature nomenclature together with observational parameters used to characterise the nature of each phase (taken from A14 and \citealt{gut17b}). A summary of the evolution of a SN~II explosion is now given, outlining the physics of SN~II light-curve morphology displayed in Fig.~\ref{LCmorph}.\\
\indent It is generally agreed that massive stars end their lives through `core-collapse' (CC) -- leading to a SN explosion. Massive stars go through successive stages of nuclear fusion \citep{hoy54}, burning
heavier elements until an iron core forms. At this stage it is no longer energetically favourable to continue this burning sequence. The iron core grows until it reaches the Chandrasekhar mass and can no longer support itself against gravitational collapse. The core collapses until the inner core reaches nuclear densities. At such densities the strong nuclear force abruptly halts the collapse leading to a shock wave that propagates out through the still collapsing outer core. It is theoretically understood that this shock wave stalls, and a successful CC~SN only ensues if the shock can be revived. The currently favoured scenario to produce a successful explosion is the `delayed neutrino driven mechanism' (see e.g. \citealt{jan17}, \citealt{pej20}, \citealt{bur21a} and \citealt{fry22} for reviews of CC~SN theory - also see \citealt{sok19} for discussion of an alternative scenario). Once sufficient neutrino energy is captured, the shock wave is revitalised and can continue to propagate out through the rest of the star. From a physical perspective this can be considered the beginning of the SN ($t_0$). However, at this stage no electromagnetic radiation can be seen by the observer\footnote{Although neutrinos and gravitational waves may be observed at times before the first electromagnetic emission is seen.}.\\ 
\indent As the shock propagates outwards it accelerates and heats the stellar envelope. Once the shock arrives to the stellar surface, an intense flux of photons is able to leak out, producing the first electromagnetic radiation signal observed, known as the `shock break-out' (SBO; e.g. \citealt{che76,fal77,ens92}, see review by \citealt{wax17}). At SBO the SN undergoes a rapid increase in its luminosity. The duration of this transient feature is very short and depends on the size of the progenitor star. In the case of SNe~II -- with RSG progenitors -- it is thought to last up to a few hours. While this is longer than for other SNe (with more compact progenitors), it is still an observationally difficult timescale to detect, that requires an especially designed survey with very high observational cadence (we do not observe this phase in any event in our sample), with only a small number of claimed detections \citep{sod08,sch08,gez15,gar16,ber18,tom19,val21}. The time of SBO is shown on the extreme left of the light curve phase diagram in Fig.~\ref{LCmorph}. It is observable at all wavelengths but its brightness is much higher at bluer bands (however the SBO is expected to be smeared out in the presence of significant circumstellar material, CSM, see \citealt{mor11} and further discussion below). After this initial phase, rapid expansion leads to cooling of the ejecta.\\
\indent The second phase in SN~II light curves is often referred to as `shock cooling' and is shown in Fig.~\ref{LCmorph} to last more than 10 days. During shock cooling the peak in the optical light curves is observed. These peak luminosities are a consequence of the peak of the SN spectral energy distribution (SED) progressively moving to longer wavelengths (lower temperatures) -- thus SNe~II peak earlier in bluer bands (Fig.~\ref{LCmorph}, see also \citealt{gon15}). After light curve maximum ($M_{max}$), SNe~II show an initial decline in their brightness ($s_1$). This decline continues until the third phase of the light curve is reached -- the plateau.
The time between initial explosion and the end of this shock-cooling phase has been referred to as $C_d$ (the `cooling duration'; \citealt{gut17b}), and lasts several weeks.
Until recently, it was thought that SNe~II explode into a relatively clean environment, and thus their observed properties were not significantly affected by interaction between the ejecta and any surrounding material. However, there is now mounting evidence that many (and possibly the majority) of SNe~II undergo interaction between the ejecta and CSM or an extended envelope (e.g. \citealt{kha16,ter16,yar17,bru21}) that is sufficiently strong to affect their early time light curves \citep{gon15,mor17,for18}. Hence, in Fig.~\ref{LCmorph} it is noted that the shock-cooling phase may also be powered by CSM interaction (in which case it is expected to last significantly longer).\\
\indent At the end of the shock-cooling phase and the onset of the plateau, the outermost regions of the ejecta have reached temperatures of around 6000\,K and the initially ionised hydrogen envelope starts to recombine (thus the plateau is also referred to as the ‘recombination phase’; see \citealt{far19} for an analytic discussion), and the decline rate ($s_2$) starts to slow down considerably.
During this phase hydrogen is recombined at different layers of the SN~II ejecta. The recombination reduces the opacity -- dominated by electron scattering -- allowing the energy to diffuse outwards more easily. Recombination is the physical mechanism responsible for the escape of radiation. However, the energy released by the recombination process itself is only a small fraction of the radiated energy.
The radiated energy is dominated by the energy stored in the SN ejecta initially deposited by the shock wave during its propagation through the stellar interior.
Additionally, the late-plateau phase can be partially powered by the emission from the radioactive decay of $^{56}$Co. The exact moment when this contribution starts to affect SN~II light curves depends on the degree of mixing of the radioactive material into the ejecta (see \citealt{ber11,koz19}).\\
\indent At this stage in the description a note is required. The plateau phase is often referred to in the literature as `a phase of almost constant luminosity...'. However, a minority of SNe~II show such behaviour (see detailed discussion in A14 and Sect.~\ref{diss}). Instead, SNe~II show a large range in their plateau decline rates ($s_2$) -- some even increasing in optical luminosity (e.g. SN~2006bc, as will be shown in Fig.~\ref{06bclcspec}), and others declining by up to several optical magnitudes before the end of the plateau (see Fig.~\ref{07ablcspec}, later in the paper). Thus, it is more accurate to refer to the plateau phase ($P_d$ in Fig.~\ref{LCmorph}) of SN~II evolution as `a phase of slowly changing luminosity relative to previous and subsequent phases, that lasts several months'. (Indeed, the example shown in Fig.~\ref{LCmorph} displays a clear plateau that declines relatively quickly.) This plateau is more prominent in the redder optical bands (Fig.~\ref{LCmorph}), and in some cases is observed as a slight increase in brightness (e.g Fig.~\ref{06bclcspec}).\\
\indent As a SN~II evolves through its plateau 
the hydrogen recombines within increasingly inner layers of the envelope, and the photosphere moves inwards in mass coordinates following the recombination wave. At the end of the plateau (where one may also characterise the SN~II through its magnitude $M_{end}$) the recombination wave has travelled to the inner edge of the hydrogen envelope. When this happens a sharp drop is observed in the light curve. This starts the fourth phase in Fig.~\ref{LCmorph} -- the `transition', where the powering mechanism transitions from being dominated by shock energy released by the recombination process to radioactivity on the tail. (The total phase from explosion to the end of the plateau is also referred to as the photospheric phase, defined as $OPTd$ in A14\footnote{To measure $OPTd$, the `end of the plateau' is defined as when the extrapolation of the $s_2$ decline-rate straight line becomes more than 0.1 magnitudes brighter than the light curve (measured here in the $V$-band).}.) This transition from a plateau to tail phase is observed in nearly all SNe~II with sufficiently well observed light curves (see discussion in A14 and \citealt{val15}). Thus, there are very few (if any) true fast-declining SNe~II that show light curves following the initial IIL description defined by \cite{bar79}. Even the classic prototypes, SN~1979C \citep{dev81} and SN~1980K \citep{but82,bar82}, show evidence for a transition phase (see fig.\,18 in A14).\\
\indent During the last phase shown in Fig.~\ref{LCmorph}, radioactive decay of $^{56}$Co (as the daughter product of $^{56}$Ni produced in the explosion) is the dominant power source\footnote{The half-life of $^{56}$Co is 77.2 days and the half-life of $^{56}$Ni is 6.1 days.}. Emission from radioactive decay is trapped by the ejecta and thus the $s_3$ tail follows the decline rate expected by the decay rate of $^{56}$Co (at least in the majority of cases, although see A14 and \citealt{mar22a}). During this phase the luminosity of a SN~II (characterised by $M_{tail}$) is a direct indicator of the $^{56}$Ni mass ejected by the SN. In addition to changes in the light-curve shape, spectral properties also significantly transition as the continuum decreases in strength and the dominant features change (see \citealt{gut17a} for further details). During the radioactive tail a SN~II gradually enters a `nebular phase' where the inner core of the ejecta is revealed, the continuum becomes very weak, and spectra become dominated by emission lines.\\
\indent At much later times, radioactivity from other elements starts to dominate the observed luminosity (e.g. $^{44}$Ti at around 1500 days post explosion, \citealt{sun92}), and additionally dust formation \citep{mat15} may produce a change in SN colour with NIR excesses observed (typically observed several hundred days post explosion). However, our data set does not extend to such late epochs and hence we end our light curve phase discussion here. In the next section the properties of our SN~II photometric data sample are outlined.

\section{The CSP SN~II sample} 
\label{data}
\subsection{The Carnegie Supernova Project: Aims and follow-up criteria}
The data presented in this paper were acquired through the CSP over two distinct observational SN follow-up programmes. CSP-I \citep{ham06} operated between 2004 and 2009 and obtained data for 74 SNe~II.
The overall aim of CSP-I was to obtain a foundational dataset of optical and NIR light curves of all types of SNe in a well-defined and understood photometric system\footnote{The release of the current dataset and those of other SN types published in previous works means that the CSP has attained its aim of establishing a fundamental data set of SNe. Data releases for CSP-I SNe~Ia can be found in \cite{kri17a}, Stripped-Envelope events (SE-SNe: type IIb, Ib and Ic) in \cite{str18a}, SNe~IIn in \cite{tad13a}, and data for two 1987A-like events in \cite{tad12a}.} \citep{ham06}.
CSP-I SN follow-up was initiated on SNe discovered before or near to the epoch of maximum brightness, that occurred at low redshifts (i.e. $z <  0.08$), at declinations of less than 20$^{\circ}$ (in order to be observed from the Las Campanas Observatory, LCO), and that were estimated to reach an apparent optical magnitude of 18 or lower. 
Once follow-up was initiated, optical and NIR photometry together with visual-wavelength spectroscopy were obtained with relatively high cadence (see Sect.~\ref{prop_sample}) until each SN was no longer sufficiently bright to continue observations, or it went behind the Sun.\\ 
\indent CSP-II \citep{phi19a} observed between 2011 and 2015 with a focus on obtaining optical and NIR photometry for SNe~Ia in the smooth Hubble flow. Thus, it was not a major aim of CSP-II to observe SNe~II. However, when nearby well-positioned (on the sky) SNe~II were discovered, CSP-II obtained data for 20 such events. A new focus in CSP-II was to obtain NIR SN spectroscopic sequences for a statistically meaningful sample of SNe~Ia \citep{hsi19a}.
NIR spectral sequences were also obtained for a number of SNe~II and were analysed and discussed by \cite{dav19a}: we display one such sequence in Sect.~\ref{diss} (Fig.~\ref{15bblcspec}). CSP-II photometric follow-up 
of suspected SNe~Ia in the Hubble flow often started before spectroscopic typing. Therefore, in a few cases relatively high-redshift SNe~II were observed (see Fig.~\ref{zdist}), but follow-up was stopped once their nature was confirmed leading to only a small amount of data per SN (e.g. for LSQ11gw and LSQ12fui). However, the majority of CSP-II SNe~II are at lower redshift where high-quality optical and NIR light curves were obtained and in some cases NIR spectral sequences.\\
\indent Details of the combined CSP-I and CSP-II SN~II samples, their discovery and classification, and additional information can be found in a Table on Zenodo\footnote{\url{https://zenodo.org/}} (see the link in the data availability section). 
Discussion of the properties of the SN~II sample is presented in Sect.~\ref{prop_sample}.
Now we 
summarise how CSP SN~II optical and NIR photometry were obtained, reduced, and calibrated.

\subsection{Data: Observations and processing}
All photometric data presented here were obtained at LCO.
The CSP instrumental set-up, SN and standard-star imaging acquisition, data reduction, and photometric calibration has been extensively outlined and analysed elsewhere (see below). We therefore do not repeat much of the detail in this paper. Here, we briefly describe all of the important concepts, but refer the reader to the relevant previous CSP data-release papers for specific details.\\
\indent All CSP-I SN~II optical imaging ($uBgVri$) was obtained with the 1\,m LCO Swope telescope and the LCO Ir\'en\'ee du Pont (`du Pont') 2.5\,m telescope. 
Most optical imaging was obtained using the direct CCD camera `SITe3' mounted on the Swope, with a small number of images obtained using the facility `Tek5' CCD camera mounted on the du Pont. The majority of host-galaxy template images (obtained once each SN had faded beyond detection) were obtained with the du Pont. NIR SN images ($YJH$) were mainly obtained with `RetroCam', a NIR imager built specifically for the CSP-I and mounted on the Swope telescope. A small number of NIR images were obtained with the Wide Field IR Camera (WIRC) on the du Pont telescope. All host-galaxy template imaging in the NIR was obtained using WIRC.
\cite{ham06} outlined the CSP-I instrumental setup in detail, while further presentation and discussion of system response functions can be found in \cite{kri17a} and \cite{str18a}. Following \cite{str18a}, we note that the $V$-band filter was changed for the Swope+SITe3 system during the course of CSP observations (on January 25$^{th}$ 2006), however the same colour term was shown to be applicable to both filters. Meanwhile, two different $J$-band filters were used on the Swope+RetroCam system: $J_{RC1}$ before January 15$^{th}$ 2009, and $J_{RC2}$ afterward. The derived colour terms for these filters were found to be sufficiently distinct to warrant publication of SN photometry on two different systems.\\
\indent The systems used to obtain photometry for CSP-II were very similar to CSP-I, but with a few important differences that are now outlined (see \citealt{phi19a} for a full description). During the first two observing campaigns (yearly observing seasons) of CSP-II, the same Swope+SITe3 system was used to obtain optical photometry as during CSP-I. Then, for the third and fourth campaigns the detector was replaced by a new e2V deep-depletion CCD (see \citealt{phi19a} for a comparison of the system throughputs in combination with the employed filters). CSP-II also used the du Pont+Tek5 system to obtain a small number of follow-up images, together with the optical host-galaxy reference images. At the start of CSP-II, the NIR imager RetroCam was moved from the Swope to the du Pont, and thus CSP-II, NIR imaging was obtained using du Pont+RetroCam. In addition, during CSP-II NIR host-galaxy template images were obtained mostly using FourStar \citep{per13} on the \textit{Magellan} Baade telescope, with a small number obtained using du Pont+RetroCam.\\
\indent The full CSP data workflow is detailed by \cite{con10} and \cite{str11} (the first and second SN~Ia data releases, respectively).
All images (science, standard star, host-galaxy templates) were processed in the standard manner, including bias subtractions, flat-field corrections, linearity correction, and correction for a shutter time delay to the exposure time. Host-galaxy template images -- obtained when the SN had faded from detection and in excellent observing conditions -- were used to subtract the contribution of galaxy light from the SN. Details of this process and a discussion of possible systematic errors involved can be found in \cite{con10} and \cite{fol10}. Photometry was then computed for local sequence stars within the field of view of each SN image (in the optical or NIR), and these sequences were calibrated with respect to standard-star fields observed on at least three photometric nights. This strategy and procedure is outlined by \cite{kri17a}. Photometry of local sequence stars is on the standard systems of \cite{smi02a} for $ugri$, \cite{lan92a} for $BV$, and \cite{per98a} for $JH$, while $Y$ is calibrated to the system defined in \cite{kri17a}.\\
\indent All CSP definitive photometry has been and will be released in the natural system of the telescope and instrumental setup of the observations. This is done to avoid major issues in using colour terms derived from stars to transform SN photometry into a standard photometric system. To achieve the former, local sequence photometry is transformed to the natural system using the formulation presented by \cite{kri17a}. Definitive natural-system point spread function photometry for SNe~II is then derived with respect to these natural local sequences. Final photometric uncertainties are estimated by combining the instrumental magnitude error of each individual SN photometric point with that on the nightly zero point. Our published photometric errors do not therefore include systematic uncertainties such as those arising from host galaxy flux subtraction.
The reader is again referred to \cite{con10} and \cite{kri17a} for a full description of the origin of photometric uncertainties and analyses of the accuracy and precision of CSP photometry.\\
\indent Photometry for the full, final optical and NIR light curves of 94 SN~II from the CSP, together with photometry of local sequences is discussed in the Sect.~\ref{fulllcs}. We first outline how we estimate explosion epochs for our sample in the next subsection. Then, we present general properties of our SNe~II that elucidate the nature of our sample, and additionally summarise previous analyses achieved with these SN~II data sets, before presenting our SN~II photometry in Sect.~\ref{fulllcs}.\\ 

\subsection{SN explosion epochs}
A key parameter to analyse any sample of SNe is the assumed date of explosion. This is important for (e.g.) SN modelling and extraction of physical parameters of their explosions, and for the use of SNe~II to measure distances. To estimate explosion epochs for our SNe~II we follow the same methodology as \cite{gut17a} and apply this to both the CSP-I and CSP-II samples. \citealt{gut17a} already estimated explosion epochs for most of the CSP-I sample. We re-check those values in this work and the majority of epochs published here are consistent with that previous work (see Table on Zenodo through the link given on the data availability section). If a constraining nondetection exists\footnote{For a nondetection to be constraining it must be deeper than the discovery photometry. This would appear obvious, however it is not unusual for SN nondetections to be published that are shallower than the discovery magnitude.} less than 20 days before discovery then this is used to estimate the time of explosion. The explosion epoch from this `N', nondetection method is then taken to be the mid point between the nondetection and the first detection, with the error being that time range divided by two.\\ 
\indent It is important to note that the uncertainties presented on our `N' explosion epochs are not Gaussian. Without any additional knowledge of (for example) the photometric rise to maximum or the exact magnitudes of the nondetection limit and the discovery detection, such errors should be understood to mean that there is an equal probability that the SN exploded at any time $t$ between the nondetection and the first detection (i.e. the explosion time\,$\pm$\,the error). However, for each individual SN~II, one does have additional information, the most relevant of which (for the current sample) is the difference in magnitude between the nondetection and the first detection. If the nondetection limit is close in magnitude to the first detection, then it is not particularly constraining, leaving the possibility that the SN exploded close in time or even before the nondetection. The deeper the nondetection the more constraining it becomes that the event exploded after that limit. While we do not use such information to more precisely constrain explosion epochs in this work, details of this photometry are listed in the Table available on Zenodo (see the link in the data availability section) so that the reader can assess each specific SN~II constraint on the merits of its data.\\
\indent In the case where no constraining nondetection exists (i.e. any such data are more than 20 days before the first detection), then spectral matching (`S') is used to constrain explosion epochs. Spectral matching refers to the technique of comparing a spectrum or a spectral sequence of a given SN to those from a library of template spectra of observed SNe with well-constrained explosion epochs. Several tools have been developed to perform this task, and here we use the Supernova Identification (SNID) code \citep{blo07a}. For a full overview of our use of SNID please refer to section 4 of \cite{gut17a}.\\
\indent Similarly to the nondetection method, there are also caveats to the use of spectral matching to obtain accurate explosion epochs. Libraries of spectral templates are often lacking in spectra of peculiar objects meaning that constraints for any such objects may be less reliable. \cite{gut17a} compared the differences between explosion epochs estimated through the nondetection (N) and spectral matching (S) techniques for objects with good constraints from both. Good agreement was found, with a mean absolute error between techniques of less than four days and a mean offset of 0.5 days (the latter indicates that there is no systematic offset between the techniques in one particular direction). For a subset of the SNe~II presented in this work, the explosion epochs from the above techniques were compared to those obtained from hydrodynamical modelling of their bolometric light curves and spectral velocities in \cite{mar22b}. The mean offset between our `observational' explosion epochs and those estimated through modelling is close to zero, indicating that there is no systematic offset between the different methodologies\footnote{While the prior in the explosion epoch modelling of \cite{mar22b} was defined to be within the observational explosion limits, it does not follow that the mean offset has to be close to zero. Additionally, in \cite{mar20} such priors were relaxed (allowing explosion epochs outside the observational limits) and negligible changes were observed in the resulting explosion epochs.}. Following the above, while for any specific SN there may be some peculiarities in its data (photometry, spectra) meaning that its explosion epoch is somewhat in error, no systematic issues have been found in applying our employed explosion epoch estimation methodology.

\subsection{Properties of the CSP SN~II photometry sample}
\label{prop_sample}
\begin{figure}
\includegraphics[width=8.5cm]{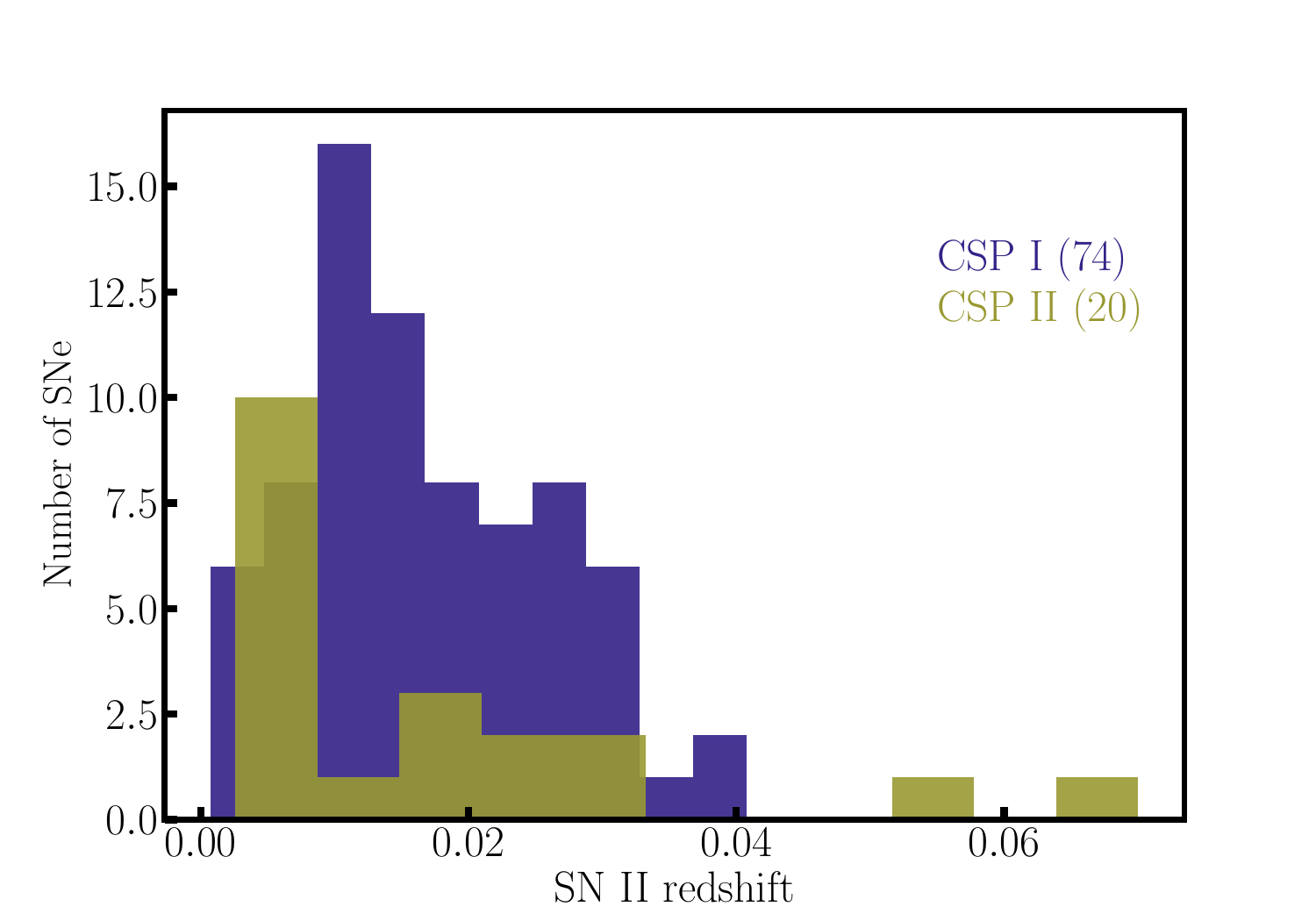}
\caption{Redshift distributions of the CSP-I and CSP-II SN~II samples. The number of events in each subsample is given in brackets.}
\label{zdist}
\end{figure}
\begin{figure}
\includegraphics[width=8.5cm]{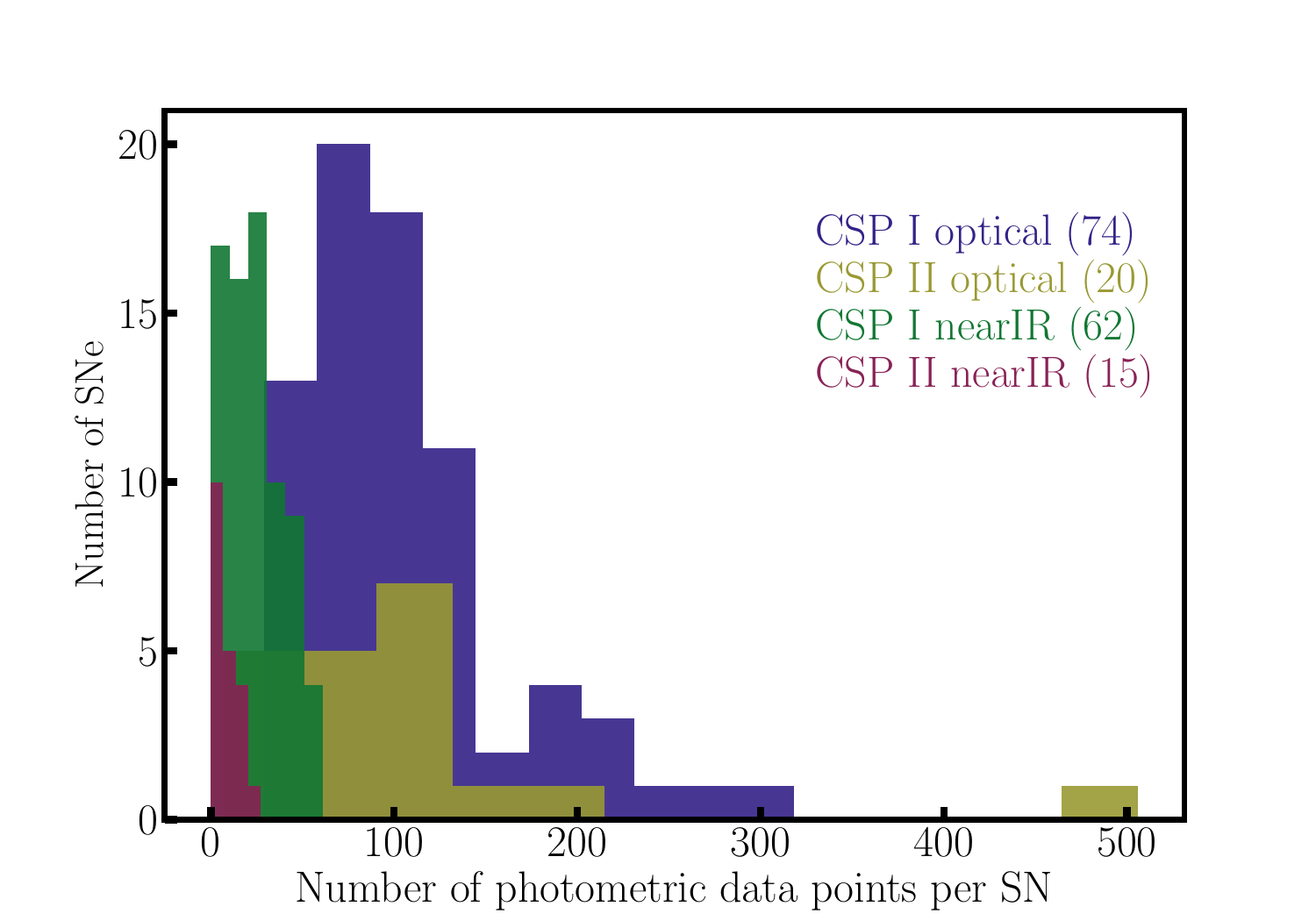}
\caption{Number of optical ($uBgVri$) and NIR ($YJH$) photometric data points per SN for the CSP-I and CSP-II SN~II samples. The number of events in each subsample is given in brackets.}
\label{Npointsopt}
\end{figure}

\begin{figure}
\includegraphics[width=8.5cm]{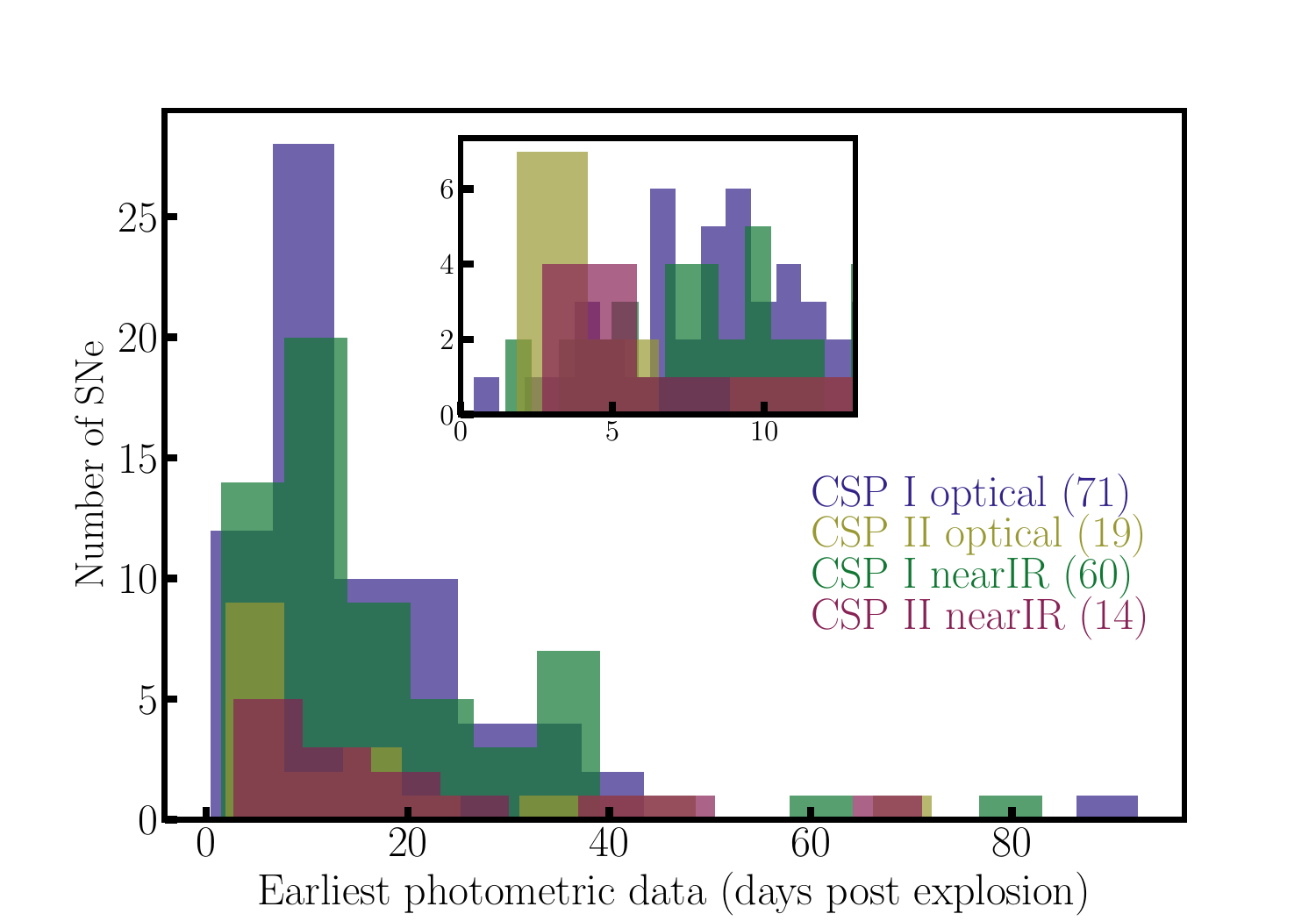}
\caption{Distribution of the epochs of the earliest optical and NIR photometry obtained (with respect to explosion epoch) for each SN in our sample. The number of events in each subsample is given in brackets. The inset shows a zoom into the first 13 days post explosion. }
\label{earlydist}
\end{figure}

\begin{figure}
\includegraphics[width=8.5cm]{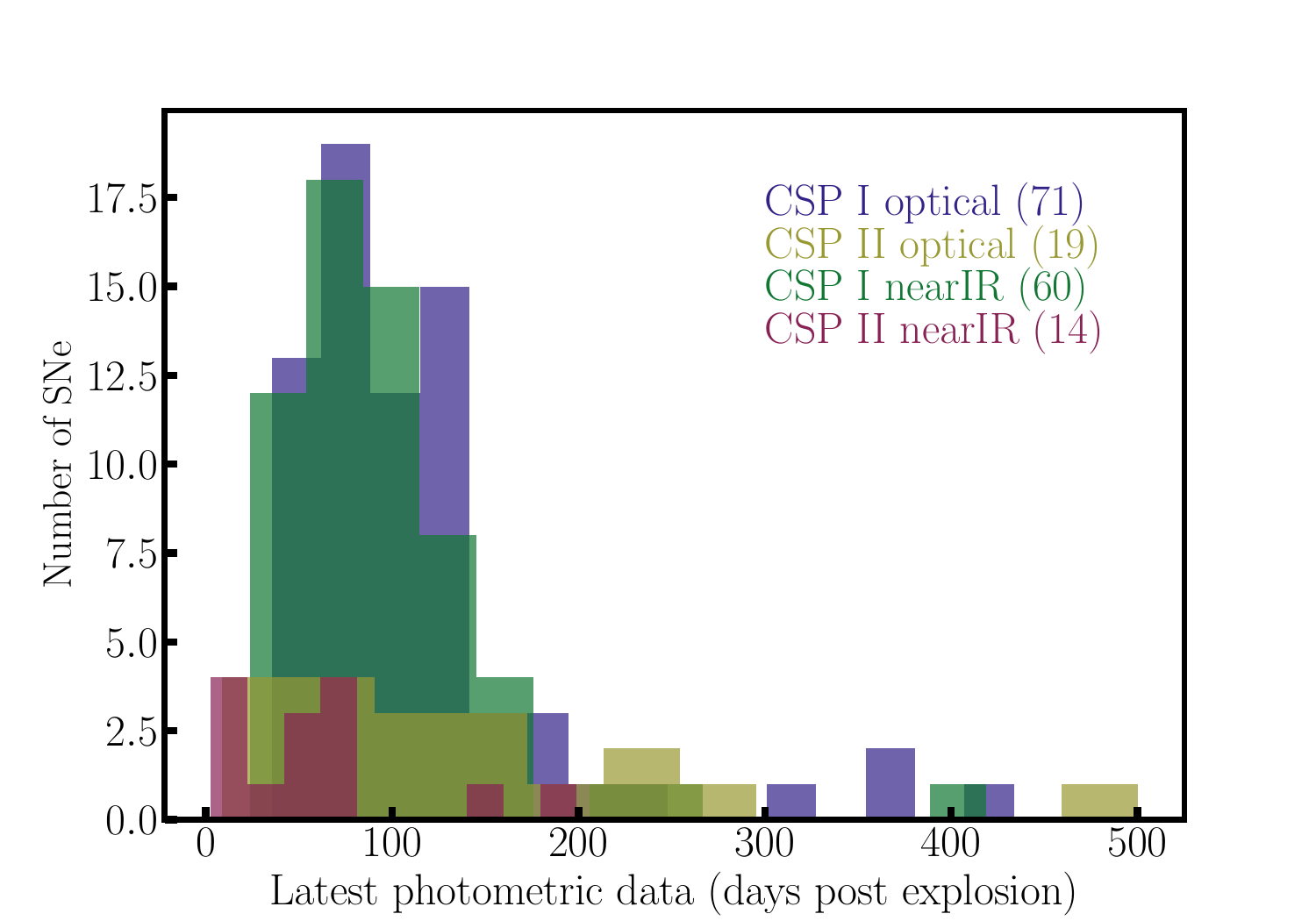}
\caption{Distribution of the epochs of the latest optical and NIR photometry obtained (with respect to explosion epoch) for each SN in our sample. The number of events in each subsample is given in brackets.}
\label{latedist}
\end{figure}

Information on the SNe~II contained within this CSP data release, discovery and classification references, together with the number of data per SN are given in a Table on Zenodo (see the link in the data availability section). 
As outlined above, a SN~II was considered for follow-up by both CSP campaigns if it was discovered before or close to maximum light and if it was bright enough for high-quality data to be acquired. This defines the sample as magnitude limited, while SN discoveries were provided by a large number of different surveys. 
SNe~II observed by CSP-I were mostly discovered by galaxy-targeted surveys (88\%), as during the 2004--2009 discovery epoch most nearby SNe were still discovered by such programmes (see appendix of A14). This led to a sample with very few SNe~II in low-luminosity galaxies (see fig.\,22 of A14). 
SN type and property distributions can change significantly when found in different luminosity galaxies (see, e.g. \citealt{arc10a}). Indeed,
SNe~II found within lower luminosity hosts (i.e. lower than the distribution of the A14/CSP-I sample) show intriguing differences compared with those events found in brighter galaxies (see, e.g. \citealt{tad16a,gut18a,and18a,sco19a}). Such differences are possibly linked to the effects on stellar evolution of SNe~II arising from lower metallicity progenitors (implied from the lower luminosity of their hosts). CSP-II meanwhile was executed when a number of large-field-of-view surveys had come on line, that scanned the sky in an untargeted fashion. The CSP-II sample contains only 40\%\ of objects discovered by targeted surveys. SN~2015bs -- part of the CSP-II sample -- is one clear example of a SN~II that would have been missed by galaxy-targeted surveys given its extreme low-luminosity host. This object had very weak metal lines in its optical-wavelength spectra -- indicative of a low-metallicity progenitor, together with very strong oxygen lines at nebular times -- indicative of a relatively massive progenitor \citep{and18a}.
It is important to keep the above sample selection issues in mind when comparing the current sample to other SN~II data sets in the literature.\\
\indent In Fig.~\ref{zdist} we plot the redshift distribution of the sample, separating events into CSP-I and CSP-II. 
The full sample has a median redshift of 0.015 with a maximum of 0.07 (LSQ12fxu, that was initially observed as part of the `SN~Ia cosmology sample' for CSP-II), and
a minimum of 0.00075 (SN~2008bk). As can be seen in Fig.~\ref{zdist}, apart from a few higher redshift events (initially suspected of being SNe~Ia), the CSP-II sample generally concentrates at lower redshift than CSP-I. This was by design, as CSP-II only initiated significant SN~II follow-up for the closest events -- where NIR spectral sequences could be obtained.\\
\indent Here we present a total of 9817 optical ($uBgVri$) photometric measurements for 94 SNe~II (7716 for CSP-I and 2101 for CSP-II).
Meanwhile, in the NIR ($YJH$), 1872 data points are presented (1705 for CSP-I and 167 for CSP-II). 
For the full (CSP-I + CSP-II) sample, SNe~II have a median of 93 optical photometric points and a median number of NIR points of 18. 
All 94 SNe~II have at least some optical photometry, while 77 (62 CSP-I and 15 CSP-II) have NIR light curves.
Figure~\ref{Npointsopt} displays the distribution of the number of distinct photometric data points.
The largest optical data set is of 506 photometric measurements (SN~2013hj), while the minimum is of 7 (LSQ12fui). In the NIR, SN~2008ag has the most data points (61), while there are a number of events without any data at NIR wavelengths. The sample has a median $V$-band cadence of 4.7 days -- over a median optical-photometry coverage of 79.9 days. In the NIR the median $J$-band cadence is 7.2 days over a NIR coverage of 58.9 days.\\
\indent The epochs of the earliest and latest optical and NIR photometry per SN are displayed in Figs.~\ref{earlydist} and~\ref{latedist} respectively.
The earliest photometry obtained at optical and NIR wavelengths is within two days of the estimated explosion epoch (SN~2008gr). The inset in Fig.~\ref{earlydist} shows a zoom of the first 13 days post explosion. For CSP-I only a small number of SNe~II have their first photometric point within less than five days from explosion. However, for CSP-II almost half the sample have such early time data. This is a consequence of the evolution of transient surveys -- during CSP-I it was still relatively rare to discover SNe within a few days of explosion. Post CSP-I (and during CSP-II) survey capabilities significantly improved and more SNe were (and are) found at such early epochs.
The latest optical photometry was obtained for SN~2013hj at 500 days post explosion, while in the NIR data were obtained at 419 days post explosion for SN~2008bk.
The majority of the CSP-I sample have accompanying visual-wavelength spectral sequences (presented and analysed by \citealt{gut17b,gut17a}), while the CSP-II SNe~II have at most one or two optical spectra, but several have well-sampled NIR spectroscopy \citep{dav19a}. The number of spectra and the range of their epochs for each SN~II is listed in a Table on Zenodo (see the link in the data availability section), while the epochs of the spectra are depicted on each light curve that are also presented on Zenodo.

\subsection{Previous analyses achieved with the CSP SN~II samples}
\label{previous}
Before presenting and discussing the multiwavelength light curves released in this work, here we briefly summarise the previously presented analyses that have used these data (and in some cases also published preceding releases of photometry\footnote{The photometry published in the current work supersedes photometry previously released to the community.}). We do this to (a) aid the interested reader in surveying the CSP SN~II literature, and (b) give context to the subsequent qualitative discussion in the latter sections of this paper. Papers are discussed in chronological order.\\
\indent As described above, \cite{ham06} presented the first instalment of the CSP -- describing the survey and reduction methodology while also presenting example photometry from the first few years of the project. Specifically, light curves and spectra were presented for the type II SN~2004fx (see Fig.~\ref{04fxlcspec} in  the current paper). \cite{tak14a} analysed the `intermediate luminosity' SN~2009N, constraining its progenitor properties through hydrodynamical modelling and additionally using the SN observations to constrain the distance to its host galaxy, NGC 4487.\\ 
\indent The current data release is a natural succession to the $V$-band data release in A14. A14 characterised the $V$-band light curves of SNe~II, investigating the observed diversity within a large sample (more than 100 events when including the data published by \citealt{gal16}). That work defined different light-curve properties including, absolute magnitudes, decline rates, and durations of different phases -- concluding that SNe~II form a continuum in their light-curve properties with no evidence for distinct photometric groups. \cite{gut14} tied these photometric properties to the morphology of the dominant \ha\ spectral line, finding that faster declining SNe~II displayed much weaker \ha\ absorption than slowly declining explosions. \cite{des14} compared observations to model spectra of distinct progenitor metallicity suggesting that most of the current sample arose from stars with solar metallicity, with a lack of low-metallicity events, which is unsurprising given that the CSP-I objects are mainly SNe~II discovered by means of targeted surveys. \cite{and14b} again concentrated on the trademark SN~II spectral feature -- \ha. There, we demonstrated that the peak emission of this feature is generally blueshifted with respect to zero rest velocity, a property that is reproduced by the models of \cite{des13} and can be understood as being tied to the steep density profiles of SN~II ejecta.\\
\indent CSP-I photometry was used by \cite{dej15} to show that luminosity distances can be calculated when only using SN~II photometry -- an important finding in the current and future eras of wide-field surveys where many more transients will be detected than one can follow-up spectroscopically. Following the work of \cite{des14}, in \cite{and16} the CSP-I SNe~II spectral and photometric properties were compared to oxygen abundances of host \hii\ regions. SN~II metal-line strength was shown to correlate with environment elemental abundance, strengthening the case for the use of SN~II to measure galaxy metallicities. The Photometric Color Method (PCM) introduced by \cite{dej15} was later used by \cite{dej17a}, where CSP SNe~II were combined with higher redshift samples to produce a Hubble diagram and measure cosmological parameters.\\
\indent As previously noted, visual-wavelength spectroscopy of the CSP-I SN~II sample was published by \cite{gut17a}. Subsequently, \cite{gut17b} compared a large range of spectral-line measurements to photometric parameters. This analysis showed that SNe~II with higher expansion velocities are more luminous, and had more rapidly declining light curves, shorter plateau durations, and higher $^{56}$Ni masses.\\
\indent CSP-II photometry of SN~2015bs was published by \cite{and18a}, where photospheric- and nebular-phase spectra were used to show that its progenitor had the lowest metallicity and highest mass of any SN~II studied to-date. \cite{dej18a} argued that dispersion seen in SN~II observed colours is dominated by intrinsic colour differences of the explosions and not by the effects of host-galaxy extinction. \cite{pes19} compared the optical light curves of SNe~II (from the CSP-I) to literature SNe~IIb, showing that there is no continuum between the properties of the fastest declining hydrogen-rich SNe~II and SNe~IIb -- suggesting distinct progenitor channels for the two groups.\\
\indent The CSP-II obtained NIR spectra for a significant fraction of the sample SNe~II. These were analysed -- together with photometry -- by \cite{dav19a}. Distinct spectroscopic families were found, somewhat in contradiction to the clear continuum of events observed in photometry (A14). 
CSP data of SN~2008bm, SN~2009aj, and SN~2009au were published by \cite{rod20a}, who found that these three events were initially powered by significant ejecta-CSM interactions, boosting their luminosity while slowing down their expansion velocities.
Finally, data of SN~2006Y (see our Fig.~\ref{06Ylcspec}) and SN~2006ai (Fig.~\ref{06ailcspec}) were presented by \cite{hir21}, where it was concluded that RSGs with enhanced mass loss are required to produce their light curves (specifically their `short plateaus'; see also \citealt{mar22c}).\\

\indent Together with this data release, we have also published an additional three studies by \cite{mar22a,mar22b,mar22c} (respectively M22a, M22b and M22c henceforth) using the CSP-I SN~II sample. First, in M22a we present bolometric light curves for all events and outline a methodology for accurately estimating the bolometric flux for any given SN~II. This paper shows the importance of using NIR photometry for such estimates and gives corrections to use when such data are unavailable. Together with the spectral velocities published by \cite{gut17a}, bolometric light curves are used to determine progenitor and explosion parameters of the SN~II sample in M22b through comparison to hydrodynamical models (following the procedure of \citealt{mar20}). Specifically, we derive ZAMS masses for the sample, finding a bias towards low progenitor mass. Finally, in M22c we correlate physical parameters from the best-fit models to observed transient parameters to further understand the underlying physics driving SN~II diversity. In Sect.~\ref{diss}, we quote the results of these studies when comparing specific events in our sample to the general
SN~II phenomenon.

\section{Ninety-four SN~II multiband light curves}
\label{fulllcs}
In this section, we present an example SN~II: showing SN and local-sequence photometry, a finder chart, and the $uBgVriYJH$ optical to NIR wavelength light curves.
Complete electronic, machine readable tables of SN and local-sequence photometry can be downloaded from the CDS (Strasbourg astronomical Data Center)\footnote{\url{https://cds.unistra.fr//}}$^{,}$\footnote{The same data can also be downloaded from the CSP website: \url{https://csp.obs.carnegiescience.edu/data}.}, and finder charts and light curves for our full sample of SNe~II can be found on Zenodo through the links in the data availability section.
After showing an example case, we characterise the CSP SN~II sample in terms of its observed parameters in the $V$ band (following A14), and define a small SN~II comparison sample to put our sample within the context of the literature.\\
\indent Throughout the rest of the paper, light curves are depicted with respect to their explosion epochs (while tabulated photometry is with respect to Julian date). Explosion epochs (and their uncertainties) together with information used to constrain them are listed for each SN~II in a Table on Zenodo (see the link in the data availability section). 
Light curves in figures are displayed in both apparent and absolute magnitudes. Apparent and absolute magnitudes are corrected for Milky Way extinction taken from \cite{sch11}, assuming the extinction law from \cite{car89} with $R_V = 3.1$ (but published photometry is not corrected). Absolute magnitudes are calculated using distance moduli taken from A14 and \cite{dav19a} (or derived using the same methodology). Light curves are not corrected for host galaxy extinction as we believe that in most cases these are small ($<$0.1\,mag) and any corrections employed are generally unreliable (see \citealt{dej18a}). Uncertainties on individual photometric measurements are shown in the figures, but are often too small to be visible on the plots.

\subsection{An example of CSP SN~II photometry: SN~2007oc}

\begin{figure}
\includegraphics[width=8.5cm]{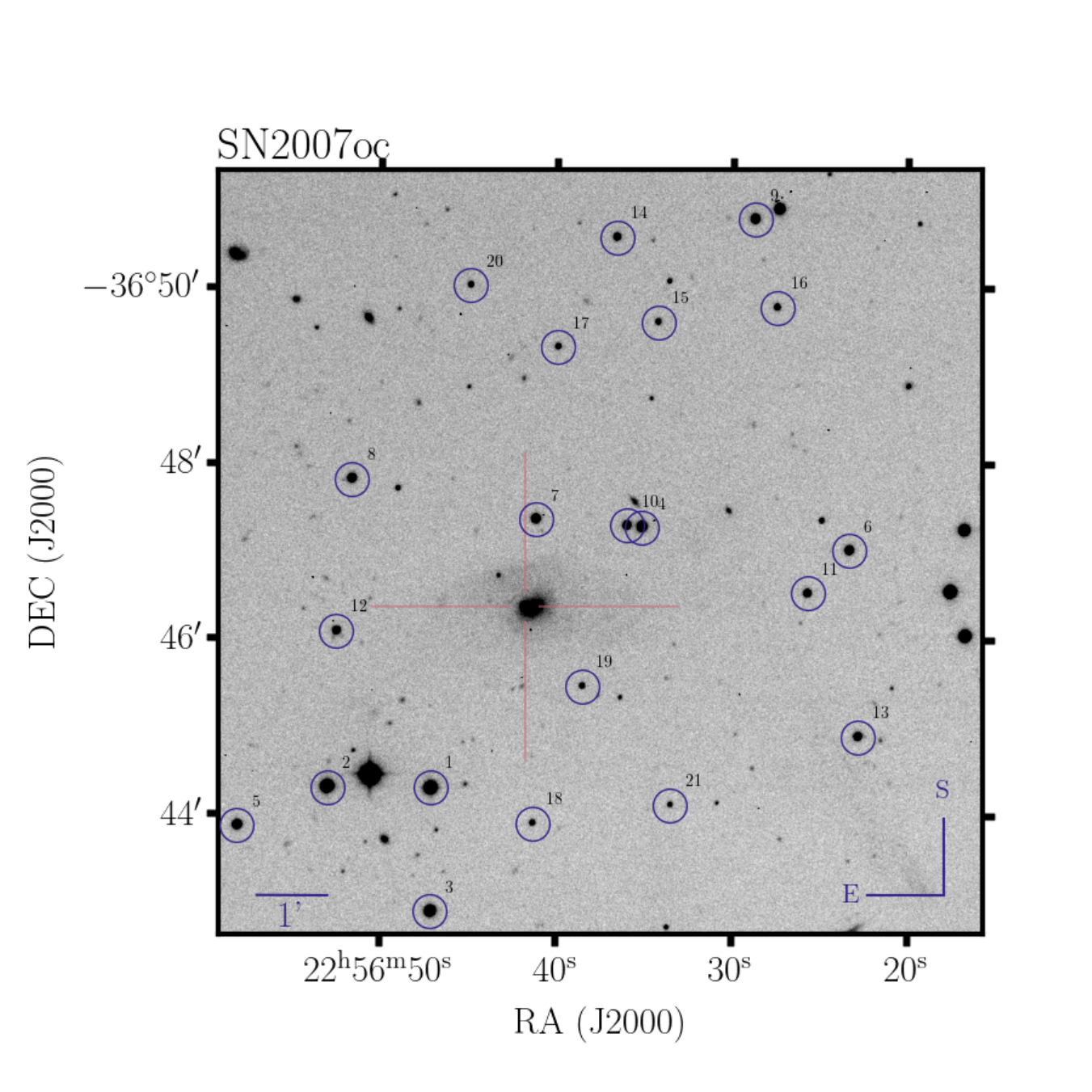}
\caption{Finder chart of SN~2007oc with the SN position marked with a red cross, while optical local sequence stars are marked with blue circles.
(Such finder charts for all SNe~II in our sample can be found on Zenodo through the link in the data availability section.)}
\label{finder07oc}
\end{figure}
\begin{figure}
\includegraphics[width=8.5cm]{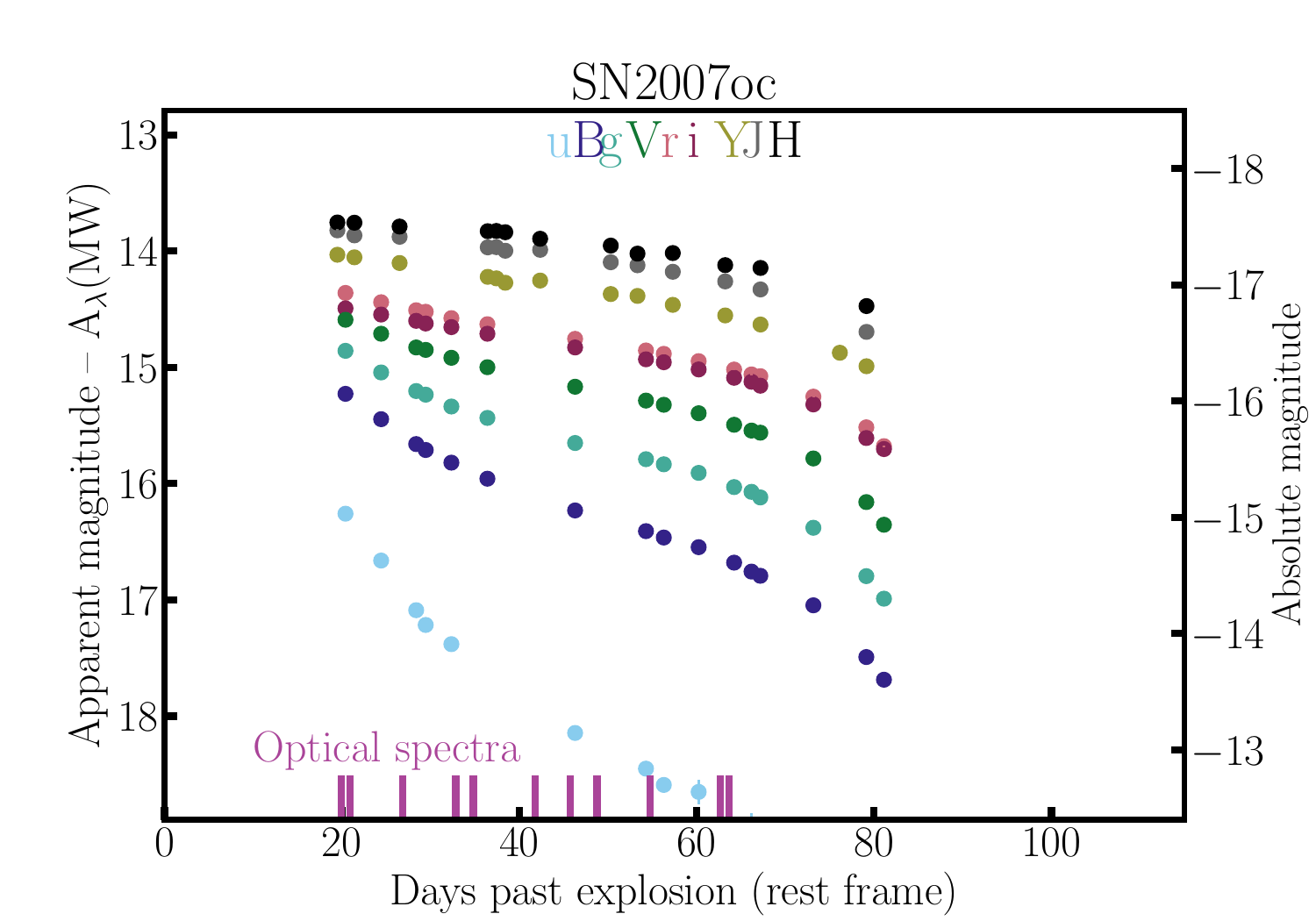}
\caption{$uBgVri$ and $YJH$ apparent and absolute magnitude light curves of SN~2007oc plotted relative to the explosion epoch. Epochs when visual-wavelength spectra were obtained are indicated on the abscissa.
(Such light curves for all SNe~II in our sample can be found on Zenodo through the link in the data availability section.)}
\label{07oclc}
\end{figure}

Figure~\ref{finder07oc} displays the finder chart for SN~2007oc, with optical local sequence stars labelled. Local-sequence optical and NIR photometry are listed  in Tables~\ref{oploc07oc} and ~\ref{nloc07oc}, respectively. This local-sequence photometry is published in the \textit{standard} system of \cite{smi02a} for $ugri$, the \textit{standard} system of \cite{lan92a} for $BV$, the \textit{standard} system of \cite{per98a} for $JH$ and the \textit{natural} system of the Swope+RetroCam setup for $Y$ (defined as the `standard' system for $Y$-band photometry by \citealt{kri17a}). The apparent and absolute magnitude light curves of SN~2007oc are shown in Fig.~\ref{07oclc}. SN photometry is published in the \textit{natural} CSP systems, and optical photometry is given in Table~\ref{opphot07oc}, while NIR photometry is given in Table~\ref{nphot07oc}.\\
\indent SN~2007oc has extremely well-sampled optical and NIR photometry between around 25 to 80 days post explosion, but is missing data at early and late times. This SN~II has a relatively fast declining light curve with an $s_2$ of 1.8$\pm$0.01\,mag\,/100\,d and a relatively short photospheric phase with an $OPTd$ of 71$\pm$5\,days (see Figs.~\ref{s2} and ~\ref{optd} for reference, and see below for a description of how these parameters are defined and measured).

\subsection{Sample characteristics and comparison-sample definition}
\label{sampchar}
\begin{figure}
\includegraphics[width=8.5cm]{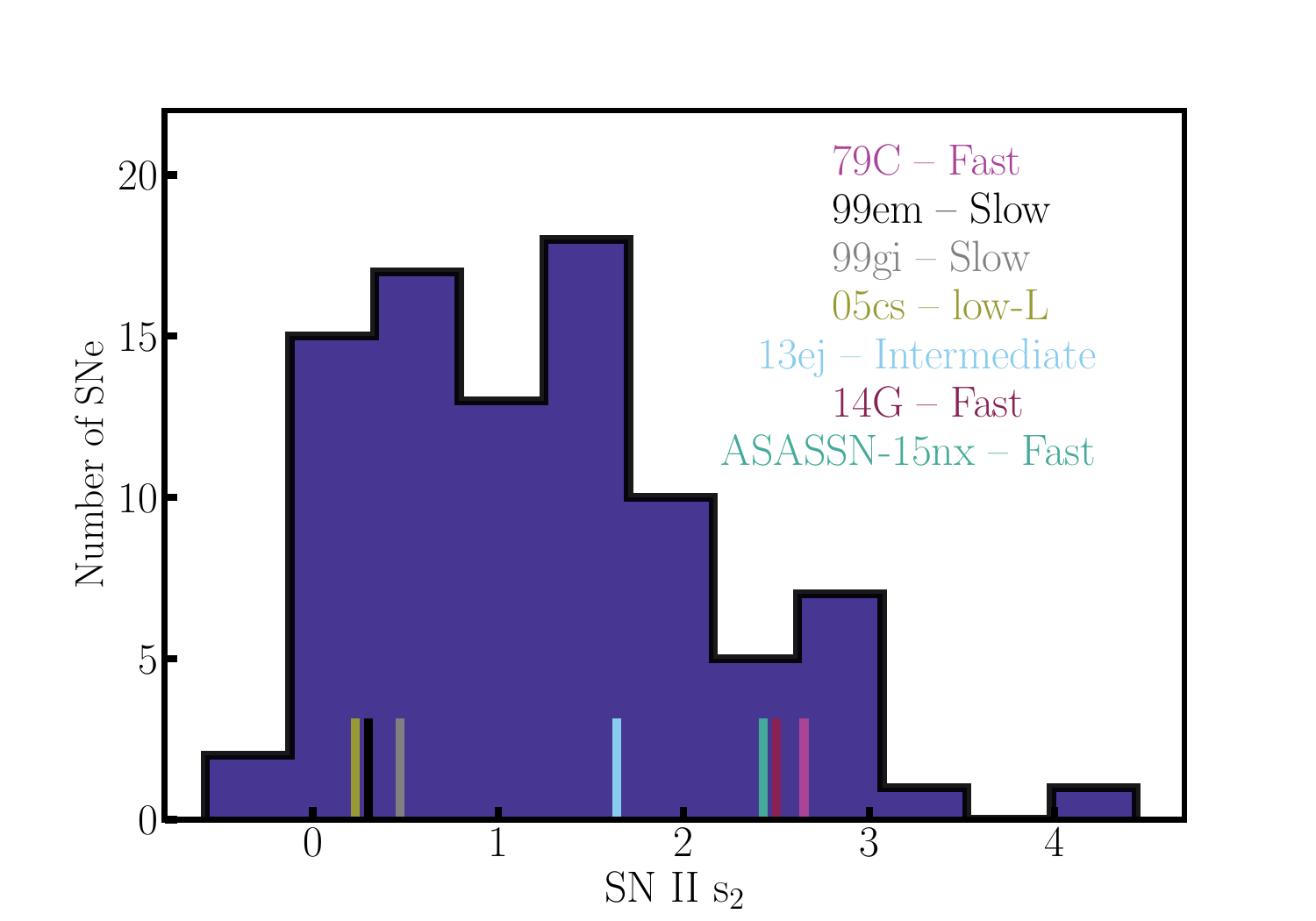}
\caption{$V$-band $s_2$ distribution for the CSP and comparison samples. The comparison sample values are shown as ticks on 
the abscissa.}
\label{s2}
\end{figure}

In A14 we presented a characterisation of the $V$-band light curves of SNe~II, including many of the events for 
which we publish photometry here. A14 attempted to capture the main parameters of SN~II light curves
through defining a number of absolute magnitudes at different times, decline rates at different epochs, and durations of different
phases. In the current work we also use the $V$ band for easy comparison, and concentrate on two parameters that have historically defined the discussion of SN~II light-curve morphology: the plateau decline rate, parameterised here through $s_2$ (the decline rate of the second, slower decline in the light curve post maximum light), and the duration of the plateau, parameterised here through $OPTd$ (the time from explosion to the end of the plateau)\footnote{In A14 and \cite{gut17a} we also defined `$Pd$' (`plateau duration') -- the time from the $s_1$-to-$s_2$ inflection point until the end of $OPTd$. However, that parameter is harder to measure and thus we use $OPTd$ here to provide a greater sample size for comparison.}, both measured in the $V$ band. Following A14, $s_2$ is measured by fitting a piecewise linear model with four parameters: the decline rates $s_1$ and $s_2$, the epoch of the inflection between these points $t_{tran}$, and the magnitude offset. The start of $s_1$ and the end of $s_2$ are visually defined and the values of the decline rates and their inflection point are determined through weighted least squares minimisation. The Bayesian Information Criterion (BIC, \citealt{sch78}) is used to evaluate if the data are better fit by one slope (that is then assigned to be $s_2$) or two ($s_1$ and $s_2$). The SN~II parameters $s_2$ and $OPTd$ are listed for all SNe in a Table on Zenodo (see the link in the data availability section).\\
\indent Throughout the subsequent discussion we select a group of SNe~II from the literature for comparison.
These events are selected by being either well-known literature SNe~II or because they have specific characteristics that enable us to put the current sample within the context of the observed diversity of SNe~II. While the reader should note that this selection is subjective, the comparison sample is not used to derive strong conclusions on the nature of SNe~II and thus its exact nature is somewhat unimportant. In order of discovery, the comparison sample is, SN~1979C, SN~1999em, SN~1999gi, SN~2005cs, SN~2013ej, SN~2014G, and ASASSN-15nx. SN~1979C \citep{dev81} is often cited as the historical example of a SN~IIL fast decliner, and thus we include it here. This event was extremely bright; it was more luminous than any SN~II in the current sample. SN~1999em (e.g. \citealt{ham01,leo02}) and SN~1999gi (e.g. \citealt{leo02_2}) are prototypical slow decliners (IIP) showing all the common characteristics that are discussed as belonging to the SN~II phenomenon. Specifically SN~1999em is often used as the standard event for model comparison (e.g. \citealt{utr07,ber11,des13}). SN~2005cs is one of the best-observed `low-luminosity' SNe~II (\citealt{pas06}; see \citealt{spi14} for a sample analysis). Such events generally show low ejecta velocities in combination to their dimness. SN~2013ej has been called both a IIP and IIL in the literature (e.g. \citealt{val14b,bos15,hua15,dhu16,mau17}), as it shows a relatively fast declining light curve but also a clear plateau (as do nearly all hydrogen-rich SNe~II). We therefore include this in our comparison sample as an `intermediate' SN~II. SN~2014G \citep{ter16} is included as another example of a fast decliner (but which again shows a clear plateau). Finally, ASASSN-15nx is possibly the only hydrogen-rich SN~II in the literature with a true `linear' decline post maximum light \citep{bos18}\footnote{ASASSN-15nx may be considered the only true SN~IIL thus far observed if one uses a literal interpretation of the initial IIP, IIL classification scheme \citep{bar79}.}. We therefore include this event for comparison as the most extreme hydrogen-rich SN~II: a bright SN~II, with a fast declining light curve and no apparent plateau phase.\\
\indent Figure~\ref{s2} shows the $V$-band $s_2$ distribution of the CSP sample together with our comparison sample. The majority of SNe~II decline between zero and two magnitudes per 100 days before the end of the plateau. A few events have negative $s_2$ values -- they increase in optical brightness during this phase. A small number of events decline faster than 2.5 mag\,/100\,d. Further discussion of the fast decliners is presented below.\\
\indent Figure~\ref{optd} shows the $OPTd$ distribution. A strong peak is observed around 90 days and the vast majority of events arrive to the end of their plateau between 75 and 100 days post explosion. A tail is seen at low $OPTd$ (see discussion in Sect.\,5), and a few SNe~II do not drop from the plateau until almost 120 days after explosion.\\
\indent Figure~\ref{uViH} displays CSP SN~II $uViH$ absolute magnitude light curves for a subsample of our SNe that are well-observed. These bands are chosen to show the different light-curve evolution as one moves from the bluest ($u$) through to the reddest ($H$) wavebands. It is clear that SN~II light-curve evolution is much faster at bluer wavelengths -- the $u$-band luminosity falls off quickly post maximum and plateau phases are almost nonexistent. In $V$ the figure shows the typical light-curve morphology of SNe~II: post maxima flat or slowly declining light curves until arriving to the end of the plateau. In $i$ one starts to observe a larger fraction of SNe~II that increase in luminosity during the plateau phase. In the NIR $H$ band, SN~II light curves show much longer rise times to maximum, and light curves display a more round and gradual evolution than observed at visual wavelengths. 

\begin{figure}
\includegraphics[width=8.5cm]{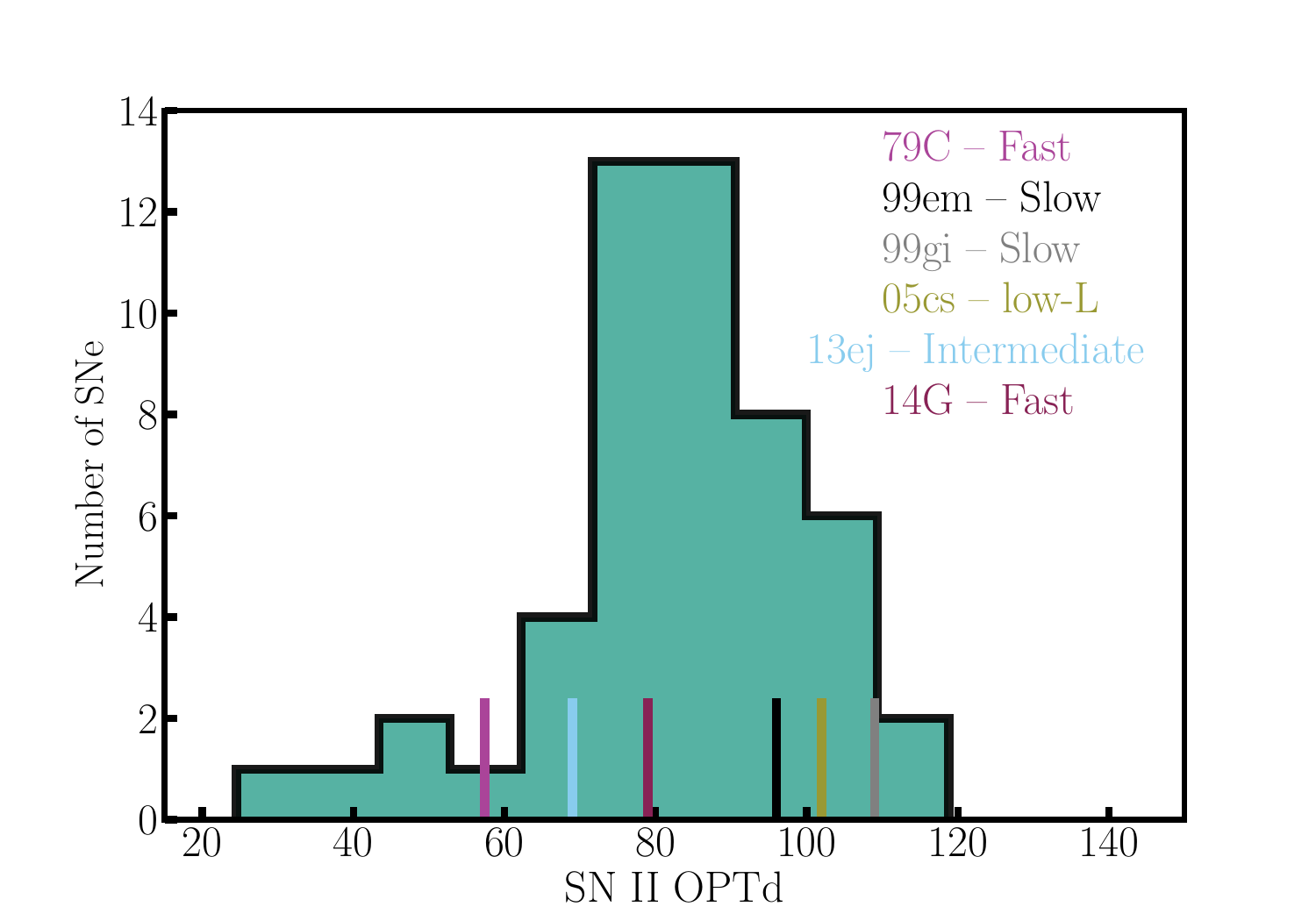}
\caption{$OPTd$ plateau duration distribution measured in the $V$ band for the CSP and comparison samples. The comparison-sample values are shown as ticks on 
the abscissa. (ASASSN-15nx is not shown here as it was not possible to define the end of any plateau phase -- there is no discernible break to the post maximum decline rate out to late times.)}
\label{optd}
\end{figure}

\begin{figure*}
\sidecaption
\centering
\includegraphics[width=12cm]{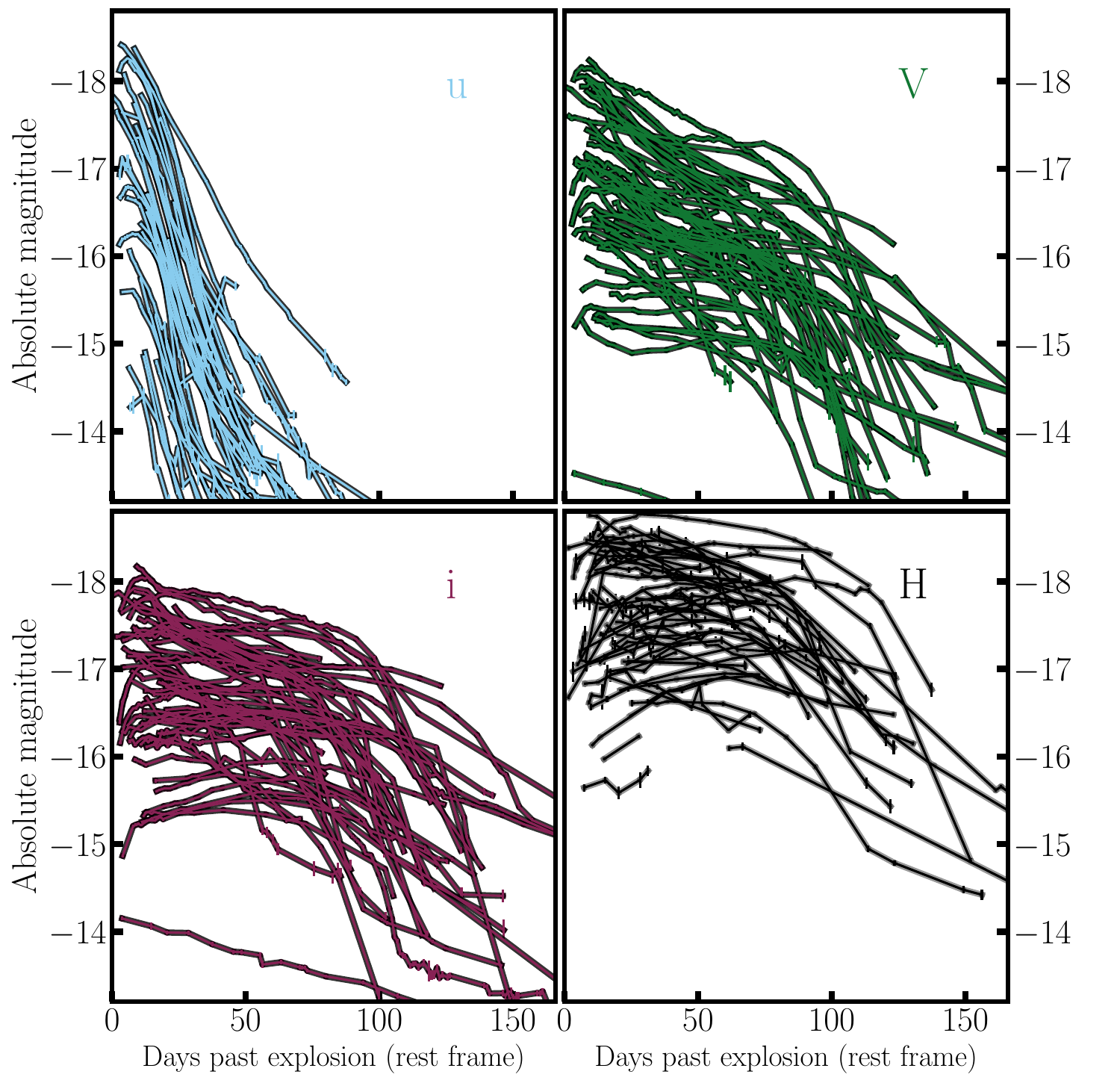}
\caption{$uViH$ absolute magnitude light curves of CSP SNe~II. Individual photometric points have been joined by lines to provide a visualisation of the different light curve morphologies observed at different wavelengths.}
\label{uViH}
\end{figure*}

\section{Discussion}
\label{diss}
This discussion section is devoted to summarising the properties of a number of specific SNe~II in the CSP samples, helping to put these events -- and indeed literature SNe~II -- in some context.
In the following, we present absolute magnitude light curves and spectral sequences 
for different explosions that capture the main diversity of
SNe that form the SN~II class. In addition, we highlight a number of nonstandard SNe~II and discuss their unusual properties. Visual-wavelength spectral sequences of the CSP-I sample were presented by \cite{gut17a}, while for CSP-II SNe~II we have at most a few such spectra. On the contrary, there are no NIR spectra of the CSP-I SN~II sample, but many CSP-II SNe~II have well-observed NIR spectral sequences (see \citealt{dav19a}).  
We discuss these example events in terms of their $V$-band light-curve morphologies and spectral ejecta velocities, and any other
peculiarities and/or interesting features that each SN displays. The comparison sample introduced in the previous section is again highlighted, where it aids the reader in comprehending each example SN. As we have not made any host-galaxy extinction corrections to our CSP SN~II sample, we also ignore such effects for the literature comparison SNe for internal consistency. \\
\indent Where relevant the explosion and progenitor estimates for our SNe~II from M22b are also stated below. In this regard, it is important to note that the modelling in M22b assumed standard single-star stellar evolution to produce pre-SN progenitors that were then exploded. Changes in such assumptions can lead to distinct estimates on progenitor and explosion properties compared to observations (see e.g. \citealt{des19}). In addition, the modelling in M22b did not include the effects of interaction between the SN ejecta and any surrounding CSM (see \citealt{mor18} who included such material in hydrodynamical modelling of SNe~II, and \citealt{hil19} for an analysis of how differences in progenitor, explosion, and CSM properties may produce observed diversity in SN~II light curves and spectra). M22b only modelled SN~II light curves after 30 days post explosion, where it is assumed that any effects from earlier ejecta--CSM interaction are unimportant. Given that our sample also generally lacks data within the first few days post explosion, we also ignore discussion of CSM interaction below. However, the reader should be aware that such interaction may still be affecting the very-early time data (when available) and its interpretation, given the mounting evidence for such interaction in the majority of `normal' SNe~II (e.g. \citealt{for18,bru21} and references therein). It is also possible that CSM interaction may be the source of some of the peculiar objects discussed at the end of this section.\\
\indent Throughout this discussion we compare specific SNe~II to the general population of light-curve and spectral properties using median values from \cite{gut17b}. Specifically, we compare to the absolute magnitude at maximum light in the $V$ band ($M_{max}$, with a median value -16.7 mag), the plateau decline rate ($s_2$, 1.1 mag/100\,d), the duration of the plateau phase ($OPTd$, 87 days), the ejecta velocity from the absorption component of \ha\ at 50 days post explosion (\ha$_{50}$, 7270\,km/s), and the ejecta velocity from the Fe 5169\AA\ line (Fe$_{50}$, 3940\,km/s). All individual SN~II values are taken from \cite{gut17a}, or are measured following the same methodology (however here we present $s_2$ and $OPTd$ values in the rest frame of each SN - in \citealt{gut17a} values were presented in the observer frame). We also use the physical parameter distributions from M22b to further elucidate the nature of each specific object.

\subsection{Slowly declining SNe~II}
We define `slowly declining SNe~II' as those events showing a clear, relatively flat plateau that would generally be considered `normal' SNe~II by the community. Prototypical slowly declining events can be found in the literature as the well-known SNe~1999em and 1999gi, as already described in the previous section. In the following, we highlight CSP SNe~II showing similarly classical SN~II light-curve morphologies. However, we also note the diversity displayed even within this smaller subsample.

\begin{figure}
\includegraphics[width=8.5cm]{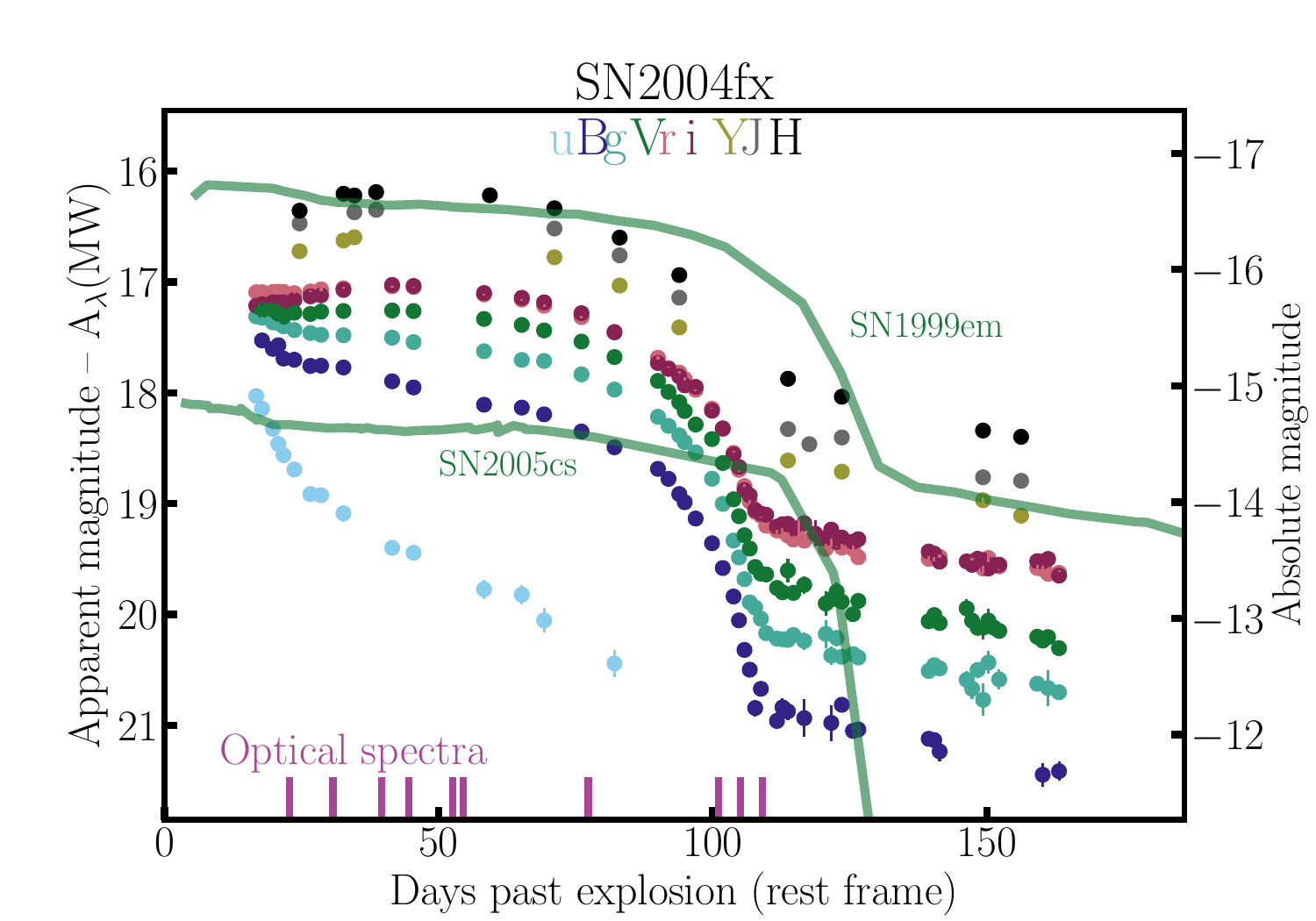}
\includegraphics[width=7.5cm]{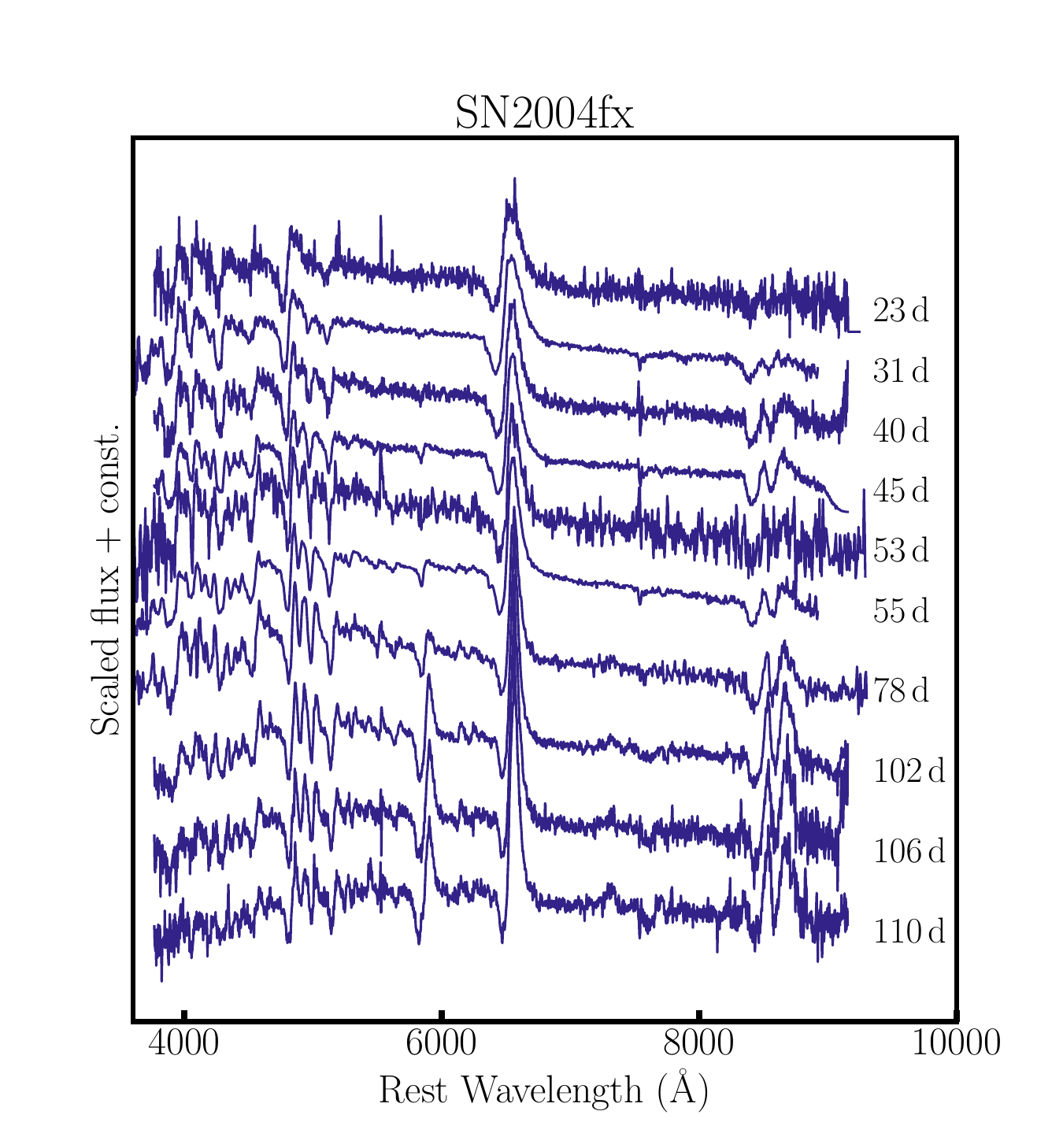}
\caption{Absolute and apparent magnitude $uBgVri$ and $YJH$ light curves (top) and visual-wavelength spectra (bottom) of SN~2004fx. The $V$-band absolute magnitude light curves of SN~1999em and SN~2005cs are also displayed for comparison.}
\label{04fxlcspec}
\end{figure}

\begin{figure}
\includegraphics[width=8.5cm]{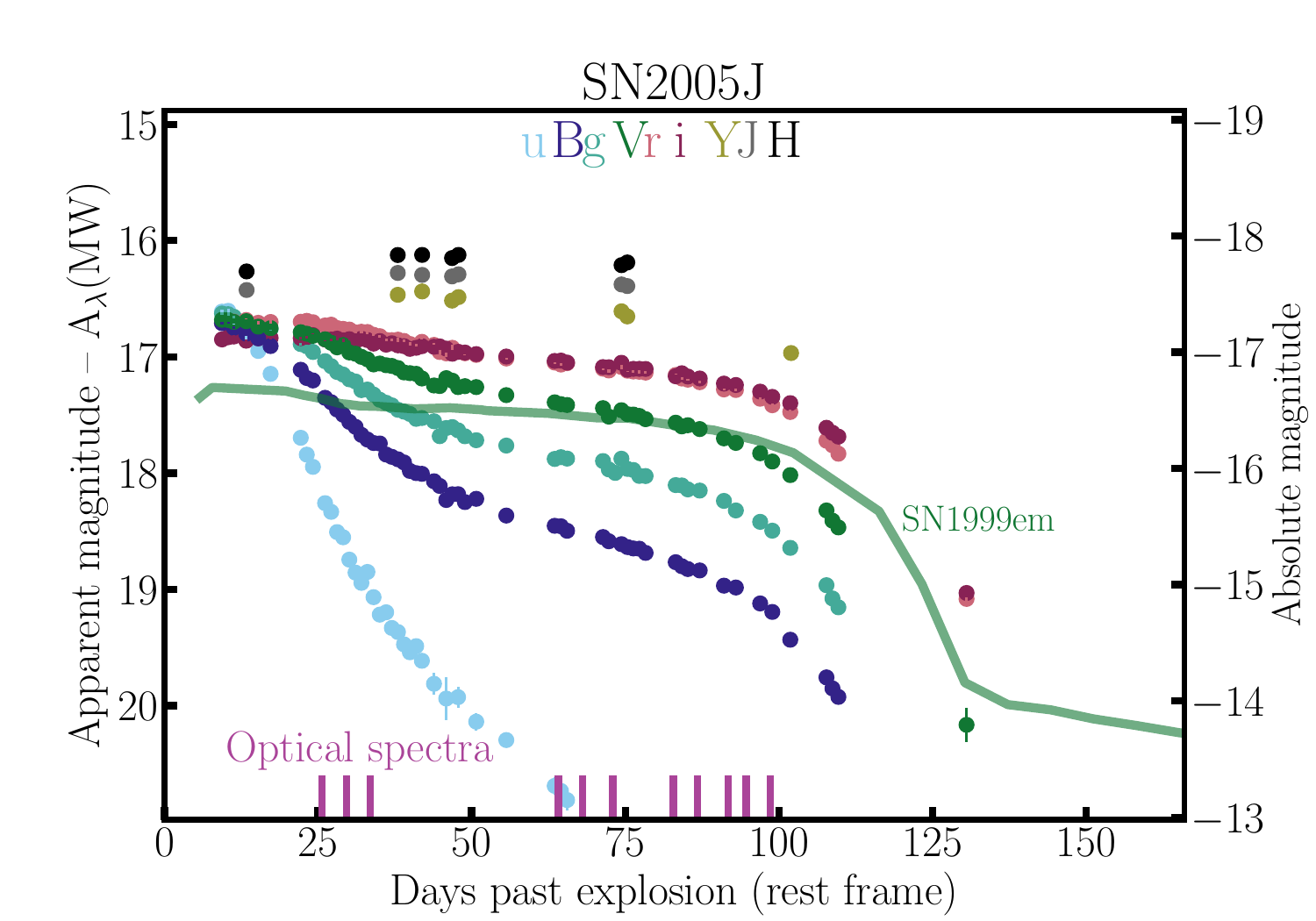}
\includegraphics[width=7.5cm]{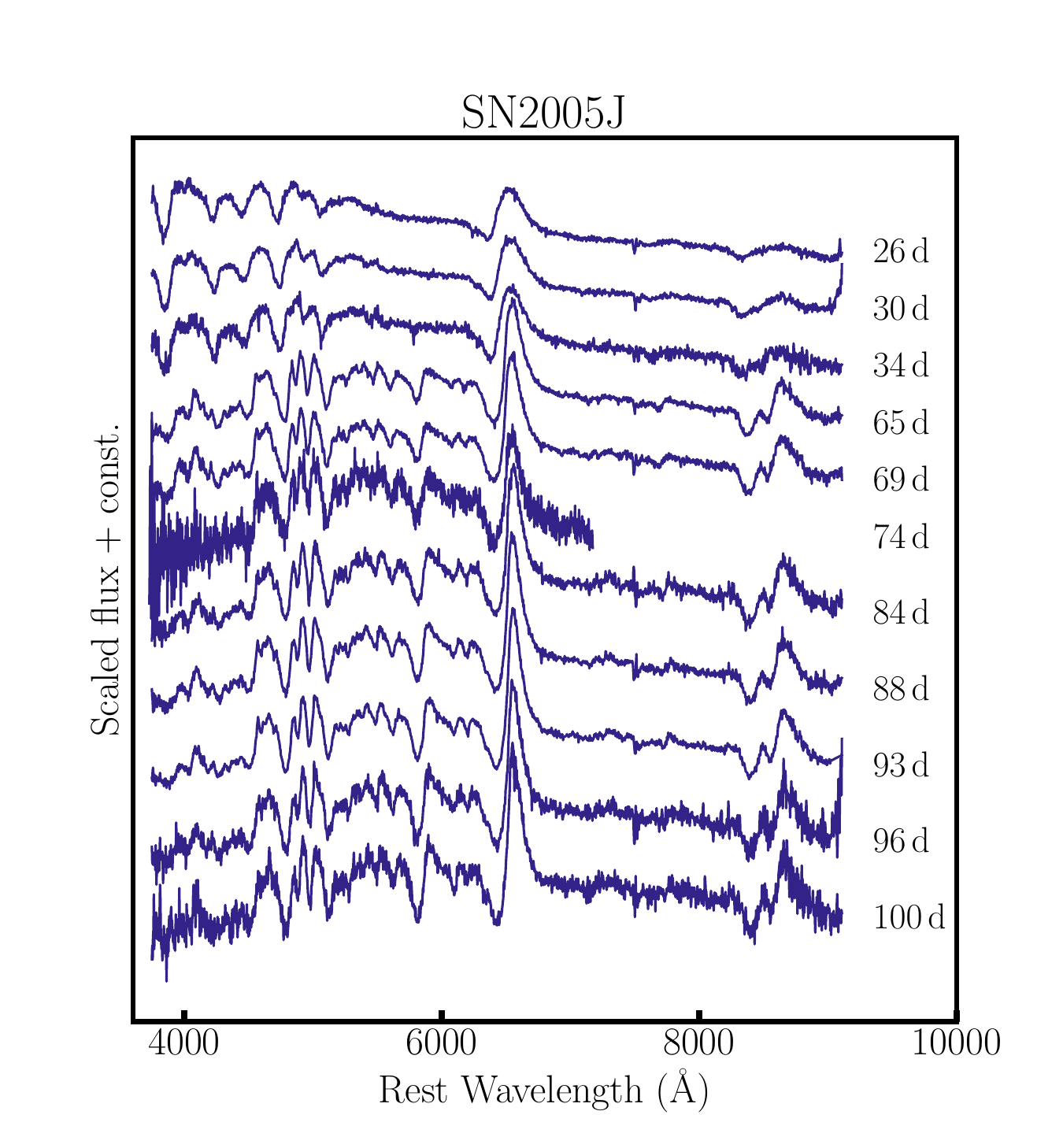}
\caption{Absolute and apparent magnitude $uBgVri$ and $YJH$ light curves (top) and visual-wavelength spectra (bottom) of SN~2005J. The $V$-band absolute magnitude light curve of SN~1999em is also displayed for comparison.}
\label{05Jlcspec}
\end{figure}

\begin{figure}
\includegraphics[width=8.5cm]{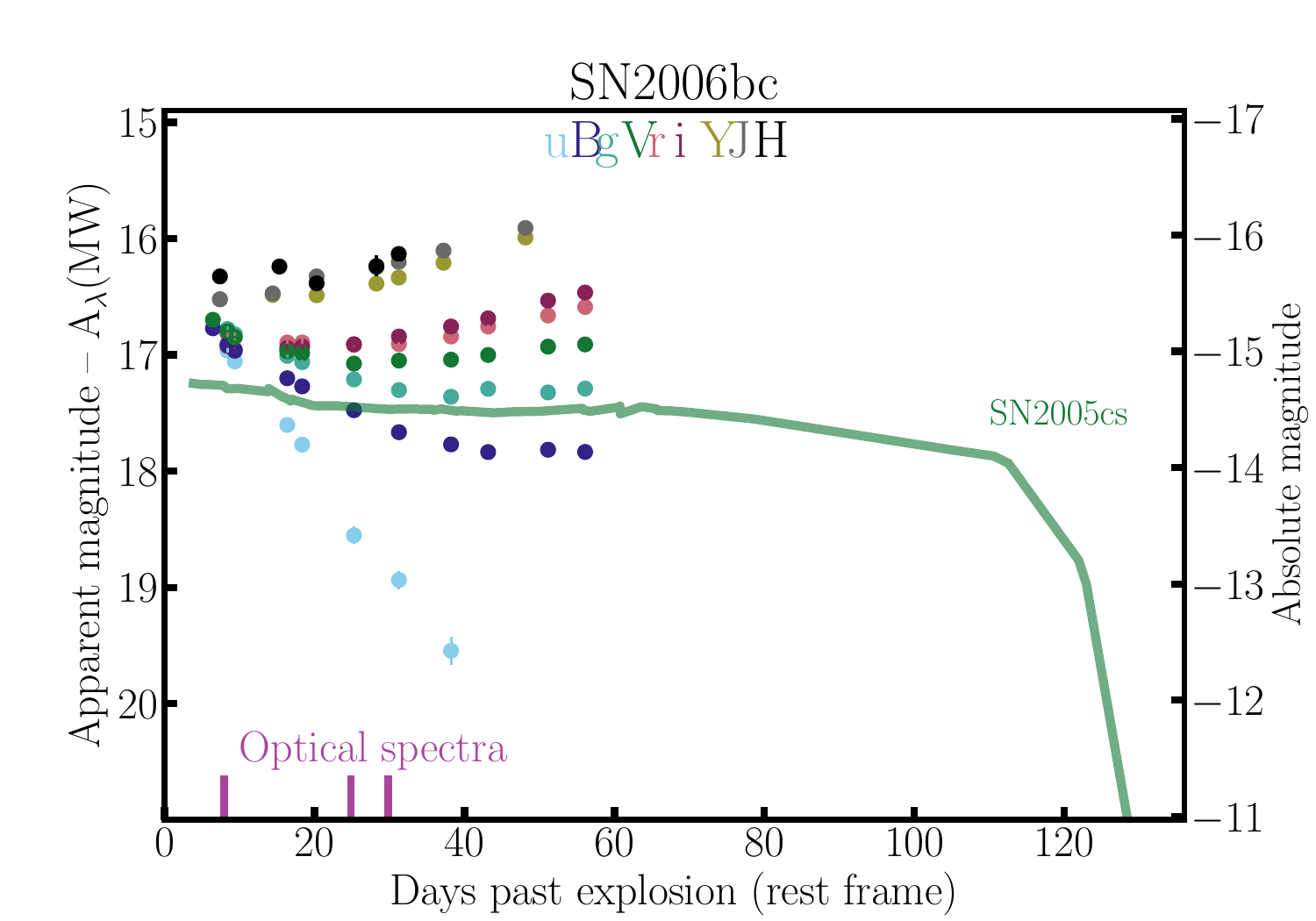}
\includegraphics[width=8.5cm]{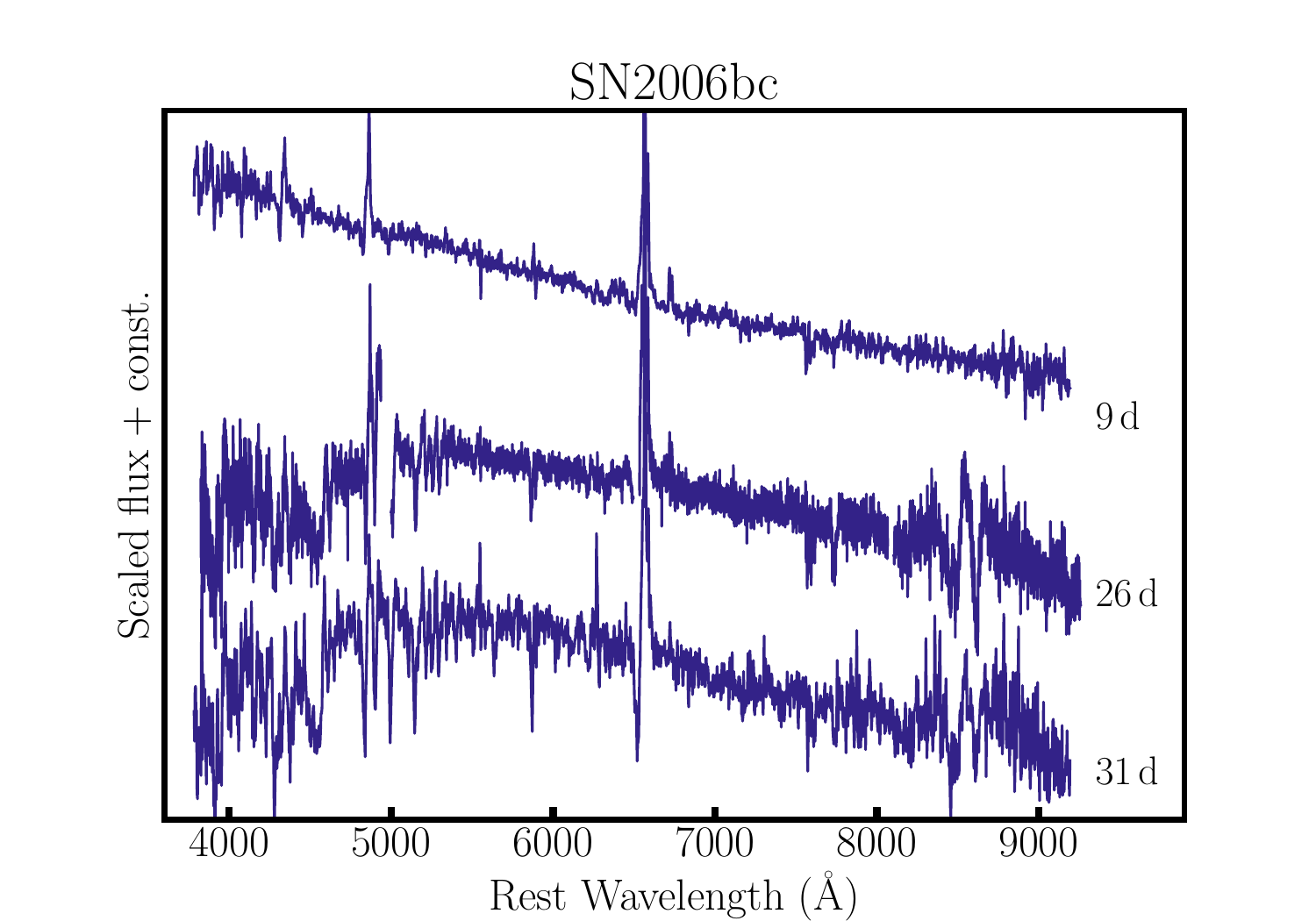}
\caption{Absolute and apparent magnitude $uBgVri$ and $YJH$ light curves (top) and visual-wavelength spectra (bottom) of SN~2006bc. The $V$-band absolute magnitude light curve of SN~2005cs is also displayed for comparison.}
\label{06bclcspec}
\end{figure}

\begin{figure}
\includegraphics[width=8.5cm]{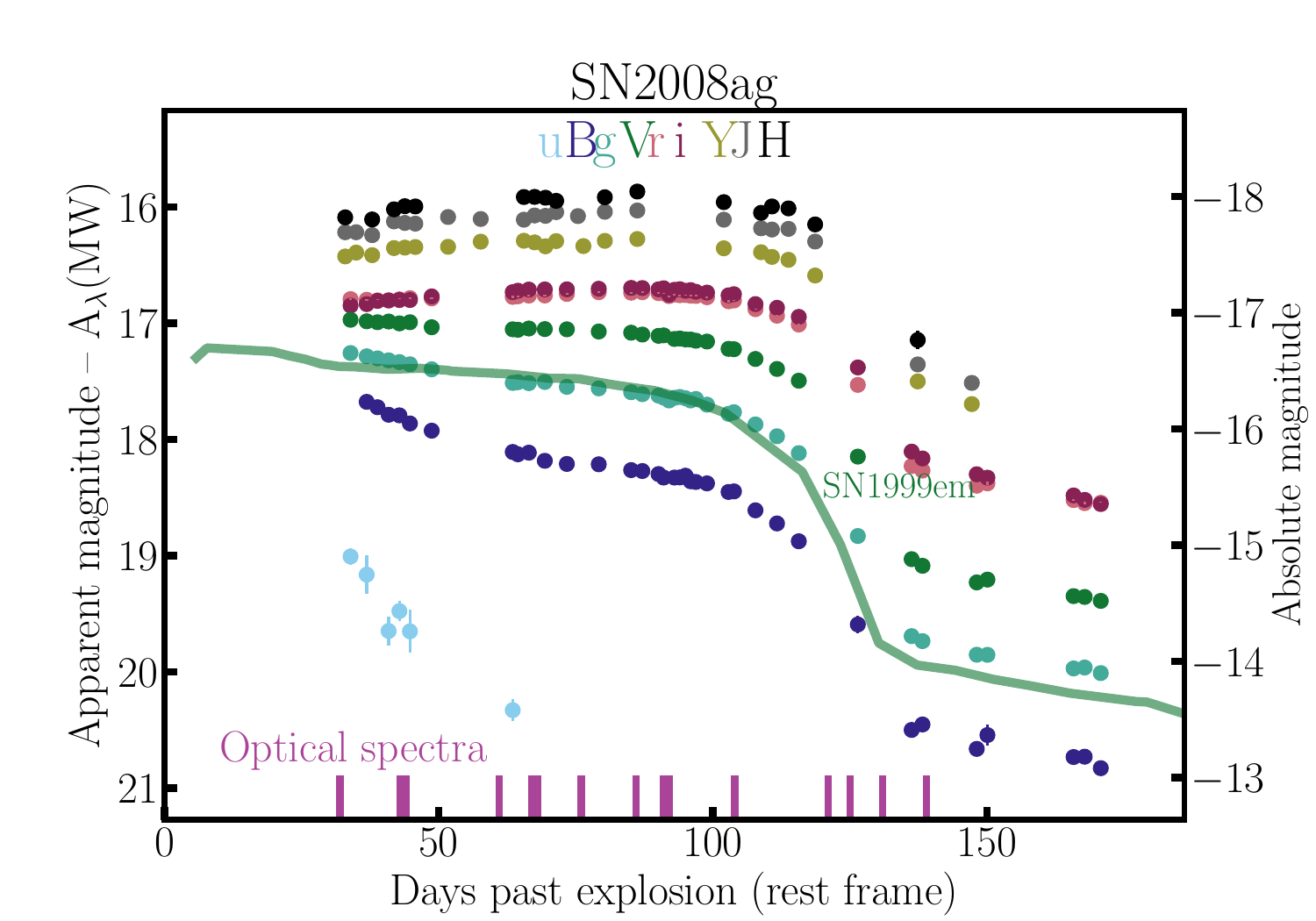}
\includegraphics[width=7.5cm]{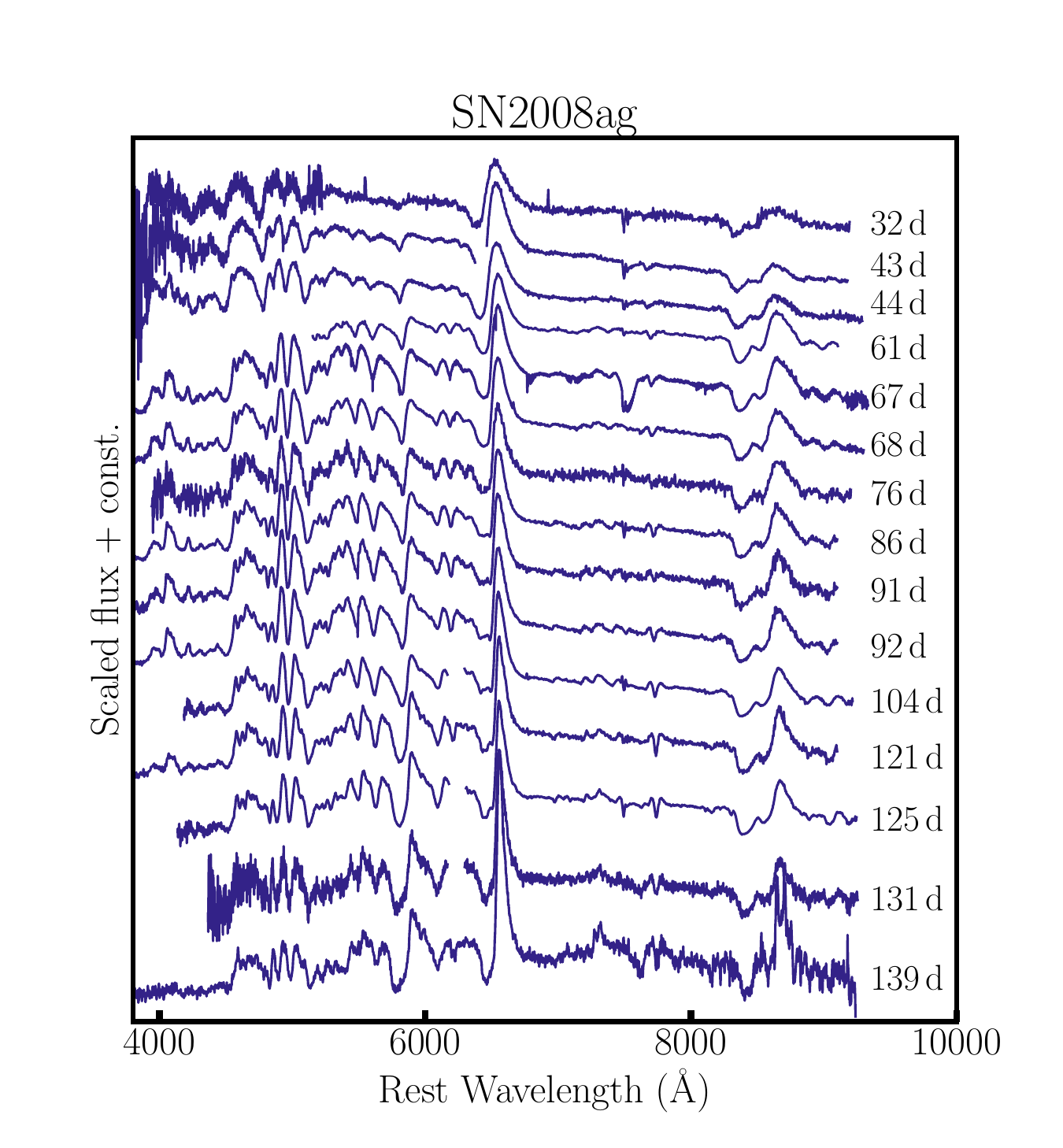}
\caption{Absolute and apparent magnitude $uBgVri$ and $YJH$ light curves (top) and visual-wavelength spectra (bottom) of SN~2008ag. The $V$-band absolute magnitude light curve of SN~1999em is also displayed for comparison.}
\label{08aglcspec}
\end{figure}

\begin{figure}
\includegraphics[width=8.5cm]{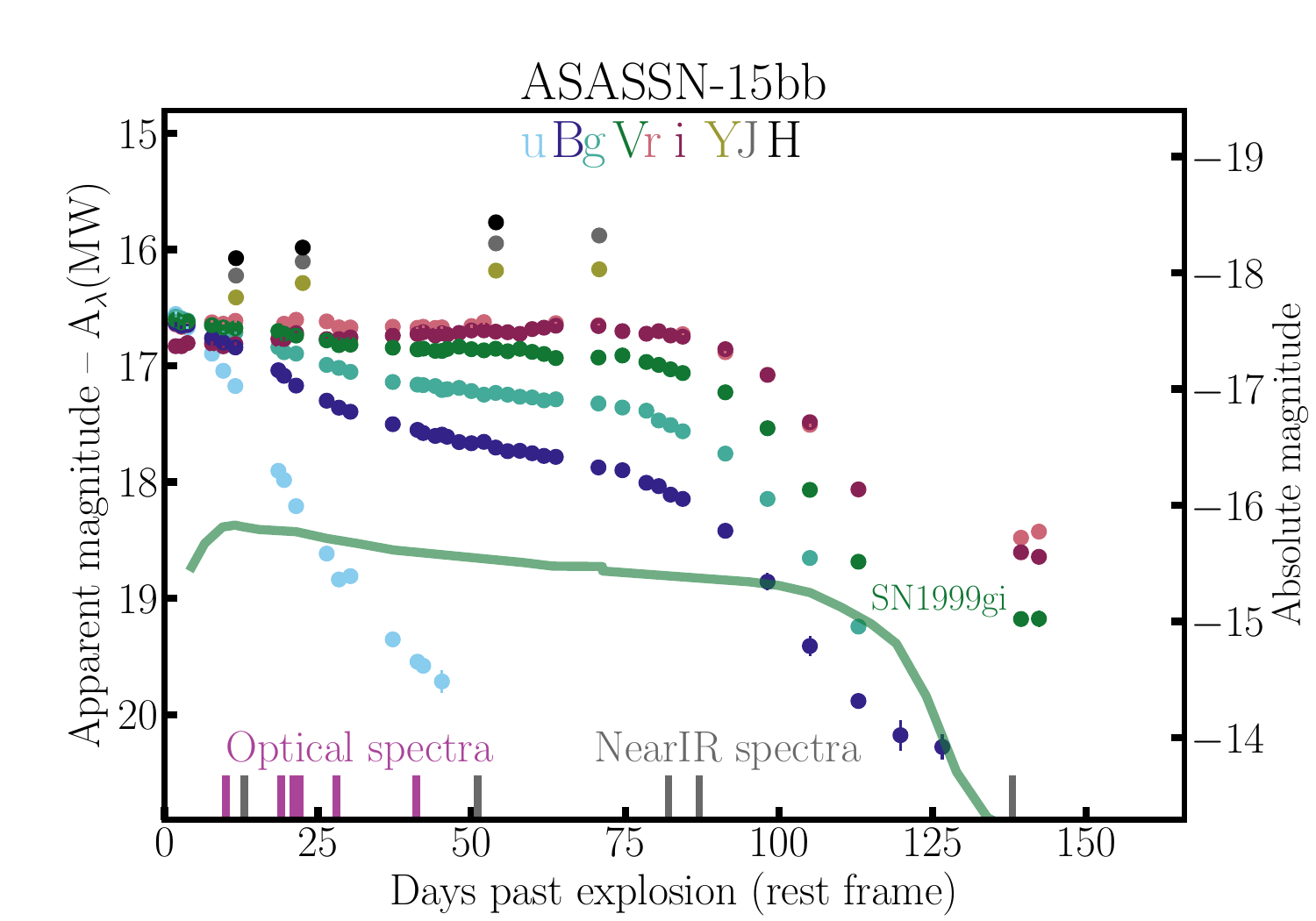}
\includegraphics[width=7.5cm]{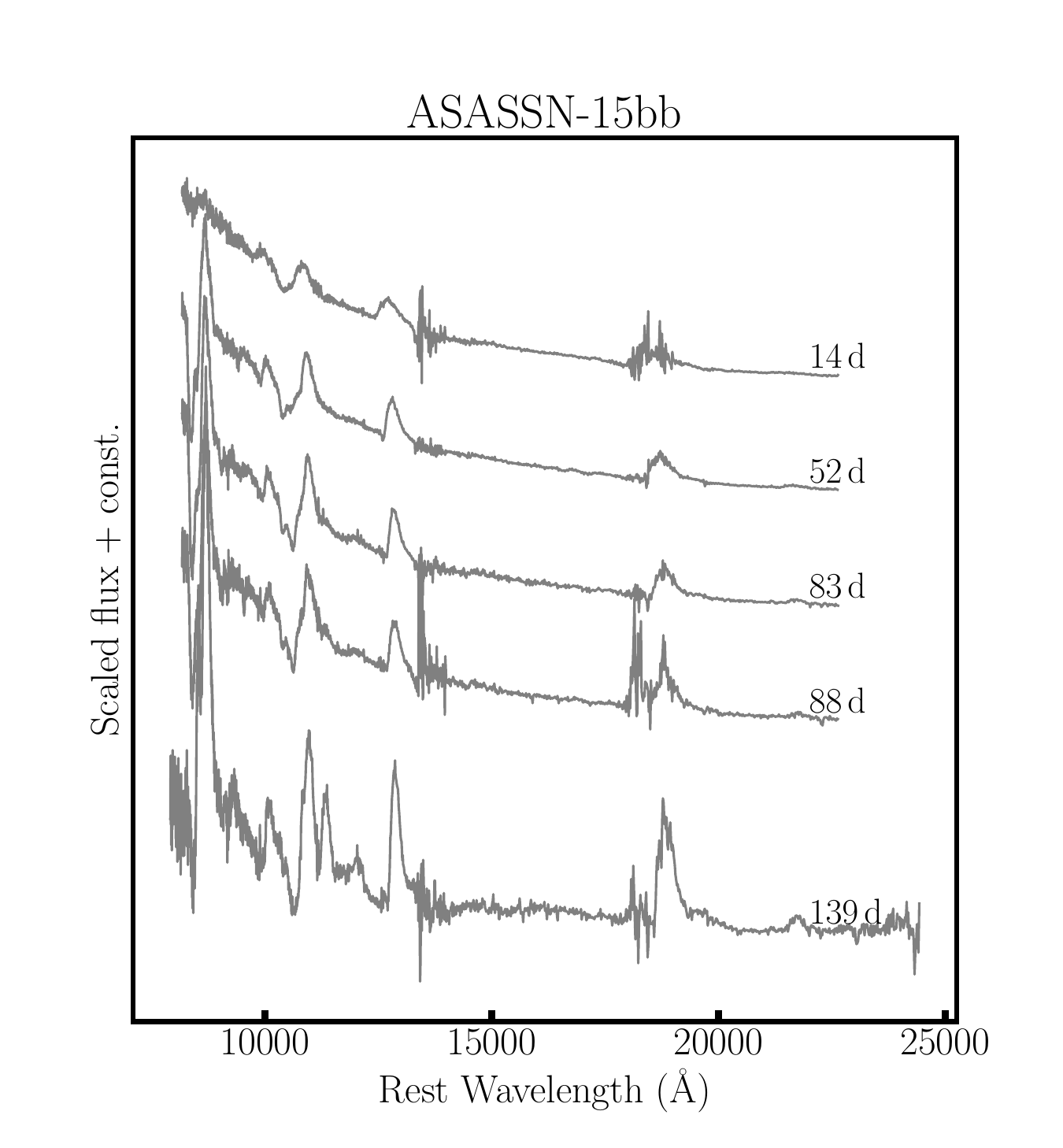}
\caption{Absolute and apparent magnitude $uBgVri$ and $YJH$ light curves (top) and NIR wavelength spectroscopy (bottom) of ASASSN-15bb. The $V$-band absolute magnitude light curve of SN~1999gi is also displayed for comparison.}
\label{15bblcspec}
\end{figure}

SN~2004fx:
Optical and NIR photometry and the visual-wavelength spectral sequence of SN~2004fx are displayed in Fig.~\ref{04fxlcspec}. This SN~II exhibits a standard slowly declining light-curve morphology, with a very well defined plateau, transition and tail. The epoch of peak luminosity appears to have been missed (a common property of CSP-I SNe~II) and any definition of an initially steeper post-maximum decline ($s_1$) is unclear. The decline rate is quite slow, with an $s_2$ of 0.25$\pm$0.02\,mag/100\,d: much slower than the 1.1 \,mag/100\,d median. SN~2004fx appears slightly subluminous for a slowly declining event, falling between SN~1999em and the low-luminosity SN~2005cs (Fig.~\ref{04fxlcspec}). The light curve shows a slow transition from the plateau to the tail. 
The $OPTd$ is noticeably shorter than in other slow decliners, with a value of 68$\pm$6\,days (compared to the 87 day median for the full sample). The spectral sequence displays typical features of SNe~II -- with strong formation of the \ha\ P-Cygni profile in the first spectrum at 23 days post explosion (+23\,d).
The modelling presented by M22b suggests a relatively low explosion energy (0.4$\pm$0.01\,foe\footnote{One `foe' equals 1.0$\times$10$^{51}$ erg.} compared to the 0.6\,foe median), a low $^{56}$Ni mass (0.012$\pm$0.001\msun, 0.036 median), and a low-mass progenitor (ZAMS mass of 10.0$\pm$0.15\msun, 10.4\msun\ median)\footnote{The errors quoted on the physical parameters are statistical and are small due to the high quality of the observations of our sample. However, they do not include uncertainties arising from choices in standard stellar evolution modelling parameters. In addition, the asymmetric errors given in M22b are averaged here for presentation purposes.}.\\
\indent SN~2005J:
Absolute magnitude light curves and the visual-wavelength spectral sequence of SN~2005J are displayed in Fig.~\ref{05Jlcspec}. This SN~II is more luminous than most other events, with an $M_{max}$ of $-$17.3$\pm$0.1\,mag. While the tail is not observed, this event still displays a typical light-curve morphology and has a decline rate of 1.1$\pm$0.02\,mag/100\,d. Even though both SN~2005J and SN~2004fx would be considered `typical' SNe~II in terms of light curve morphology, visually the shapes of their luminosity evolution clearly differ -- SN~2004fx displays a much more `rounded' transition phase. In addition to being slightly more luminous, SN~2005J also displays larger spectral velocities than the general population, with an \ha$_{50}$ of 8430$\pm$760\,km/s (median 7270\,km/s) and an Fe$_{50}$ of 4220$\pm$440 \,km/s (median 3940\,km/s). These properties suggest a more energetic explosion for SN~2005J, with the modelling of M22b giving an explosion energy of 0.8$\pm$0.08\,foe (and additionally a ZAMS mass of 15.4$\pm$1.1\msun).
Another interesting feature of SN~2005J is that it appears to show a long initial post-maximum decline before arriving to $s_2$ (although while this is clearly visible in the $B$ band in Fig.~\ref{05Jlcspec}, it is not statistically observed in $V$).\\
\indent SN~2006bc:
Absolute magnitude light curves and the visual-wavelength spectral sequence of SN~2006bc are displayed in Fig.~\ref{06bclcspec}.
While this SN~II is not particularly well observed (the end of $OPTd$ was not caught), one can immediately identify two defining features in its transient behaviour. After an initial slight decline in the optical photometry post peak, all wavebands redder than $g$ show a clearly increasing luminosity during its $OPTd$. The $s_2$ of SN~2006bc is $-$0.58$\pm$0.04\,mag/100\,d, the most negative such value in our sample (Fig.~\ref{s2}). The SN is relatively faint, with an $M_{max}$ of $-$15.2$\pm$0.3\,mag (A14), and the spectra displayed in Fig.~\ref{06bclcspec} show extremely narrow P-Cygni profiles (the last spectrum is at +30\,d and thus we do no have values at the comparison epoch of 50 days post explosion). These SN properties probably imply a low explosion energy event; however, the data were deemed insufficient for reliable modelling and thus no physical parameters were extracted for SN~2006bc by M22b.\\
\indent SN~2008ag:
Absolute magnitude light curves and the visual-wavelength spectral sequence of SN~2008ag are displayed in Fig.~\ref{08aglcspec}. Visually assessing this SN~II one may conclude that this is a `prototypical' slowly declining explosion. Indeed, while the first several tens of days of the transient evolution were missed, for the rest of its evolution it clearly shows features consistent with the literature definition of a normal SN~II: an $OPTd$ which lasts for several months ($OPTd$ of 104$\pm$9\,days) where the luminosity stays relatively constant ($s_2$ of 0.12$\pm$0.03\,mag/100\,d). However, the SNe~II highlighted in this subsection also clearly show that a large range of such events would be considered normal but display obvious differences in their light curves and spectra, suggesting differences in their progenitor and/or explosion properties. $OPTd$ is relatively long for SN~2008ag compared with the general population (median 87 days), while its light curve is much flatter than in most SNe~II (0.12$\pm$0.03\,mag/100\,d compared to a median of 1.1\,mag/100\,d). In addition SN~2008ag is slightly brighter ($M_{max}$ of $-$17.0$\pm$0.2\,mag, compared to the $-$16.7 mag median). The spectral sequence displayed in Fig.~\ref{08aglcspec} is typical of SN~II line-formation evolution and its velocities are close to the median values. Modelling from M22b gives a high ZAMS mass of 20.9$\pm$0.9\msun\ (the highest extracted mass in the CSP-I sample) and an explosion energy of 0.9$\pm$0.03\,foe. These values are a consequence of the long and bright $OPTd$ together with `normal' velocities of SN~2008ag.\\
\indent ASASSN-15bb: Finally, we show a SN~II from the CSP-II (ASASSN-15bb) in Fig.~\ref{15bblcspec}. This is specifically shown to advertise an example of the NIR spectral sequences available for the CSP-II SN~II sample (as presented and analysed in \citealt{dav19a}). As with SN~2008ag, ASASSN-15bb shows a classical SN~II morphology -- both visually and quantitatively. As is common for the CSP-II sample, ASASSN-15bb was observed relatively early post explosion. After an unclear peak epoch this SN~II had a relatively flat light curve during its $OPTd$, with an $s_2$ of 0.48$\pm$0.06\,mag/100\,d. Its photospheric phase had a typical length for a SN~II of 82$\pm$4 days, while it was relatively bright -- being brighter than $-$17 mag ($V$) through the duration of $OPTd$. The NIR spectral sequence displayed in Fig.~\ref{15bblcspec} is dominated by the Paschen series -- as is normal for SNe~II observed at such wavelengths and epochs.


\subsection{A new sample of fast-declining SNe~II}
\begin{figure}
\centering
\includegraphics[width=8.5cm]{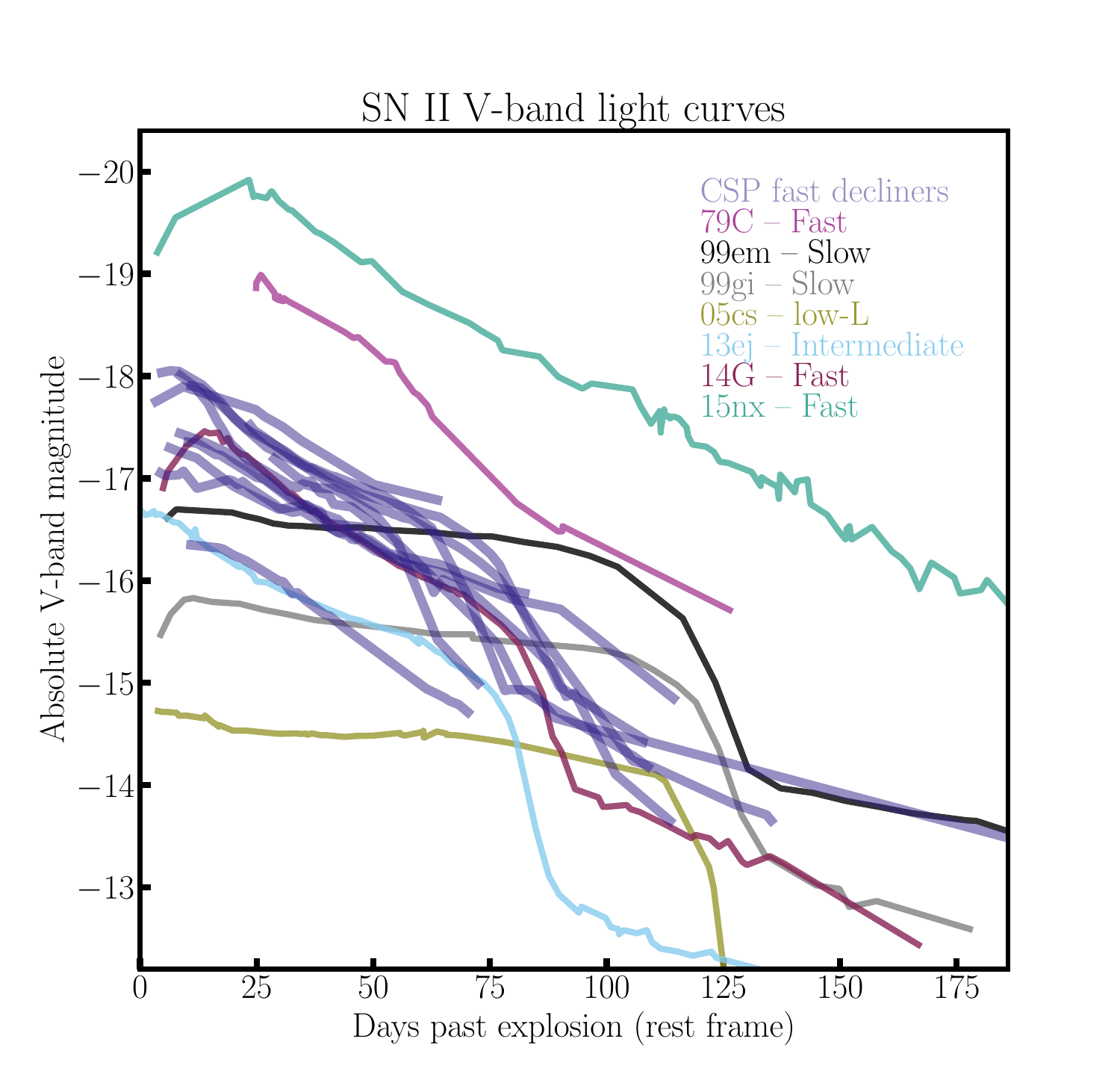}
\caption{$V$-band absolute magnitude light curves of the CSP fast-declining SN~II subsample, together with the comparison literature SNe.}
\label{fastlcs}
\end{figure}

\begin{figure}
\includegraphics[width=8.5cm]{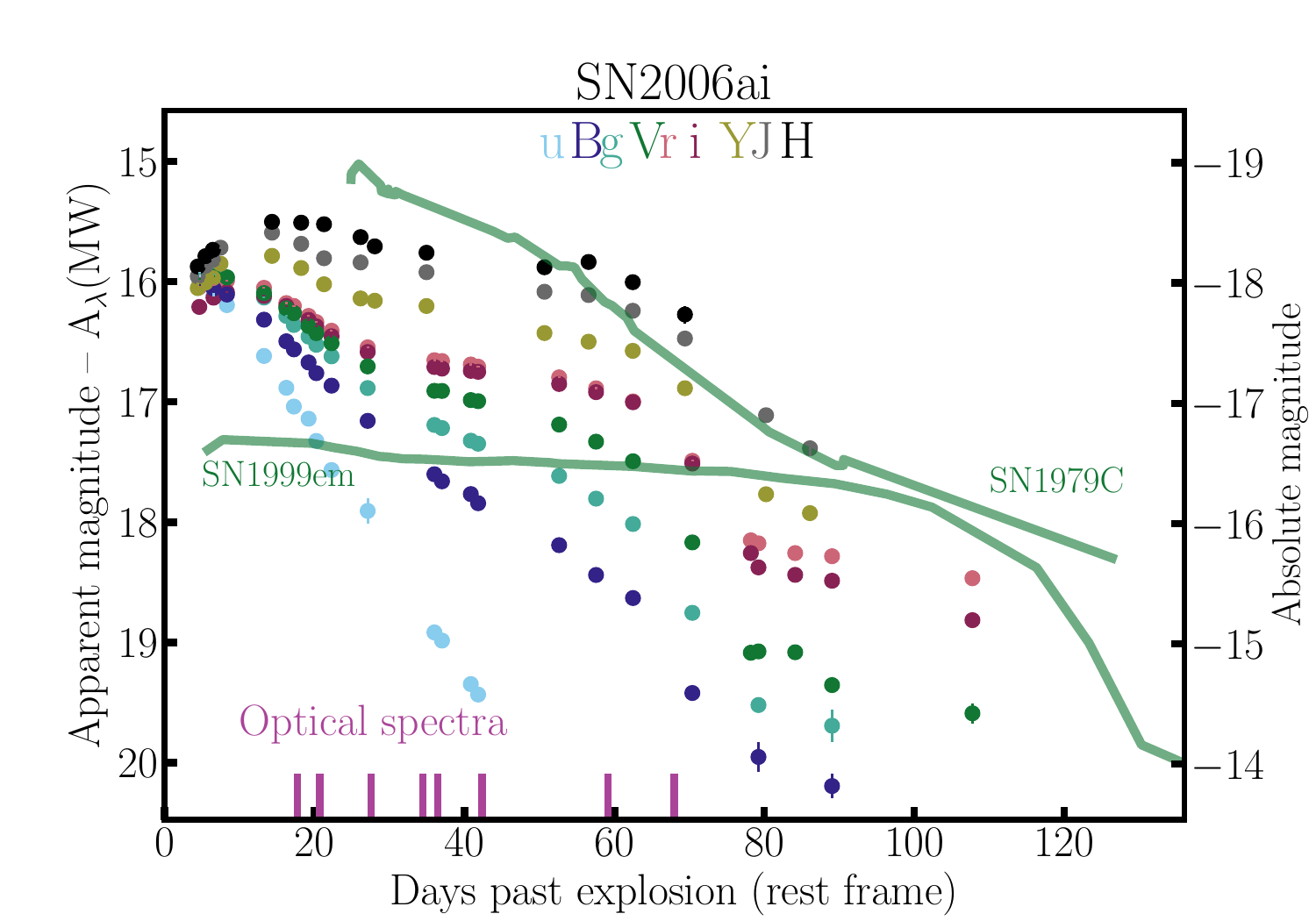}
\includegraphics[width=7.5cm]{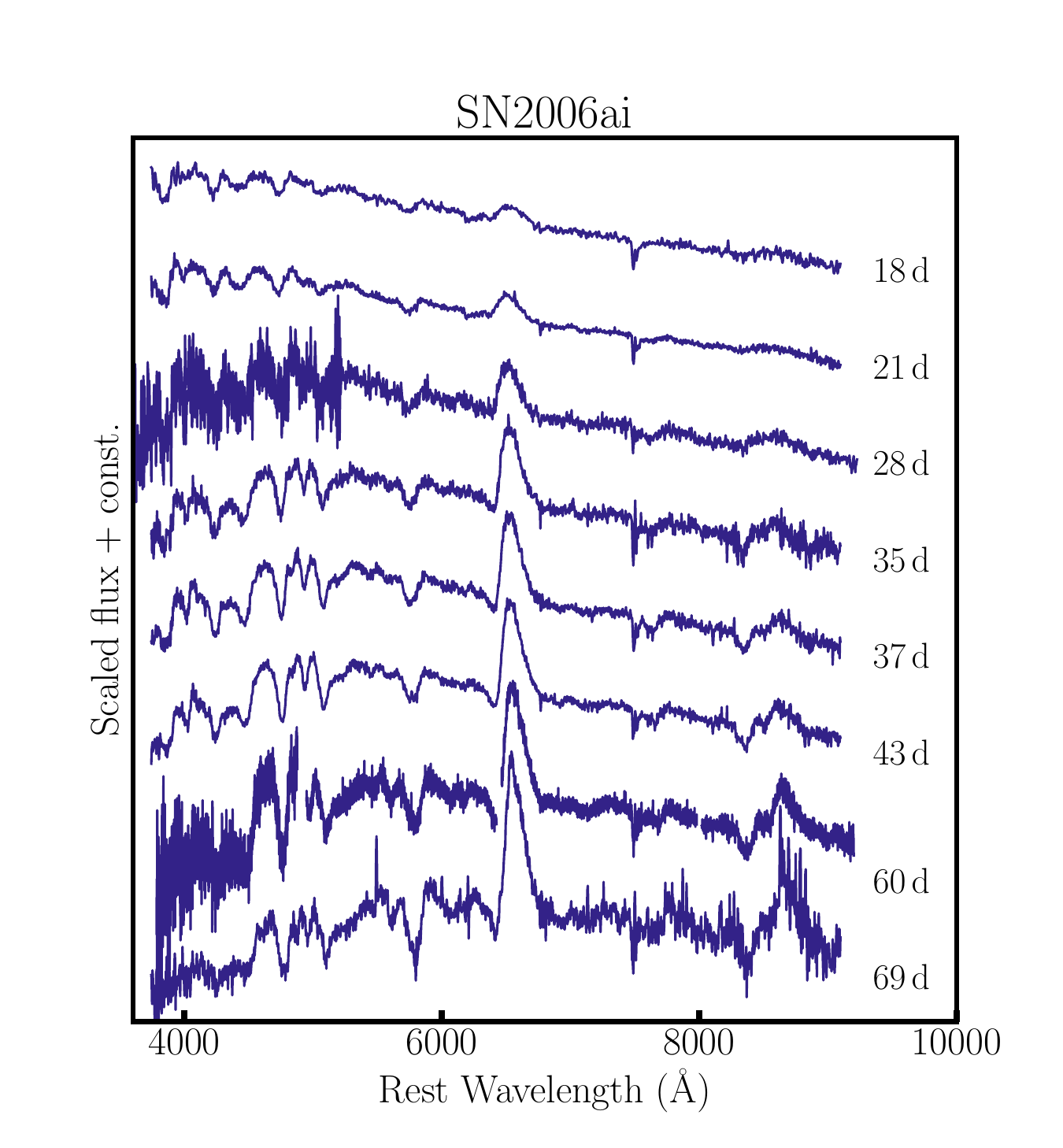}
\caption{Absolute and apparent magnitude $uBgVri$ and $YJH$ light curves (top) and visual-wavelength spectra (bottom) of SN~2006ai. The $V$-band absolute magnitude light curves of SN~1979C and SN~1999em are also displayed for comparison.}
\label{06ailcspec}
\end{figure}

\begin{figure}
\includegraphics[width=8.5cm]{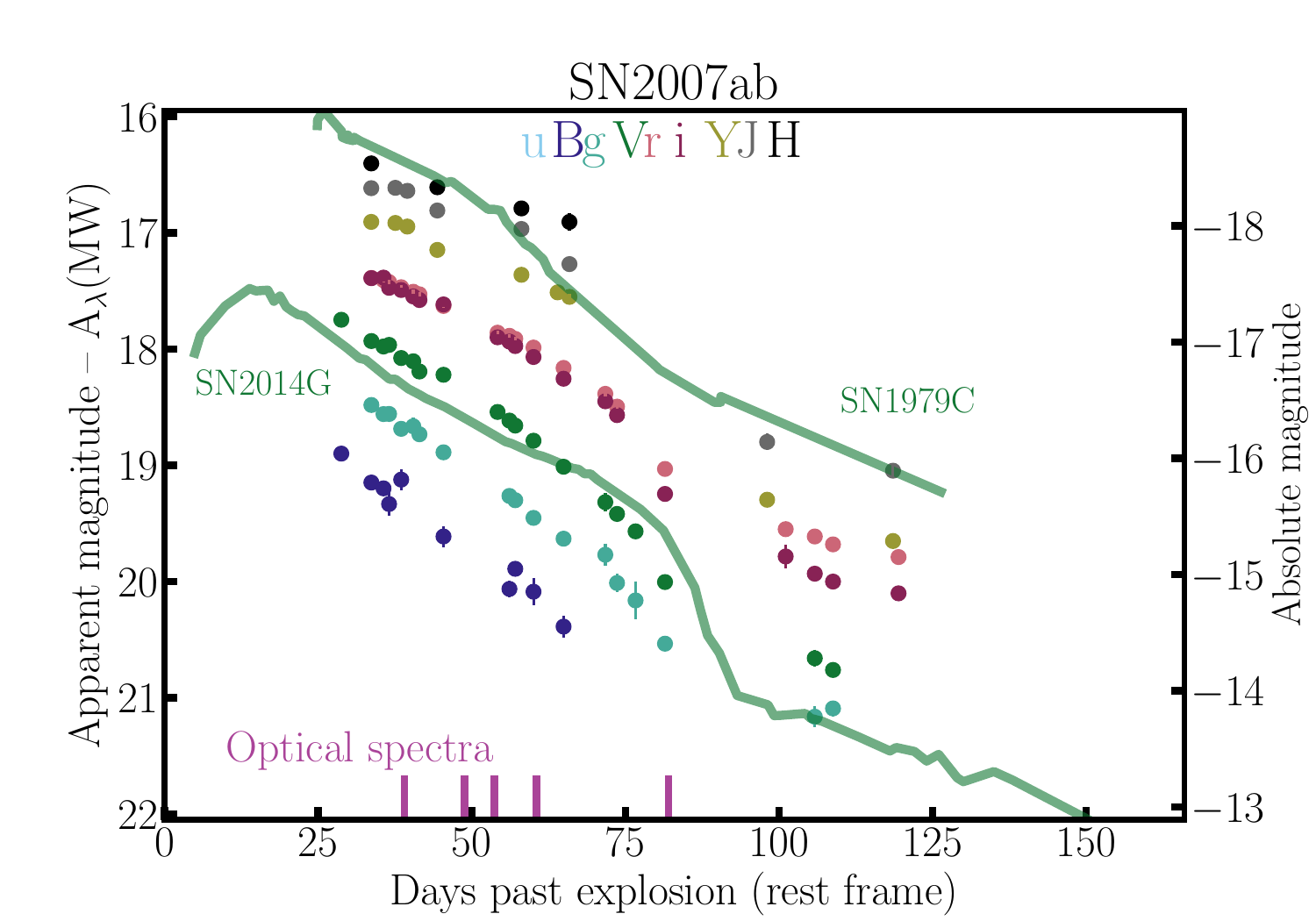}
\includegraphics[width=8.5cm]{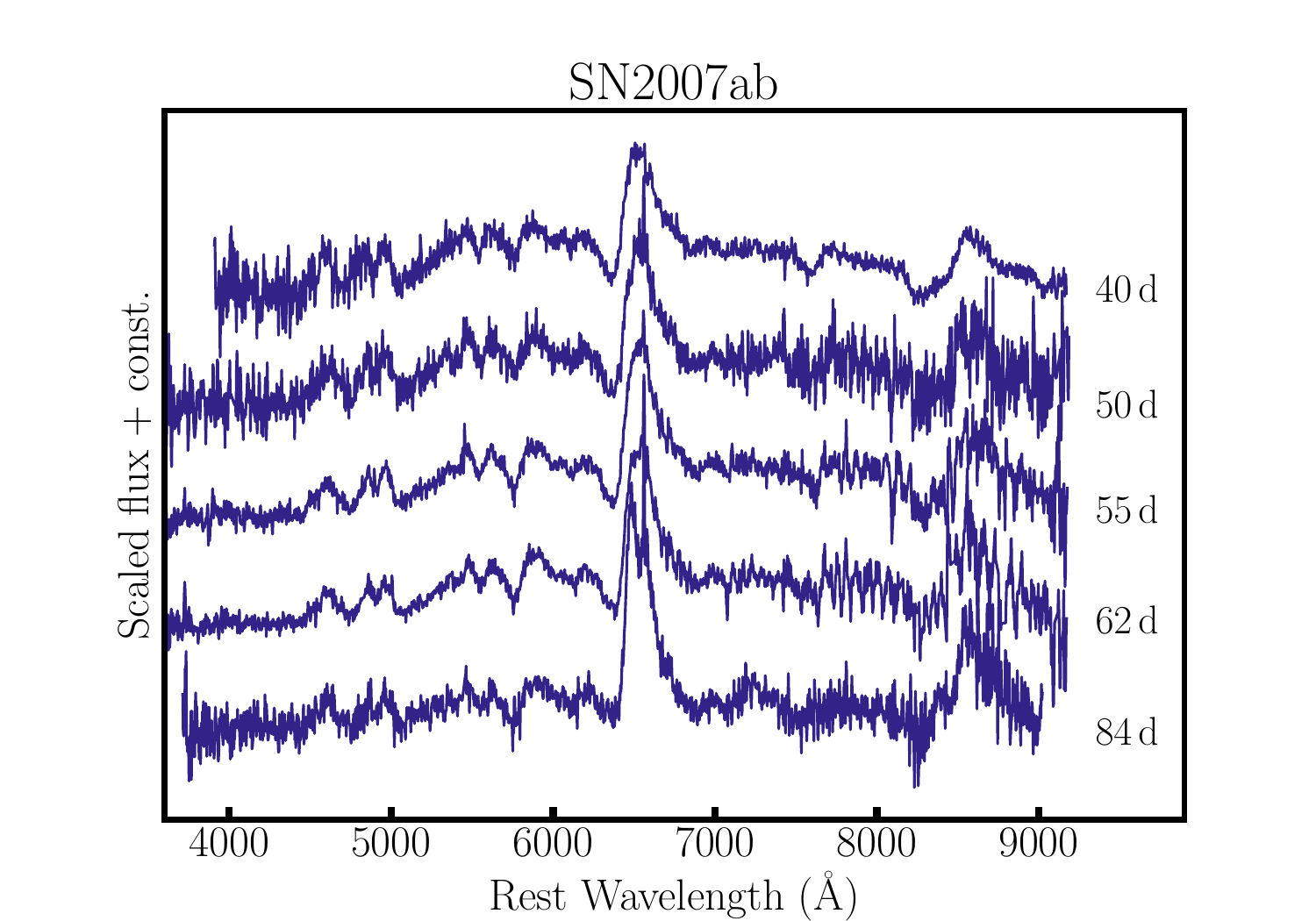}
\caption{Absolute and apparent magnitude $uBgVri$ and $YJH$ light curves (top) and visual-wavelength spectra (bottom) of SN~2007ab. The $V$-band absolute magnitude light curves of SN~1979C and SN~2014G are also displayed for comparison.}
\label{07ablcspec}
\end{figure}

\begin{figure}
\includegraphics[width=8.5cm]{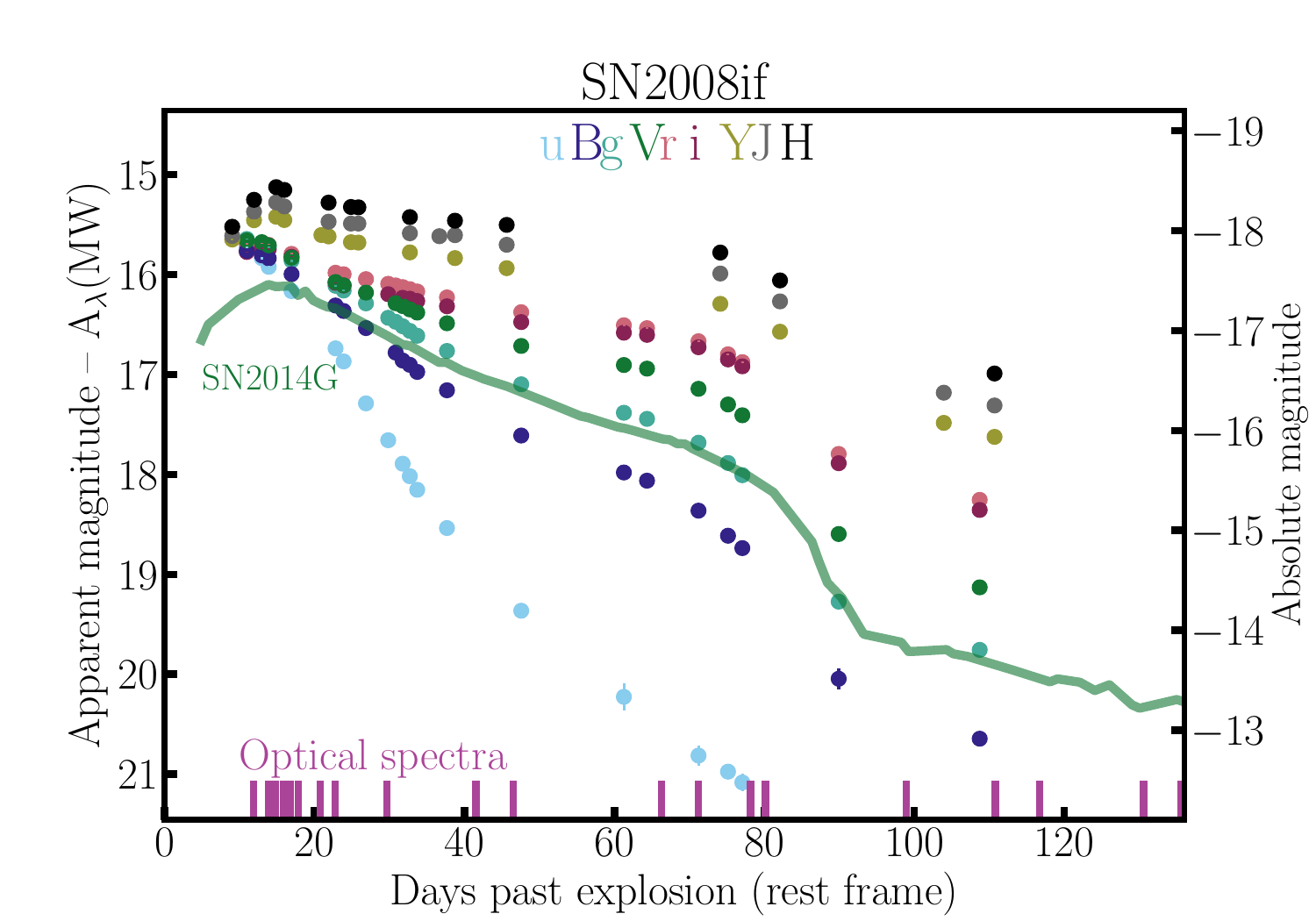}
\includegraphics[width=7.5cm]{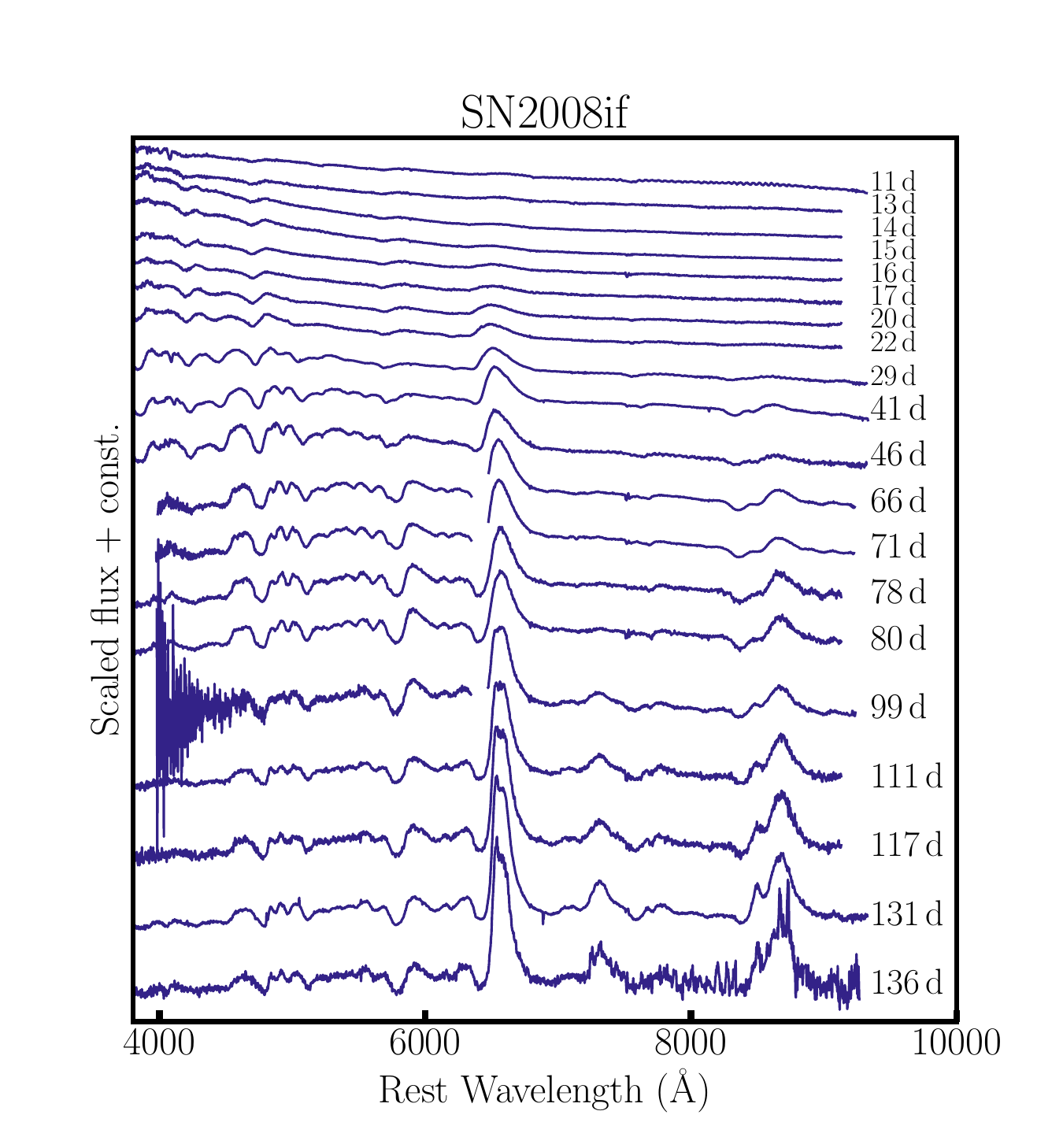}
\caption{Absolute and apparent magnitude $uBgVri$ and $YJH$ light curves (top) and visual-wavelength spectra (bottom) of SN~2008if. The $V$-band absolute magnitude light curve of SN~2014G is also displayed for comparison.}
\label{08iflcspec}
\end{figure}

\begin{figure}
\includegraphics[width=8.5cm]{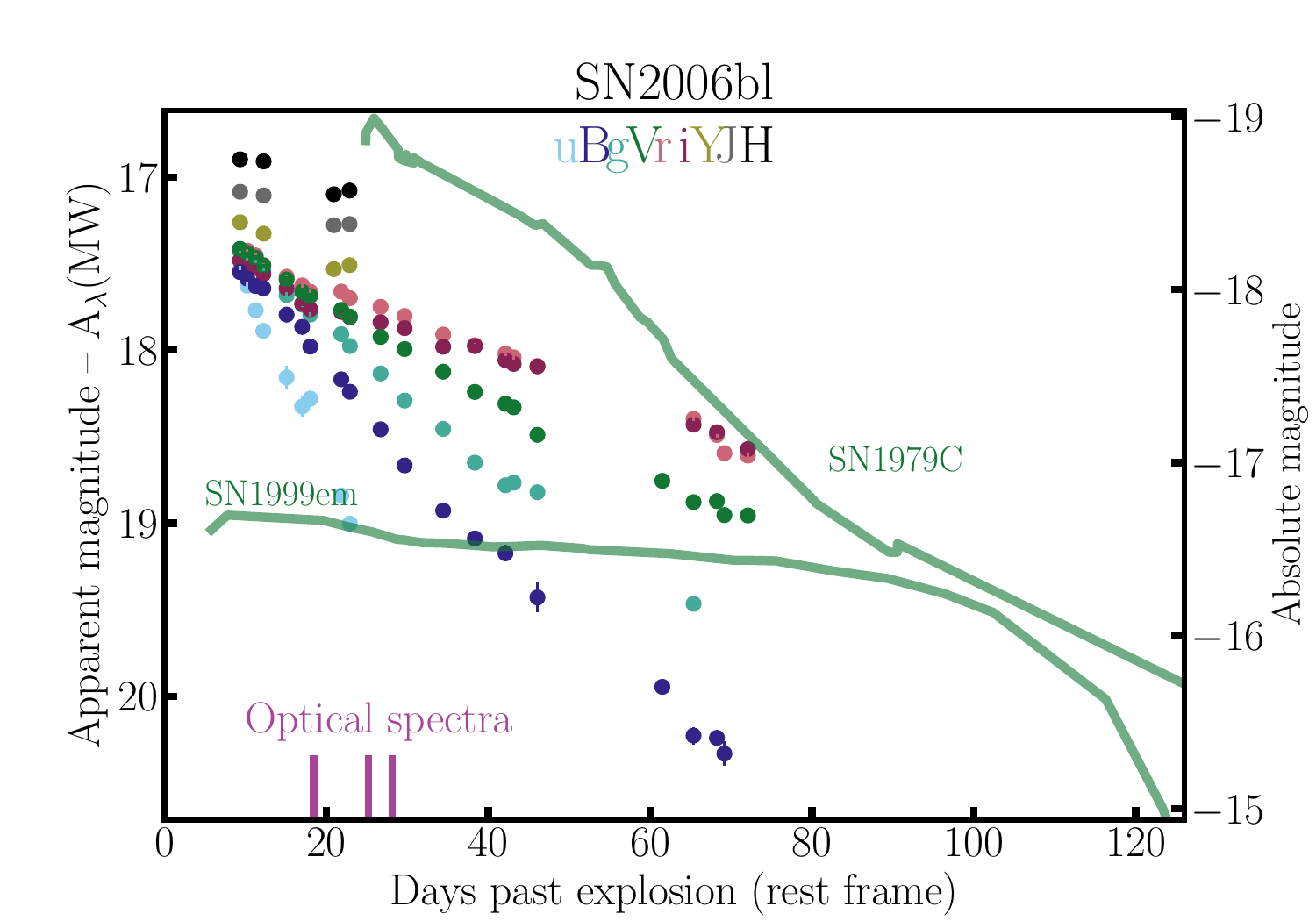}
\includegraphics[width=8.5cm]{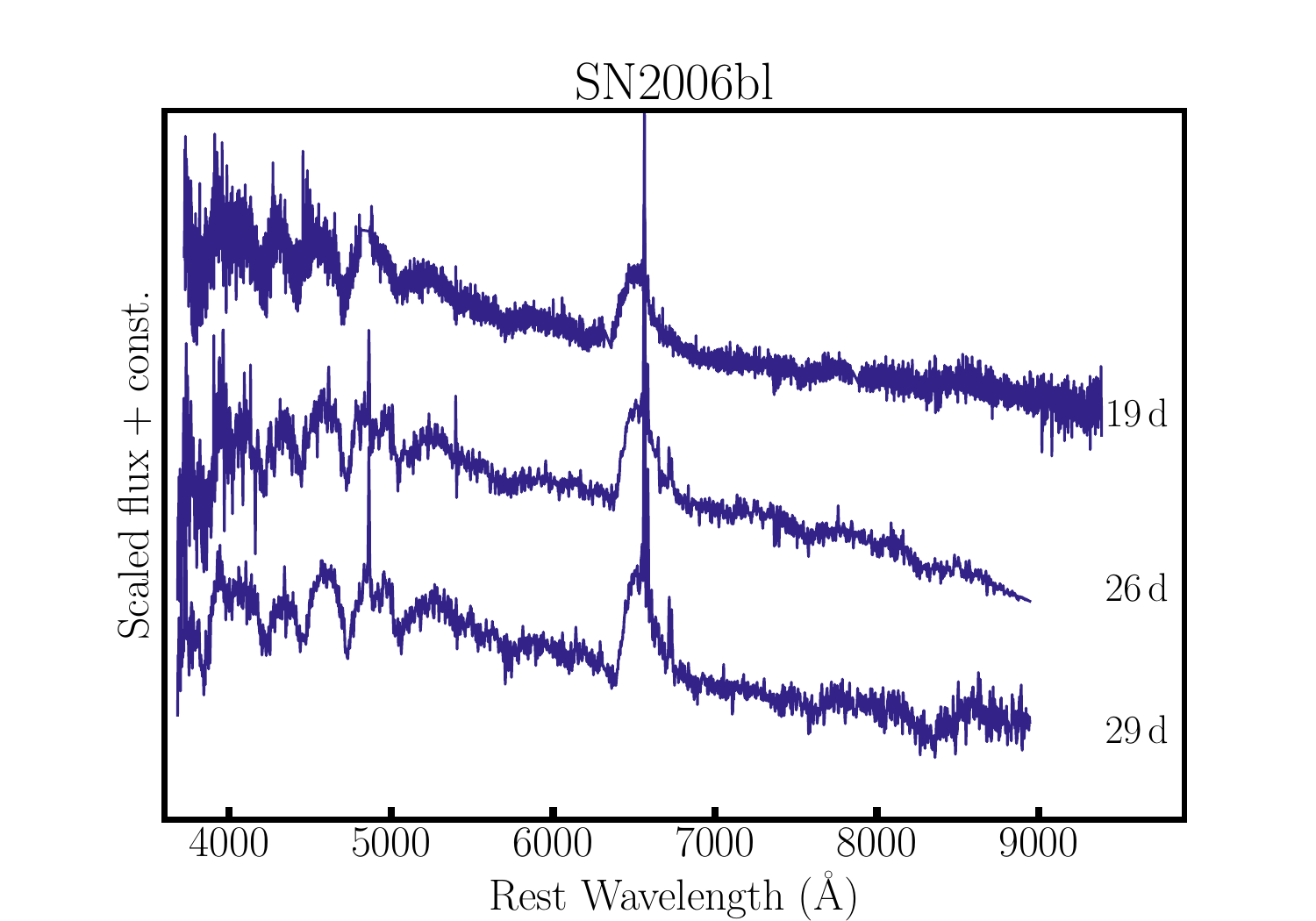}
\caption{Absolute and apparent magnitude $uBgVri$ and $YJH$ light curves (top) and visual-wavelength spectra (bottom) of SN~2006bl. The $V$-band absolute magnitude light curves of SN~1979C and SN~1999em are also displayed for comparison.}
\label{06bllcspec}
\end{figure}

SNe~II show a continuum in the vast majority of their observational properties -- there is no clear separation between slowly declining and fast-declining events (e.g. A14, \citealt{gut17b}). Thus, one may question the validity of discussing different subsets of the general SN~II population. However, while a clear continuum exists in SN~II light curve properties, the decline rate ($s_2$) of a SN~II correlates with many other photometric and spectral properties (see details in \citealt{gut17b}), and thus it is also enlightening to discuss how SNe with different decline rates differ in other observed properties.\\
\indent Here, we define a subset of the CSP SN~II sample as `fast decliners', present example cases, and discuss their properties. This is done to (a) give the reader an overview of the fast-declining population within the CSP (and indeed the literature) and (b) present data of such well-observed events that the community may wish to model and/or use as a prototypical sample of fast-declining SNe~II. This subset is defined as being true `fast decliners' with high $s_2$ values, but additionally being well observed (to allow proper characterisation and/or future modelling).\\ 
\indent We first select the fastest decliners in the CSP SN~II sample as those events falling within the highest 20\%\ of the $s_2$ distribution (the lowest value thus being 2.0$\pm$0.02\,mag/100\,d for SN~2008if)\footnote{We stress again that any cut in $s_2$ to divide a SN~II sample is completely arbitrary.}. Then, we remove SNe without any optical photometry past +50\,d and with fewer than five optical or NIR spectra. This leaves us with a sample of 10 SNe~II (listed starting with the fastest decliner): SN~2007ab, SN~2009au, SN~2008K, SN~2008gi, SN~2007ld, SN~2007U, SN~2008aw, SN~2005lw, SN~2006ai and SN~2008if. SN~2006Y also falls within this selection (and is included in the subsample); however, this event is sufficiently different from all the CSP SN~II in many observational parameters that we discuss it separately below.\\
\indent Information on this sample of fast-declining SNe~II is listed in Table~\ref{tabfast}. In addition, we also highlight and discuss the fast decliner SN~2006bl. This SN~II was identified as one of the few possible `true' SNe~IIL discussed by A14 (it did not make it into the above subsample because it only has two optical spectra available). It is immediately obvious from Table~\ref{tabfast} that these fast-declining SNe~II are nearly all relatively luminous. All but SN~2009au (see additional discussion below) are brighter than $-$17 mag (in $V$) and thus brighter than the median values of the full sample. This is consistent with the observed correlation between $M_{max}$ and $s_2$ presented by A14. In addition, all of the subsample except SN~2007ab and (again) SN~2009au have higher Fe velocities than the sample median: faster decliners are also higher velocity SNe~II (see also \citealt{gut17b}). It was only possible to measure $OPTd$ values for six fast decliners. While these are generally shorter than the SN~II sample median the number of events is too low to draw strong conclusions\footnote{\cite{gut17b} presented a weak anticorrelation between $s_2$ and $OPTd$.}.\\
\indent Figure~\ref{fastlcs}, displays the $V$-band absolute magnitude light curves of our fast decliners, together with those for the comparison sample defined in the previous section. The literature SNe~II 1979C and ASASSN-15nx are clearly not `normal' hydrogen-rich events. Both are much brighter than the most luminous SNe~II in our sample. SN~1979C shows a light-curve morphology which one may argue is consistent with having a steeply declining plateau phase ending around 50 days post explosion (see fig.\,18 in A14), although the data are not particularly constraining. However, ASASSN-15nx shows no clear light-curve break and continues to display a linear decline from maximum out to more than +170\,d. While SN~1979C is often cited as the prototypical fast decliner in the literature, there are very few similar SNe~II that have been observed (although see \citealt{rey20} for a compilation of bright and fast declining SNe~II -- their figure 4). Our fast-declining sample is dimmer by at least one magnitude, with almost zero SNe~II in our sample being brighter than --18 mag ($V$ band). This probably points to the rarity of such luminous SNe~II in the local Universe (see \citealt{pes23} for recent work identifying and analysing `bright' SNe~II).
Comparing to the other light curves of SN~II in the literature in Fig.~\ref{fastlcs}, the fast decliners are brighter and appear to evolve more quickly compared with SN~1999em and SN~2005cs (as seen quantitatively above). We now discuss a few of the CSP fast decliners in more detail.\\
\indent SN~2006ai: Absolute magnitude light curves and the visual-wavelength spectral sequence of SN~2006ai are displayed in Fig.~\ref{06ailcspec}. The figure shows a few additional features common to many of the fast decliners that have not been noted above. After peak magnitude (which is well observed for SN~2006ai), this event displays a clear fast initial decline ($s_1$) before settling onto the more slowly declining $s_2$ (that is still clearly much quicker than e.g. SN~1999em in the figure). This clear $s_1$--$s_2$ break is observed in a number of other well-observed SNe~II (both slow and fast decliners: SN~2004er, SN~2006Y, SN~2008M, SN~2008aw, SN~2008if, SN~2009N). SN~2006ai evolves quickly through the transition phase, arriving to the tail before +80\,d: the overall time evolution of this event -- together with many other of the fast decliners -- is significantly faster than that of `normal' slow decliners. The spectral sequence displayed in Fig.~\ref{06ailcspec} shows the typical features of a SN~II; however, comparing it to the slow-declining SN~2005J (Fig.~\ref{05Jlcspec}), one observes that the \ha\ absorption component is very weak in SN~2006ai. \cite{gut14} indeed showed that higher-$s_2$ SNe~II generally display lower absorption to emission strengths of \ha\ (a/e) compared to the rest of the population. M22b constrains SN~2006ai to be an extremely energetic event (compared with the rest of the sample), with an explosion energy constrained to be 1.3$\pm$0.08\,foe -- consistent with the fast evolution of the different light-curve phases.\\
\indent SN~2007ab: Absolute magnitude light curves and the visual-wavelength spectral sequence of SN~2007ab are displayed in Fig.~\ref{07ablcspec}. SN~2007ab is the fastest decliner in the CSP sample, with an $s_2$ of 3.3$\pm$0.06\,mag/100\,d. It also has a relatively short $OPTd$ of 70$\pm$10\,days. As Fig.~\ref{07ablcspec} shows, SN~2007ab displays a very shallow transition from the plateau to tail phases -- such a transition would have been undetectable if less photometry had been available. Unfortunately the first few weeks of transient evolution were missed and therefore it is unclear how bright this object was at maximum. SN~2007ab shows even weaker metal line formation than SN~2006ai, while its \ha\ feature is particularly strong. \ha$_{50}$ is 9770$\pm$510\,km/s, significantly higher than the median sample value and higher than that of SN~2006ai. Hydrodynamical modelling in M22b constrains the explosion energy to be 1.2$\pm$0.07\,foe -- twice the median sample value. Indeed, a common property of fast-declining SNe~II appears to be their high estimated explosion energies (M22c).\\
\indent SN~2008if: Absolute magnitude light curves and the visual-wavelength spectral sequence of SN~2008if are displayed in Fig.~\ref{08iflcspec}. SN~2008if is extremely well observed, both in photometry and spectroscopy. It shows a very similar light-curve morphology to SN~2006ai: a peak (most easily seen in NIR bands, as at bluer wavelengths the peak is earlier), followed by a steeper decline phase ($s_1$), which then breaks to a slower decline ($s_2$) and shows a shallow transition to the radioactive tail. With an $OPTd$ of 76$\pm$6\,days, SN~2008if again shows a shorter plateau phase than the general SN~II population. Figure~\ref{08iflcspec} displays the high-cadence spectral sequence of SN~2008if, the features of which are typical for hydrogen-rich SNe~II. However, one may again note the weak \ha\ absorption -- as compared to SN~2005J (Fig.~\ref{05Jlcspec}), a previous example of a slowly declining SN~II. Once again SN~2008if is estimated (M22b) to have a relatively large explosion energy of 1.0$\pm$0.06\,foe (together with a low progenitor ZAMS mass of 10.1$\pm$0.13\msun).\\
\indent SN~2006bl: Absolute magnitude light curves and the visual-wavelength spectral sequence of SN~2006bl are displayed in Fig.~\ref{06bllcspec}. As stated above, SN~2006bl is a clear fast decliner, with an $s_2$ of 2.5$\pm$0.06\,mag/100\,d; however, given that only three spectra were obtained it did not fall within the selection criteria for our fast-declining subsample. We highlight it here as it is one of the few CSP SNe~II that one could claim shows a true classical IIL light-curve morphology. There is no clear break in the light curve defining any transition from a photospheric to tail phase. Also, unlike some of the fast decliners discussed above, SN~2006bl does not display any clear decline-rate break post peak (no clear $s_1$ to $s_2$ transition). It is also relatively bright at --18.2$\pm$0.07\,mag (median of --16.7 mag). The spectra that were obtained show strong and broad \ha\ emission (Fig.~\ref{06bllcspec}) with weak absorption (although the last spectrum is only at +29\,d). Interestingly, modelling from M22b constrains SN~2006bl to have exploded with the highest explosion energy of the CSP-I sample (1.4$\pm$0.08\,foe) and to have also arisen from a relatively high initial mass progenitor (ZAMS mass of 15.5$\pm$2.1\msun). However, the modelling for this event is not particularly well constrained, given that it is not possible to define the end of any plateau phase.\\

\begin{table*}
\centering
\caption{Fast-declining CSP SNe~II.}
\begin{tabular}{lcccccc}
\hline
SN & Host galaxy & redshift & $s_2$ (mag\,100/\,d) & $OPTd$ (days) & $M_{max}$ (mag) & Fe$_{50}$ (km/s)\\  
\hline
2005lw & IC 672 & 0.026 & 2.1(0.04) & $\cdots$ & --17.1(0.1) & $\cdots$\\
2006Y & anon  & 0.034  & 2.2(0.20) & 46(5) & --18.0(0.1)& 5080(550)\\
2006ai & ESO 005-G009 & 0.015  & 2.1(0.04) & 62(6) & -18.1(0.1) & 4620(340)\\
2007U & ESO 552-65 & 0.026 & 2.3(0.04) & $\cdots$ & --17.9(0.1)& 5440(550)\\
2007ab & MCG -01-43-2& 0.024  & 3.3(0.1) & 70(11) & --17.0(0.1)& 4780(240)\\
2007ld & anon & 0.025 & 2.7(0.04)  &$\cdots$ & --17.3(0.1)& 3590(350)\\
2008K & ESO 504-G5& 0.027 & 2.7(0.02)  &87(7) & --17.5(0.1)& 5360(440)\\
2008aw & NGC 4939 & 0.010 & 2.1(0.05)  & 75(11) & --17.7(0.2) & 4532(650)\\
2008gi & CGCG 415-004 & 0.024 &2.7(0.11) &$\cdots$ & --17.3(0.1)& $\cdots$\\
2008if & MCG -01-24-10& 0.011 &2.0(0.02) & 76(6) & --17.9(0.2)& 4760(400)\\
2009au  & ESO 443-21& 0.009&  3.1(0.02) &$\cdots$ & --16.3(0.2)& 1470(180)\\
\hline
\end{tabular}
\tablefoot{A sample of fast-declining SNe~II from the CSP. $1\sigma$ uncertainties in measured SN~II parameters are given in brackets.}
\label{tabfast}
\end{table*}


\subsection{A few peculiar SNe~II}
\begin{figure}
\includegraphics[width=8.5cm]{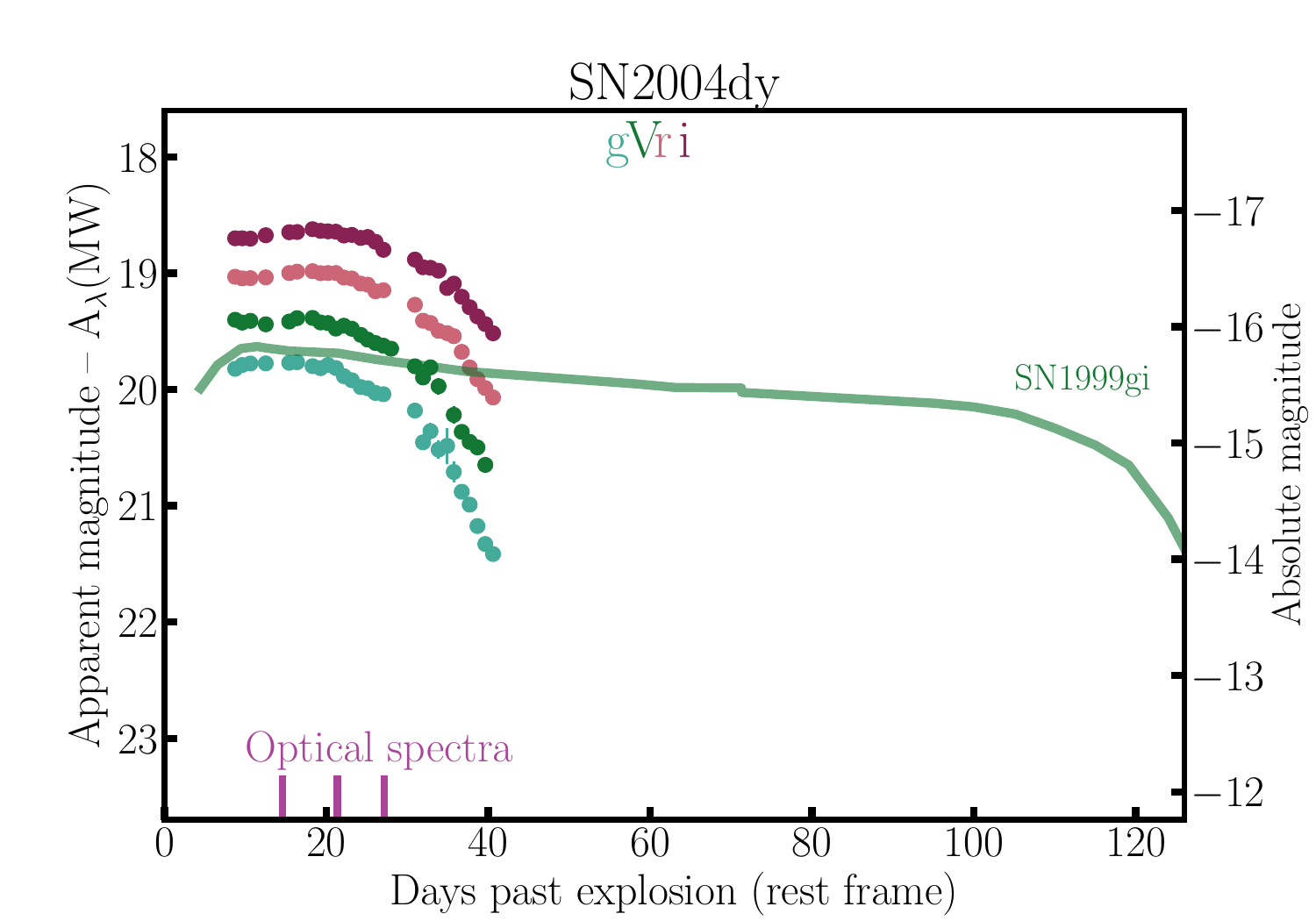}
\includegraphics[width=8.5cm]{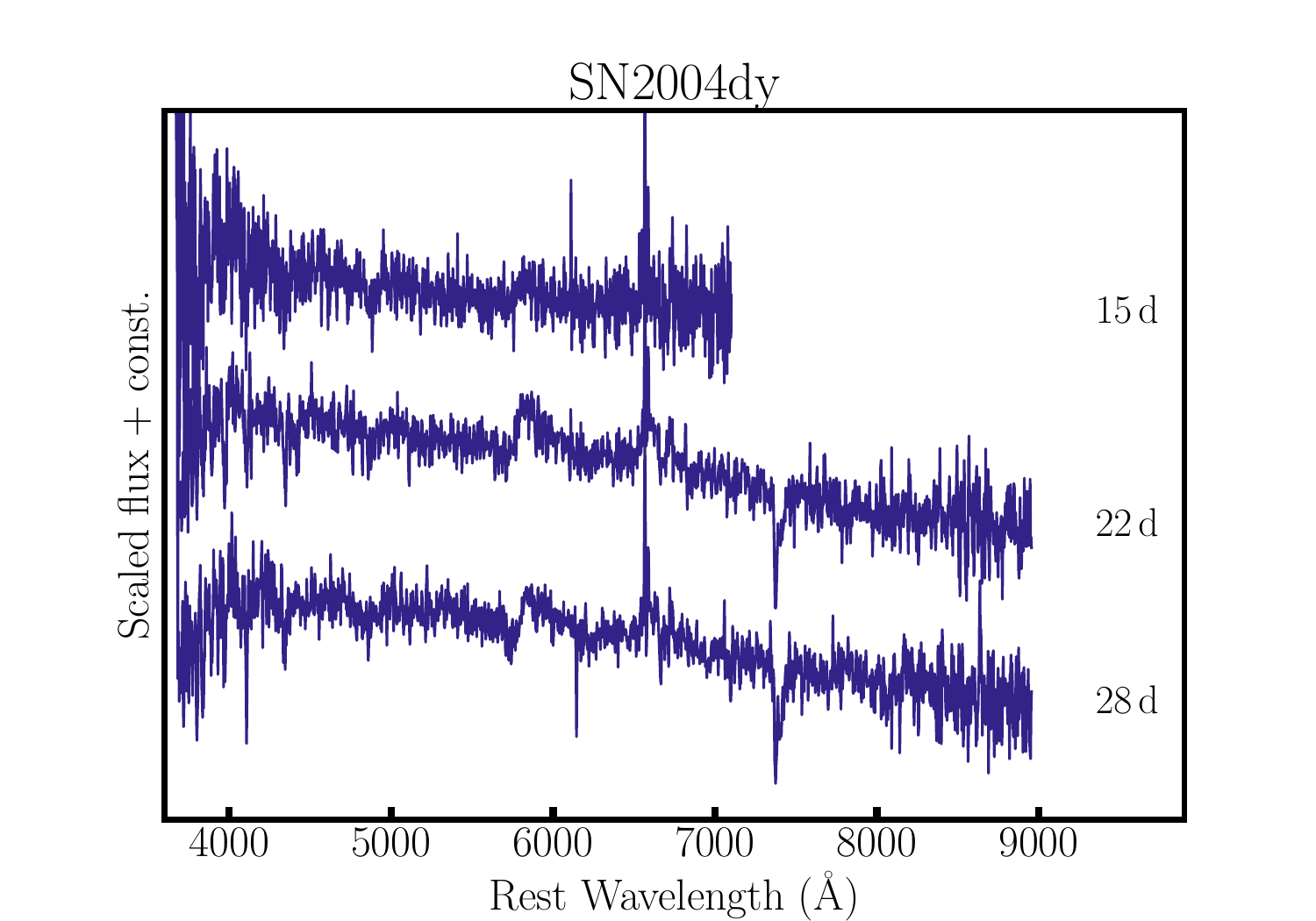}
\caption{Absolute and apparent magnitude $uBgVri$ and $YJH$ light curves (top) and visual-wavelength spectra (bottom) of SN~2004dy. The $V$-band absolute magnitude light curve of SN~1999gi is also displayed for comparison.}
\label{04dylcspec}
\end{figure}

\begin{figure}
\includegraphics[width=8.5cm]{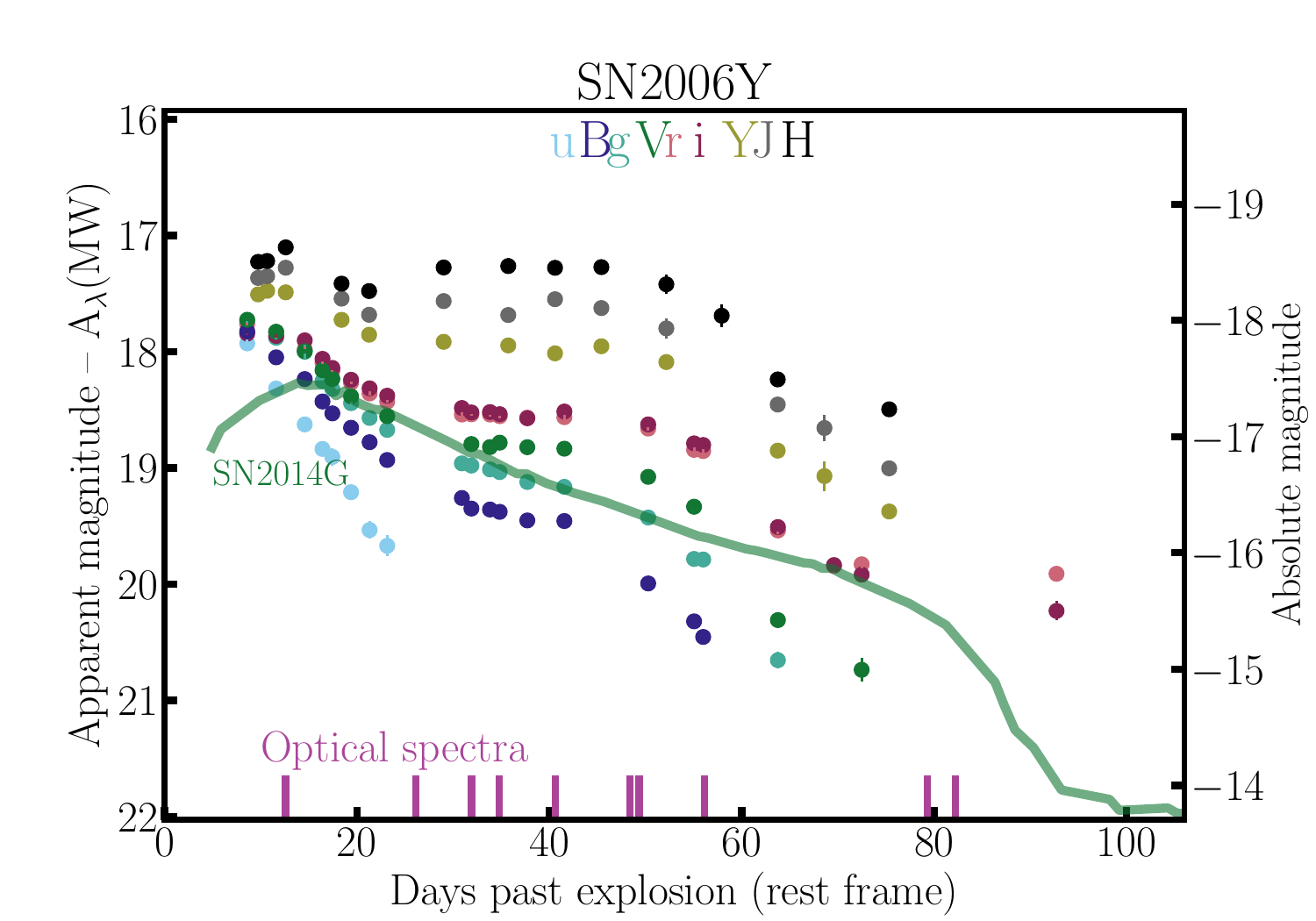}
\includegraphics[width=7.5cm]{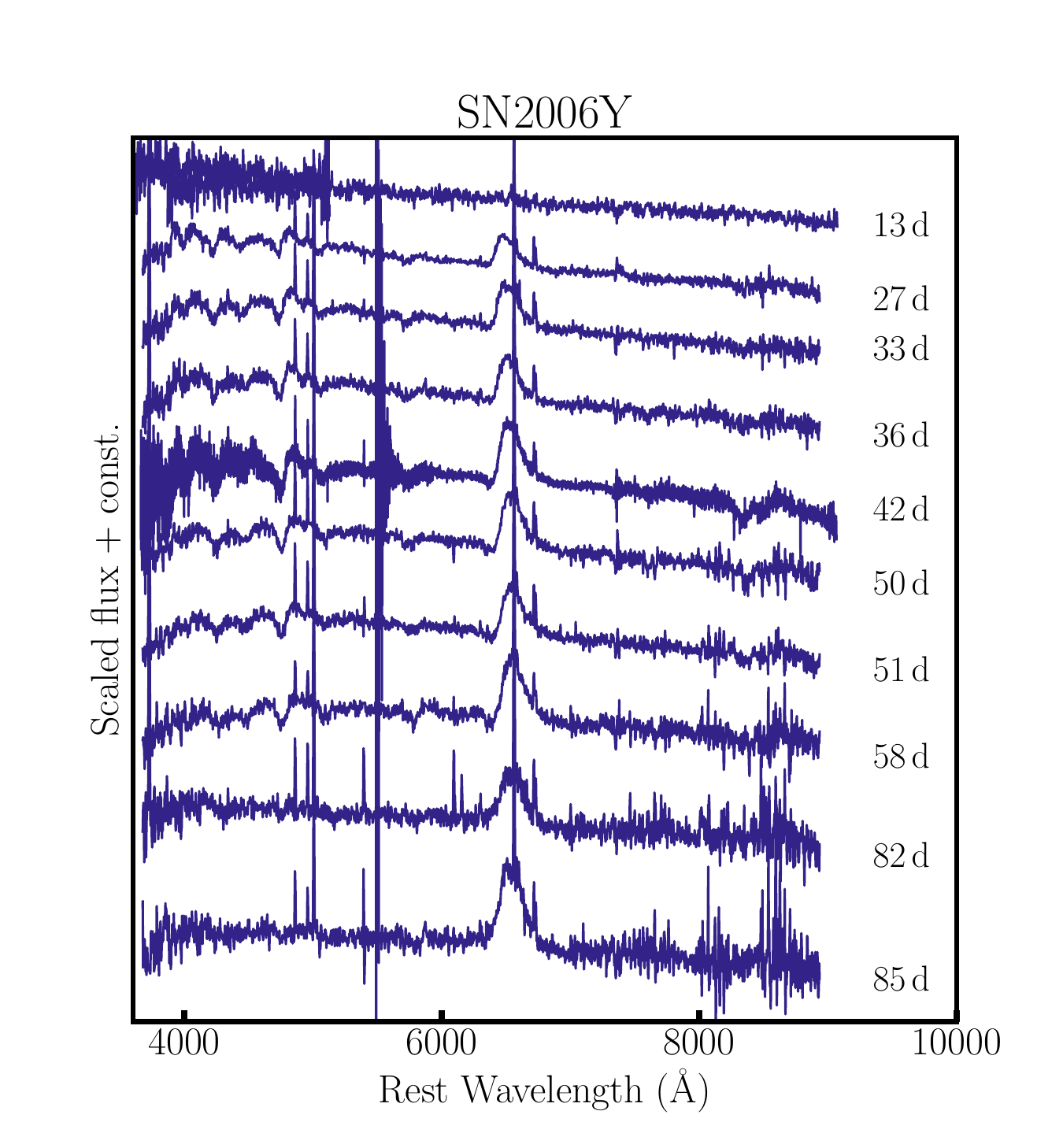}
\caption{Absolute and apparent magnitude $uBgVri$ and $YJH$ light curves (top) and visual-wavelength spectra (bottom) of SN~2006Y. The $V$-band absolute magnitude light curve of SN~2014G is also displayed for comparison.}
\label{06Ylcspec}
\end{figure}

\begin{figure}
\includegraphics[width=8.5cm]{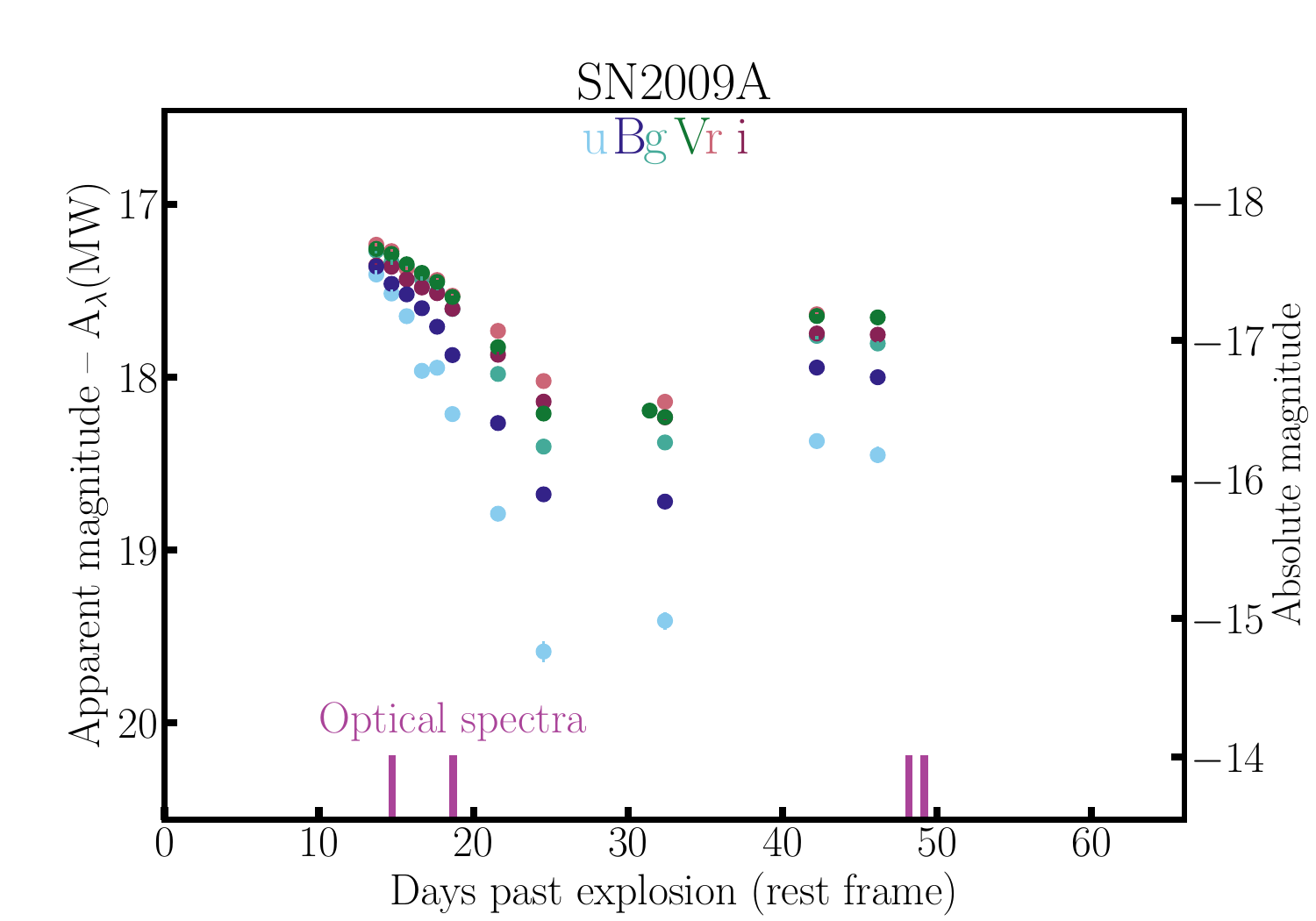}
\includegraphics[width=8.5cm]{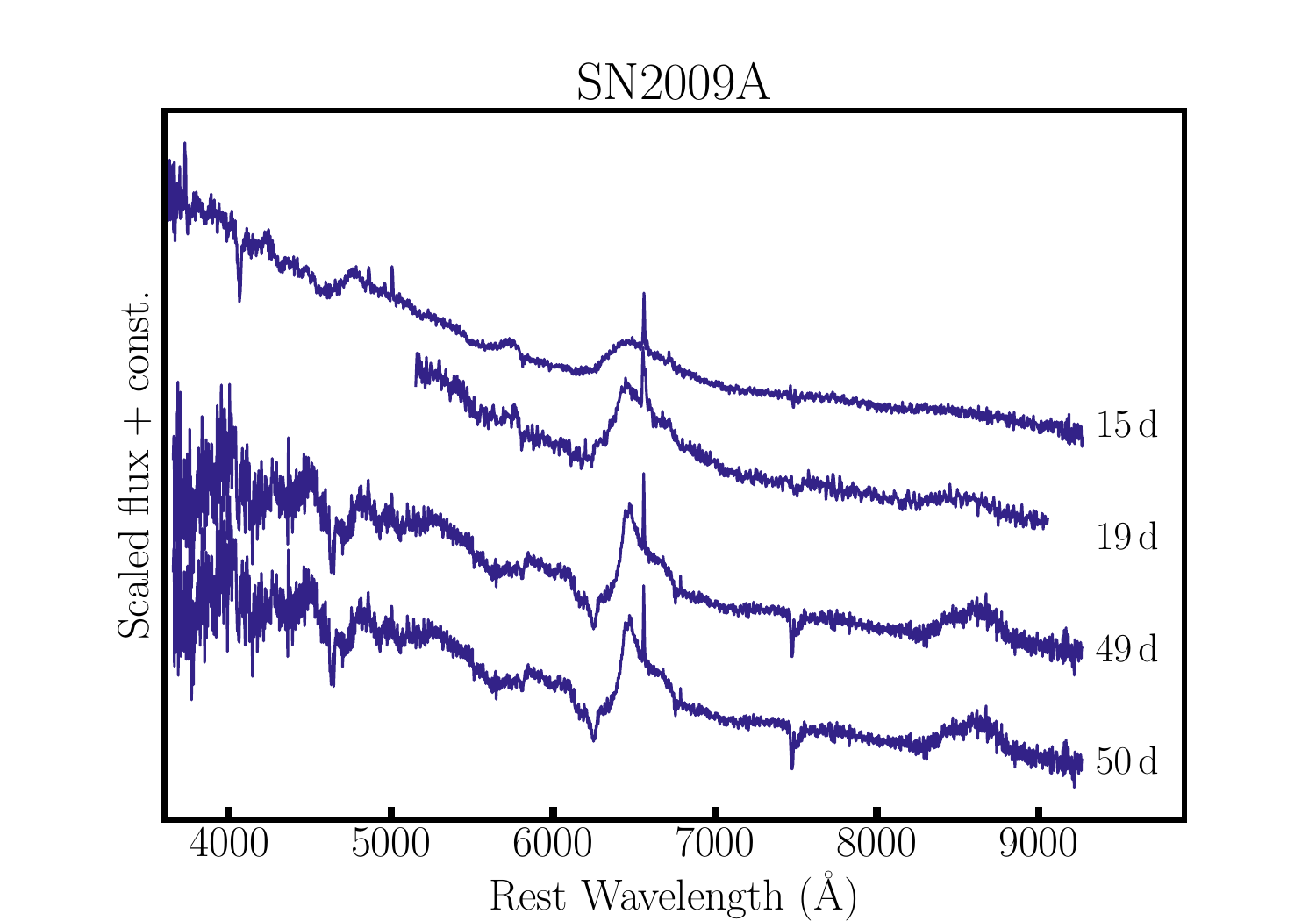}
\caption{Absolute and apparent magnitude $uBgVri$ and $YJH$ light curves (top) and visual-wavelength spectra (bottom) of SN~2009A.}
\label{09Alcspec}
\end{figure}

Finally, in this section we highlight and discuss the properties of a small number of non-standard SNe~II that are either hard to classify in the context of the previous discussion, or that show some peculiarity that sets them apart from the rest of the CSP (and indeed literature) sample.\\
\indent SN~2004dy: Absolute magnitude light curves and the visual-wavelength spectral sequence of SN~2004dy are displayed in Fig.~\ref{04dylcspec}. Unfortunately only a small amount of data were acquired for this event and thus strong conclusions on the origin of its peculiarity are not possible. The $gVri$ light curves show a flat evolution for around two weeks before the light curve starts to decline in all bands. If this is interpreted as the end of a very short $OPTd$ then such a light curve may be explained by the explosion of a progenitor with a small hydrogen envelope. Indeed, we measure an $OPTd$ of 24$\pm$6 days for SN~2004dy. However, this light-curve evolution does not match that of SNe~IIb (see, e.g. \citealt{pes19}), where the IIb class is thought to arise from progenitors with a low-mass hydrogen envelope at the epoch of explosion (inferred from their transition from hydrogen to helium line dominated spectra over the first few weeks post explosion; \citealt{fil88,fil93_2}). This difference probably implies a difference in the pre-SN progenitor structure. The spectra (Fig.~\ref{04dylcspec}) are also not particularly constraining -- they are quite noisy and end at +28\,d. The strongest line is seen just short of 6000\,\AA, and may be due to sodium or helium. \ha\ is also observed. However given the weakness of this line and the other transient properties it is not clear whether SN~2004dy is a true hydrogen-rich SN~II.\\
\indent SN~2006Y: Absolute magnitude light curves and the visual-wavelength spectral sequence of SN~2006Y are displayed in Fig.~\ref{06Ylcspec}. Comparing to the fast decliners discussed in the previous subsection, SN~2006Y may be seen as an extreme version of SN~2006ai and SN~2008if. The event discussed here is a significant outlier in most observed properties of SNe~II (see A14 and \citealt{gut17a}).
The epoch of peak brightness was probably missed (but only by a few days), then an extremely fast initial decline ($s_1$) is observed (significantly faster than in any other SN~II in the sample). It shows a quite abrupt $s_1$ to $s_2$ transition and a well-defined plateau stage. This ends at only 46$\pm$5\,days post explosion (Table~\ref{tabfast}). 
This short $OPTd$ probably implies that SN~2006Y came from a progenitor that had suffered significant envelope stripping prior to explosion (given that plateau length is associated with hydrogen envelope mass, see M22c). At the end of its $OPTd$, SN~2006Y displays a transition to a tail phase. The spectral sequence (Fig.~\ref{06Ylcspec}) is affected by significant host \hii\ region contamination. Strong \ha\ is observed that has a very weak or nonexistent absorption component. Its Fe$_{50}$ is 5080$\pm$550\,km/s, significantly higher than the sample median (3940\,km/s).
No reasonable solution was found in fitting hydrodynamical models to the observations of SN~2006Y in M22b, owing to the short plateau phase. However, in M22c additional progenitor models were used where additional mass-loss was assumed during stellar evolution, producing pre-SN progenitors with much smaller hydrogen envelopes. The best-fit model was a progenitor with a 4.4$\pm$0.36\msun\ hydrogen envelope mass (compared to 7.1\msun\ for the lowest hydrogen envelope fit to CSP-I SNe~II using standard mass-loss values). In addition, M22c constrained the explosion energy to be 1.2$\pm$0.2\,foe and the ejected $^{56}$Ni mass to be 0.08$\pm$0.004\msun. Thus, the main properties of SN~2006Y can be explained by a high-energy explosion of a progenitor with a low hydrogen envelope mass and significant ejection of radioactive material. We note that \cite{hir21} also concluded that a low hydrogen-envelope mass is required to model the data of SN~2006Y.\\
\indent SN~2009A: Absolute magnitude light curves and the visual-wavelength spectral sequence of SN~2009A are displayed in Fig.~\ref{09Alcspec}. To our knowledge, this SN~II displays a unique light curve and spectral evolution. The first photometry ($uBgVri$) was obtained just after 10 days post explosion and shows a very steep decline that is most pronounced in the bluer bands (a drop of around 1.5 mag in just over 10 days in $B$). The spectra obtained during this phase (Fig.~\ref{09Alcspec}) show strong and broad \ha\ and \hb, with the \ha\ profile evolving quite quickly between +15\,d and +19\,d -- the peak becomes sharper by the latter epoch. Between +25 and +33\,d the decline halts (unfortunately, no spectra were obtained during this phase).
Then, the light curve rises again by around a magnitude in 10 days. The next photometric epoch stays at almost the same brightness, and no more photometry was obtained. At the end of the observed light curve two additional spectra (taken only a day apart) show what seems to be a double P-Cygni \ha\ profile superimposed onto one another -- probably implying two distinct line forming regions. The emission peak of \ha\ is blueshifted by $\sim$5500\,km/s. Such blueshifts are a common feature of SNe~II (see \citealt{and14b}). However, the velocity is extremely high compared to any other SN~II. The narrower P-Cygni profile of \ha\ has an absorption component that is blueshifted by nearly 15,000\,km/s at around +45\,d. This is $\sim$6000\,km/s faster than any other CSP SN~II. In summary, SN~2009A shows a light curve evolution and spectral properties that are not observed in any other event in our sample, and that to our knowledge have not been observed in any other literature SN~II. It is unclear what progenitor and explosion properties could produce such transient behaviour, and we encourage future modelling of this event.\\


Other mentionable CSP SNe~II: Before concluding, we very briefly discuss a few other CSP SNe~II that do not appear to fall within the main observational parameter space of hydrogen-rich SNe~II. As described above, a study of three SNe~II was published by \cite{rod20a}: SNe~2008bm, 2009aj and 2009au. Those three were named `luminous for their low expansion velocities' SNe~II in reference to them being outliers in the luminosity-velocity correlation used in the application of SNe~II as distance indicators (see, e.g. \citealt{ham02}). That study concluded that the behaviour of these events can be explained by significant ejecta-CSM interaction\footnote{SNe~II where narrow emission lines are observed in spectra obtained within the first week post explosion can be explained by the presence of significant ejecta-CSM interaction.} that boosts the early luminosity and also decelerates the ejecta to produce low velocities. SN~2008bu is a fast decliner with an $s_2$ of 2.4$\pm$0.02\,mag/100\,d -- but it was not part of the above subsample owing to a lack of sufficient spectral observations. This event shows an extremely short $OPTd$ of only 44$\pm$8\,days. 
Similarly to SN~2006Y above, SN~2008bu was modelled by M22c using a progenitor with enhanced mass loss -- to enable fitting to its short plateau phase. Finally, SN~2009ao displays a strange light curve that looks somewhat similar to that of SN~2004dy presented above. It has an $OPTd$ of 42$\pm$6\,days, however it is unclear whether such a phase measurement is applicable here. Compared to SN~2004dy, SN~2009ao displays a standard spectral sequence (see \citealt{gut17a}). Given the lack of data after +60\,d, it is unclear how to interpret this event.

\section{Summary and future outlook}
\label{conclusions}
We have presented $uBgVri$ optical and $YJH$ NIR photometry for a total of 94 SNe~II from two campaigns of the Carnegie Supernova Project (CSP-I and CSP-II). This low-redshift sample is characterised by being of high quality and cadence for the first $\sim$150 days post explosion. While most of the data have been quantitatively analysed and robustly modelled elsewhere, here we present the final photometry for widespread use, and have provided an analysis of the sample properties. A descriptive analysis was presented to summarise the CSP SN~II sample and the SN~II phenomenon in general, while we have additionally highlighted a number of standard and nonstandard SNe~II that aid in understanding the diversity observed within hydrogen-rich SNe~II.\\
\indent With the advancement in survey detection and follow-up efficiency, the number of SNe classified as type II can now be counted in the thousands\footnote{See the Transient Name Server, TNS, for lists of SN discoveries and classifications: \url{https://www.wis-tns.org/}.}. The number of well-observed nearby events with multiwavelength photometry is (with the current data release) around several hundred SNe~II. Given the large samples of `standard' SNe that are now available in the literature (SNe~II as discussed here, but also SNe~Ia and SE-SNe), SN science has started to focus on observing and understanding nonstandard explosions and/or to investigate observational parameter space previously unexplored (e.g. very early times within hours of explosion). Both of these avenues of research are being aided by the increased capabilities of surveys that enable a significant fraction of the observable sky to be scanned at a frequency of one day or less (see e.g. ASASSN, \citealt{sha14,koc17}; ATLAS, \citealt{ton18,smi20}; ZTF, \citealt{bel19}; KMTNet, \citealt{afs19}). Such surveys also provide optical light curves of transients (down to their survey limits), meaning that the number of SNe with useable light curves increases rapidly each year. However, those light curves have usually at most two or three wavelength bands and thus only afford a first-order characterisation of transient behaviour. The number of events that will have spectral sequences is much lower owing to the expensive nature of spectroscopic observations. Thus, to enable the optimal science extraction from future SN~II samples, a well-defined comparison and calibration sample is required. The current CSP SN~II data release fulfils this requirement. Of particular importance for modelling SNe~II to understand physical parameters of the explosions is the estimation of bolometric light curves. The $u$ to $H$ wavelength range of the CSP enables this sample to be used to calibrate lower-quality (in terms of wavelength coverage) photometry and obtain bolometric fluxes following the methodologies outlined by M22a\footnote{M22a showed the importance of NIR observations for estimating the bolometric flux.}.\\
\indent The coming years will welcome the arrival of another leap in survey capability with the Legacy Survey of Space and Time, LSST \citep{abe19} to be undertaken by the \textit{Vera C. Rubin} Observatory (together with a number of other upcoming facilities that will also change the transient landscape). LSST will provide multiwavelength optical light curves for many thousands of SNe~II each year. These data will enable much larger population studies than currently possible. However, spectroscopy will be scarce, as will NIR photometry (as they would require separate follow-up campaigns).
Thus, a full understanding of future statistical light-curve samples will only be possible by comparison to high-quality samples such as that presented in the current work (and other examples in the literature).\\
\indent As discussed in the introduction, SNe~II -- and indeed the CSP sample -- have been used to measure distances and have been proposed to be used to measure galaxy metallicities. The full potential future use of hydrogen-rich SNe in this direction, out to high redshift, will require that the high-redshift events are calibrated against nearby samples where high-quality light curves and spectra exist, such as the CSP SN~II database.\\
\indent The utility of precise photometry also requires accurate measurements of the transmission functions which define the natural photometric system. Transmission functions are crucial for k-corrections. Type II supernovae can have strong emission lines, which if they fall near the edge of a filter function, can introduce 0.1\,mag or larger errors if the transmission function is not well calibrated.\\ 
\indent Throughout this paper, we have reiterated the high-quality nature of this data set, implying it can be used to aid in investigating a number of astrophysical and cosmological questions. However, as suggested above, while the CSP SN~II sample is of high quality in some important aspects, it does not cover the full observational parameter space of interest. The CSP-I sample (that makes up $\sim$3/4 of the full sample) has many events where follow-up was only started a week or more after explosion. During the time of that initial project, detecting SNe within a few days of explosion was relatively rare. Thus, the CSP SN~II data presented cannot be used to probe the early time evolution of these hydrogen-rich events. Such observations are of significant interest as they can elucidate the progenitor structure -- either the pre-SN radius or the extent and properties of any significant CSM close to the exploding star. Many projects are now concentrating on these early times as the scientific payoff is potentially significant. Indeed, the POISE collaboration (Precision Observations of Infant Supernova Explosions; \citealt{bur21b}) is focusing on such observations. \\
\indent While a number of studies have already been published using the current data (see Section~\ref{previous}), we envisage that additional valuable analysis is possible. M22a presented bolometric light curves that others may wish to model (together with the spectral velocities available from \citealt{gut17a}). The modelling achieved by M22b assumed standard single-star evolution parameters to create progenitors to explode and thus only sampled a constrained region of the possible progenitor parameter space.
The results of M22b and M22c should be tested by independent modelling.
The majority of the highlighted peculiar SNe discussed in the previous section warrant detailed modelling to understand which combinations of progenitor and explosion properties are able to produce such transients (of specific note is SN~2009A). There are also observational properties of SNe~II that deserve additional attention and could be addressed using this sample. For example, to date CSP SN~II studies have not analysed the shape and depth of the plateau to tail transition -- this clearly changes significantly from one SN~II to the next. It is unclear what physical parameters cause such diversity.\\
\indent In conclusion, this paper has presented a legacy data set of SNe~II: almost 100 SNe with $u$ to $H$ wavelength coverage and with high cadence and high-precision photometry from one week from explosion to more than 100 days later. The photometry is accompanied by optical and NIR spectroscopy previously analysed and released by \cite{gut17a} and \cite{dav19a}, respectively. This data set provides a baseline standard for future SN~II samples, and has the potential to be critical for the future use of SNe~II as astrophysical markers out to cosmological distances. While our understanding of SNe~II has significantly advanced in the last few decades (e.g. constraints on progenitors, constraints on CSM surrounding those progenitors), there are still many uncertainties in the exact mapping from exploding progenitors to observed transients. Additional observations will continue to further our understanding of the SN~II phenomenon, but data sets such as that presented here will provide the foundations on top of which future advances in our knowledge of the explosions of massive hydrogen-rich stars can be built.\\

\section{Data availability}
Photometry for local-sequence stars and for all SNe~II presented in this paper can be found at the CDS (Strasbourg astronomical Data Center, \url{https://cds.unistra.fr//}), and the same data can be downloaded from the CSP website: \url{https://csp.obs.carnegiescience.edu/data}.
Figures of the optical and NIR apparent and absolute magnitude light curves of the CSP SN~II sample can be found on Zenodo at the following link:
\url{https://zenodo.org/records/13743220}.
Finder charts for all SNe~II in our sample can be found on Zenodo at the following link: \url{https://zenodo.org/records/13743254}.
Tables listing important information on all SNe~II in the CSP sample can be found on Zenodo at the following link: \url{https://zenodo.org/records/13743303}.

\begin{acknowledgements}
This project was supported by the U.S.\,National Science Foundation (NSF) under grants AST-0306969, AST-0607438, AST-1008343, AST-1613426, AST-1613455, and AST-1613472.
The CSP-II was also supported in part by the Danish Agency for Science and Technology and Innovation through a Sapere Aude Level 2 grant.
M.D.S. is funded in part by  a research Project 1 grant from the Independent Research Fund Denmark (IRFD grant number 8021-00170B ) and by a VILLUM FONDEN Experiment (grant number 28021). S.G.G. acknowledges support by FCT under Project CRISP PTDC/FIS-AST-31546/2017 and FCT Project~No.~UIDB/00099/2020.
CPG acknowledges financial support from the Secretary of Universities
and Research (Government of Catalonia) and by the Horizon 2020 Research
and Innovation Programme of the European Union under the Marie
Sk\l{}odowska-Curie and the Beatriu de Pin\'os 2021 BP 00168 programme,
from the Spanish Ministerio de Ciencia e Innovaci\'on (MCIN) and the
Agencia Estatal de Investigaci\'on (AEI) 10.13039/501100011033 under the
PID2020-115253GA-I00 HOSTFLOWS project, and the program Unidad de
Excelencia Mar\'ia de Maeztu CEX2020-001058-M.
N.B.S., K.K., D.D., and S.U. acknowledge the support of the Mitchell Institute for Fundamental Physics and Astronomy. They also acknowledge the Mitchell Foundation and Sheridan Lorenz for funding workshops at the Mitchell Cooks Branch Nature Preserve. 
C.A. is supported by NSF grants AST-1907570, AST-1908952, AST-1920392, and AST-1911074.
L.M. acknowledges support from a CONICET fellowship and from grant UNRN~PI2018~40B696.
L.G. acknowledges financial support from the Spanish Ministerio de Ciencia e Innovaci\'on (MCIN), the Agencia Estatal de Investigaci\'on (AEI) 10.13039/501100011033, and the European Social Fund (ESF) "Investing in your future" under the 2019 Ram\'on y Cajal program RYC2019-027683-I and the PID2020-115253GA-I00 HOSTFLOWS project, and from Centro Superior de Investigaciones Cient\'ificas (CSIC) under the PIE project 20215AT016. 
C.G. is supported by a VILLUM FONDEN Young Investor Grant (project number 25501).
P.H. acknowledges support by NSF grant AST-856-1715133. 
A.V.F. has received support from the UC Berkeley Miller Institute for Basic Research in Science (where he was a Miller Senior Fellow), the TABASGO foundation, the Christopher R. Redlich Fund, and many individual donors.
F.F acknowledges National Agency of Research and Development' (ANID) Millennium Science Initiative through grant IC12009, awarded to the Millennium Institute of Astrophysics, ANID BASAL grant Center of Mathematical Modelling AFP-170001, ACE210010, FB210005 and ANID BASAL grant center FB210003, and FONDECYT Regular 1200710. 
B.J.S. is supported by NSF grants AST-1908952, AST-1920392 and AST-1911074.
This research has made use of the NASA/IPAC Extragalactic Database (NED) which is operated by the Jet Propulsion Laboratory, California Institute of Technology, under contract with the National Aeronautics and Space Administration.
\\
\textit{Software:} \texttt{NumPy} \citep{numpy2011}, \texttt{matplotlib} \citep{matplotlib}, \texttt{SNID} \citep{blo07d}.
\\
\textit{Facilities:} \textit{Magellan}:Baade (IMACS imaging spectrograph, FourStar wide-field
near-infrared camera, FIRE near-infrared echellette), \textit{Magellan}:Clay (LDSS3 imaging
spectrograph), Swope (SITe3 CCD imager, e2v 4K x 4K CCD imager), du Pont
(Tek5 CCD imager, WFCCD imaging spectrograph, RetroCam near-infrared imager),
Gemini:North (GNIRS near-infrared spectrograph), Gemini:South (FLAMINGOS2), La Silla-QUEST, CRTS, PTF, iPTF, OGLE, ASASSN, PS1, KISS, ISSP, MASTER, SMT.
\end{acknowledgements}

\bibliographystyle{aa}

\bibliography{Reference}

\newpage

\begin{appendix} 

\section{Additional tables}
\begin{table*}
\begin{tiny}
\centering
\caption{Optical standard local sequence photometry (mag) for SN~2007oc.}
\begin{tabular}{ccccccccc}
\hline
Star ID & RA & Dec & $u$ & $g$ & $r$ & $i$ & $B$ & $V$\\  
\hline
\hline
 1 & 22:56:47.08 & $-$36:44:18.6 & 15.075(017) & 13.653(013) & 13.211(003) & 13.073(007) & 14.040(009) & 13.374(012) \\ 
 2 & 22:56:52.97 & $-$36:44:18.3 & 15.626(023) & 13.862(012) & 13.249(003) & 12.990(006) & 14.344(010) & 13.489(012) \\ 
 3 & 22:56:47.09 & $-$36:42:53.5 & 15.423(013) & 14.317(011) & 13.954(009) & 13.817(007) & 14.642(011) & 14.097(016) \\ 
 4 & 22:56:35.13 & $-$36:47:16.2 & 16.801(055) & 15.088(011) & 14.540(012) & 14.363(016) & 15.529(014) & 14.761(012) \\ 
 5 & 22:56:58.08 & $-$36:43:52.0 & 16.902(047) & 15.445(012) & 14.913(014) & 14.713(010) & 15.889(009) & 15.134(017) \\ 
 6 & 22:56:23.35 & $-$36:47:01.0 & 16.834(043) & 15.591(010) & 15.155(021) & 15.010(010) & 15.955(020) & 15.323(017) \\ 
 7 & 22:56:41.16 & $-$36:47:21.7 & 17.755(109) & 15.839(016) & 15.164(013) & 14.912(017) & 16.334(024) & 15.471(013) \\ 
 8 & 22:56:51.63 & $-$36:47:48.7 & 16.972(073) & 15.742(019) & 15.319(014) & 15.175(030) & 16.104(026) & 15.486(018) \\ 
 9 & 22:56:28.73 & $-$36:50:46.8 &  $\cdots$ & 16.036(037) & 15.103(009) & 14.782(009) & 16.649(046) & 15.532(013) \\ 
10 & 22:56:35.97 & $-$36:47:17.4 & 18.481(106) & 16.147(022) & 15.453(018) & 15.213(037) & 16.659(026) & 15.755(023) \\ 
11 & 22:56:25.68 & $-$36:46:31.5 & 17.338(113) & 16.272(027) & 15.872(012) & 15.727(020) & 16.614(057) & 16.041(018) \\ 
12 & 22:56:52.48 & $-$36:46:05.1 & 17.545(191) & 16.221(022) & 15.795(020) & 15.623(006) & 16.629(085) & 15.974(016) \\ 
13 & 22:56:22.79 & $-$36:44:53.6 &  $\cdots$ & 16.913(024) & 15.589(016) & 14.759(012) & 17.638(069) & 16.195(014) \\ 
14 & 22:56:36.60 & $-$36:50:34.3 &  $\cdots$ & 17.521(048) & 16.142(008) & 14.958(015) & 18.234(041) & 16.735(029) \\ 
15 & 22:56:34.23 & $-$36:49:36.4 &  $\cdots$ & 17.809(042) & 17.295(035) & 17.159(028) & 18.160(120) & 17.488(083) \\ 
16 & 22:56:27.46 & $-$36:49:46.7 &  $\cdots$ & 18.179(075) & 16.847(042) & 16.127(016) & 18.772(086) & 17.455(059) \\ 
17 & 22:56:39.93 & $-$36:49:19.4 &  $\cdots$ & 18.576(082) & 17.261(022) & 16.344(023) &  $\cdots$ & 17.861(021) \\ 
18 & 22:56:41.28 & $-$36:43:54.0 &  $\cdots$ & 18.553(096) & 17.261(085) & 16.671(047) & 18.945(090) & 17.893(020) \\ 
19 & 22:56:38.49 & $-$36:45:27.8 &  $\cdots$ & 18.604(112) & 17.435(066) & 16.736(026) &  $\cdots$ & 18.064(094) \\ 
20 & 22:56:44.91 & $-$36:50:01.4 &  $\cdots$ & 18.603(106) & 17.388(027) & 16.179(020) &  $\cdots$ & 17.945(045) \\ 
21 & 22:56:33.44 & $-$36:44:06.8 &  $\cdots$ & 18.410(069) & 17.877(113) & 17.679(114) & 18.807(080) & 18.084(079) \\ 
\hline
\end{tabular}
\tablefoot{Values in brackets are $1\sigma$ statistical uncertainties. (This table and equivalent ones for all other SNe is available in electronic, machine-readable form from the CDS.)}
\label{oploc07oc}
\end{tiny}
\end{table*}

\begin{table*}
\begin{tiny}
\centering
\caption{NIR standard local sequence photometry (mag) for SN~2007oc.}
\begin{tabular}{cccccc}
\hline
Star ID & RA & Dec & $Y$ & $J$ & $H$\\  
\hline
101 & 22:57:05.05 & $-$36:44:33.8 & 13.948(025) & 13.697(036) & 13.358(015)  \\ 
102 & 22:57:08.07 & $-$36:47:50.5 & 14.781(029) &  $\cdots$ &  $\cdots$  \\ 
103 & 22:56:39.92 & $-$36:49:19.5 & 15.174(032) & 14.700(038) & 14.041(034)  \\ 
104 & 22:57:06.41 & $-$36:45:23.9 & 15.267(042) & 14.956(038) & 14.502(058)  \\ 
105 & 22:57:06.47 & $-$36:47:46.4 & 15.571(032) & 15.252(067) & 14.971(044)  \\ 
106 & 22:56:33.45 & $-$36:44:06.8 & 16.936(172) &  $\cdots$ & 16.042(115)  \\ 
107 & 22:57:03.64 & $-$36:48:23.7 & 17.571(172) &  $\cdots$ &  $\cdots$  \\ 
108 & 22:56:30.80 & $-$36:44:07.9 & 17.685(015) &  $\cdots$ &  $\cdots$  \\ 
109 & 22:56:50.32 & $-$36:46:40.0 & 17.792(149) &  $\cdots$ &  $\cdots$  \\ 
111 & 22:56:41.14 & $-$36:47:21.7 & 14.175(019) & 13.841(027) & 13.366(026)  \\ 
112 & 22:56:35.97 & $-$36:47:17.4 & 14.502(018) & 14.154(025) & 13.700(035)  \\ 
113 & 22:56:35.12 & $-$36:47:16.6 & 13.726(019) & 13.450(026) & 13.052(033)  \\ 
114 & 22:56:38.48 & $-$36:45:27.8 & 15.692(032) & 15.280(037) & 14.600(060)  \\ 
115 & 22:56:36.31 & $-$36:45:19.8 & 15.740(029) & 15.261(034) & 14.655(049)  \\ 
116 & 22:56:49.00 & $-$36:47:42.3 & 16.395(098) & 15.999(087) & 15.377(090)  \\ 
117 & 22:56:51.62 & $-$36:47:49.0 & 14.602(034) & 14.337(023) & 13.961(022)  \\ 
118 & 22:56:52.46 & $-$36:46:04.9 & 15.044(036) & 14.758(023) & 14.355(044)  \\ 
119 & 22:56:54.35 & $-$36:46:19.8 & 17.178(113) & 16.695(162) & 16.148(074)  \\ 
120 & 22:56:47.07 & $-$36:44:17.7 & 12.470(019) & 12.232(036) & 11.946(012)  \\ 
121 & 22:56:52.95 & $-$36:44:18.3 & 12.260(023) & 11.952(033) & 11.475(010)  \\ 
122 & 22:57:01.28 & $-$36:46:32.7 & 16.547(084) & 16.193(056) & 15.543(080)  \\ 
123 & 22:56:44.98 & $-$36:48:51.8 & 16.900(068) & 16.485(155) & 15.750(086)  \\ 
124 & 22:56:59.82 & $-$36:44:51.1 & 15.317(031) & 14.827(029) & 14.322(040)  \\ 
125 & 22:57:00.97 & $-$36:44:30.8 & 17.001(087) & 16.734(105) & 16.035(097)  \\ 
126 & 22:56:42.42 & $-$36:45:25.0 & 17.748(065) & 17.168(024) & 16.673(177)  \\ 
127 & 22:56:50.54 & $-$36:46:26.8 & 16.768(059) & 16.261(126) & 15.739(075)  \\ 
128 & 22:56:58.07 & $-$36:43:51.9 & 14.055(023) & 13.767(029) & 13.356(006)  \\ 
129 & 22:56:34.22 & $-$36:49:36.6 & 16.528(038) & 16.161(043) & 15.786(135)  \\ 
130 & 22:56:48.97 & $-$36:49:45.0 & 17.770(037) &  $\cdots$ &  $\cdots$  \\ 
\hline
\end{tabular}
\tablefoot{Values in brackets are $1\sigma$ statistical uncertainties. (This table and equivalent ones for all other SNe is available in electronic, machine-readable form from the CDS.)}
\label{nloc07oc}
\end{tiny}
\end{table*}

\begin{table*}
\begin{tiny}
\centering
\caption{Natural system optical photometry (mag) of SN~2007oc.}  
\begin{tabular}{cccccccc}
\hline
JD & $u$ & $g$ & $r$ & $i$ & $B$ & $V$ & telesope+instrument\\  
\hline
2454409.51 & 16.353(017) & 14.931(004) & 14.411(004) & 14.531(008) & 15.309(008) & 14.652(006) & Swope+SITe3\\ 
2454413.55 & 16.756(017) & 15.117(004) & 14.491(004) & 14.584(004) & 15.527(006) & 14.772(005) & Swope+SITe3\\ 
2454417.52 & 17.183(028) & 15.277(004) & 14.560(004) & 14.638(005) & 15.741(006) & 14.890(004) & Swope+SITe3\\ 
2454418.60 & 17.310(029) & 15.308(008) & 14.573(007) & 14.661(007) & 15.792(008) & 14.911(006) & Swope+SITe3\\ 
2454421.51 & 17.475(037) & 15.410(004) & 14.628(004) & 14.692(004) & 15.901(007) & 14.979(005) & Swope+SITe3\\ 
2454425.59 &  $\cdots$ & 15.508(005) & 14.680(004) & 14.749(006) & 16.039(009) & 15.060(006)   & Swope+SITe3\\ 
2454435.53 & 18.239(052) & 15.724(004) & 14.806(004) & 14.867(004) & 16.312(008) & 15.228(004) & Swope+SITe3\\ 
2454443.56 & 18.546(035) & 15.863(005) & 14.906(004) & 14.969(005) & 16.490(009) & 15.347(005) & Swope+SITe3\\ 
2454445.58 & 18.686(041) & 15.907(006) & 14.935(004) & 14.994(007) & 16.545(009) & 15.383(006) & Swope+SITe3\\ 
2454449.53 & 18.746(105) & 15.981(005) & 14.998(004) & 15.056(005) & 16.628(009) & 15.456(004) & Swope+SITe3\\ 
2454453.55 &  $\cdots$ & 16.103(006) & 15.070(005) & 15.128(007) & 16.761(010) & 15.555(005)   & Swope+SITe3\\ 
2454455.52 & 19.067(138) & 16.144(009) & 15.112(006) & 15.162(008) & 16.838(012) & 15.605(006) & Swope+SITe3\\ 
2454456.52 &  $\cdots$ & 16.192(009) & 15.127(007) & 15.196(008) & 16.874(011) & 15.623(009)   & Swope+SITe3\\ 
2454462.53 & 19.386(115) & 16.453(006) & 15.303(004) & 15.357(004) & 17.128(021) & 15.845(009) & Swope+SITe3\\ 
2454468.53 & 19.652(191) & 16.869(024) & 15.567(008) & 15.646(011) & 17.573(018) & 16.220(008) & Swope+SITe3\\ 
2454470.53 & 19.892(198) & 17.063(022) & 15.729(008) & 15.741(009) & 17.768(020) & 16.415(008) & Swope+SITe3\\ 
\hline
\end{tabular}
\tablefoot{The last column gives the telescope+instrument combination used to obtain the photometry. Values in brackets are $1\sigma$ statistical uncertainties.  
 (This table and equivalent ones for all other SNe is available in electronic, machine-readable form from the CDS.)}
\label{opphot07oc}
\end{tiny}
\end{table*}

\begin{table*}
\begin{tiny}
\centering
\caption{Natural system NIR photometry (mag) of SN~2007oc.} 
\begin{tabular}{ccccc}
\hline
JD & $Y$ & $J$ & $H$ & telescope+instrument\\  
\hline
2454408.59 & 14.055(010) & 13.838(011) & 13.764(009) &Swope+RetroCam\\ 
2454410.52 & 14.077(008) & 13.881(010) & 13.766(011) &Swope+RetroCam \\ 
2454415.64 & 14.126(009) & 13.893(008) & 13.799(006) &Swope+RetroCam\\ 
2454425.63 & 14.245(008) & 13.984(009) & 13.839(012) &duPont+WIRC\\ 
2454426.61 & 14.258(008) & 13.982(009) & 13.837(012) &duPont+WIRC\\ 
2454427.62 & 14.296(008) & 14.013(009) & 13.848(012) &duPont+WIRC\\ 
2454431.57 & 14.277(016) & 14.005(009) & 13.904(007) &Swope+RetroCam\\ 
2454439.57 & 14.393(006) & 14.112(008) & 13.963(008) &Swope+RetroCam\\ 
2454442.60 & 14.409(008) & 14.138(009) & 14.032(013) &Swope+RetroCam\\ 
2454446.60 & 14.486(007) & 14.194(008) & 14.027(013) &Swope+RetroCam\\ 
2454452.54 & 14.577(007) & 14.277(008) & 14.131(008) &Swope+RetroCam\\ 
2454456.54 & 14.655(008) & 14.346(009) & 14.155(012) &duPont+WIRC\\ 
2454465.53 & 14.898(010) &  $\cdots$ &  $\cdots$     &Swope+RetroCam\\ 
2454468.55 & 15.014(008) & 14.711(009) & 14.483(012) &duPont+WIRC\\ 
\hline
\end{tabular}
\tablefoot{The last column gives the telescope+instrument combination used to obtain the photometry. Values in brackets are $1\sigma$ statistical uncertainties. (This table and equivalent ones for all other SNe is available in electronic, machine-readable form from the CDS.)}
\label{nphot07oc}
\end{tiny}
\end{table*}

\section{Updated data release for SNe~IIn and SN~1987A-like events from CSP-I}
Photometry (and spectroscopy) for five SNe~IIn were released by \cite{tad13a} and for two SNe~IIn by \cite{str12}. The full sample of CSP-I SNe~IIn consists of SNe\,2005ip,\,2005kj,\,2006aa,\,2006bo,\,2006jd,\,2006qq,\,2008fq. In addition, data were released for two SN~1987A-like SNe~II by \cite{tad12a}: SNe~2006V\,and\,2006au.
During the subsequent years a number of small iterative changes were made to the CSP reduction and calibration procedures meaning that new updated photometry for these SNe is available in electronic, machine-readable format from the CDS, and can also be downloaded from the CSP website: \url{https://csp.obs.carnegiescience.edu/data}. These changes are very minor, and thus do not affect the results or conclusions published by the above works.

\end{appendix}
\end{document}